\documentclass[a4paper,fleqn]{cas-dc}

\usepackage[square,numbers,sort&compress]{natbib}

\usepackage{csquotes}
\usepackage{xcolor}
\usepackage[colorinlistoftodos,prependcaption]{todonotes}
\usepackage[inline]{enumitem}
\usepackage{xspace}
\usepackage{subcaption}
\usepackage{longtable}
\usepackage{afterpage}
\usepackage{amssymb}
\usepackage[nohyperlinks,printonlyused]{acronym}
\usepackage{amsmath}
\usepackage{pifont}
\newcommand{\cmark}{\ding{51}}%
\newcommand{\xmark}{\text{\ding{55}}}

\def\tsc#1{\csdef{#1}{\textsc{\lowercase{#1}}\xspace}}
\tsc{WGM}
\tsc{QE}
\tsc{EP}
\tsc{PMS}
\tsc{BEC}
\tsc{DE}

\usepackage{float}
 \usepackage[usenames,dvipsnames]{pstricks}
\usepackage{pstricks-add}
 \usepackage{epsfig}
 \usepackage{pst-grad} 
 \usepackage{pst-plot} 
 \usepackage[space]{grffile} 
 \usepackage{etoolbox} 
 \makeatletter 
 \patchcmd\Gread@eps{\@inputcheck#1 }{\@inputcheck"#1"\relax}{}{}
 \makeatother
\usepackage{tikz}
\usetikzlibrary{mindmap,trees}
\usepackage{verbatim}
\usepackage{multicol}
\usepackage[ruled,lined,noresetcount ]{algorithm2e}
\usepackage{algpseudocode}
\usepackage{mathtools}

\DeclarePairedDelimiter\norm{\lVert}{\rVert}

\hypersetup{
   colorlinks=true,
}
\begin{document}
\definecolor{ElsevierBlueish}{RGB}{0, 128, 172}
\hypersetup{
    linkcolor=ElsevierBlueish,
    citecolor=ElsevierBlueish,
    urlcolor=ElsevierBlueish
}

\let\WriteBookmarks\relax
\def\floatpagepagefraction{1}
\def\textpagefraction{.001}


\shorttitle{Deep transfer learning for intrusion detection in industrial control networks}

\shortauthors{Kheddar, Himeur, and Awad}



\title [mode = title]{Deep transfer learning for intrusion detection in industrial control networks: A comprehensive review}

\vskip2mm

\author[1]{Hamza Kheddar\corref{cor1}}
\ead{kheddar.hamza@univ-medea.dz}
\cormark[1]
\credit{Conceptualization; Methodology; Data Curation; Resources; Investigation; Visualization; Writing original draft; Writing, review, and editing}
\author[2]{Yassine Himeur}
\ead{yhimeur@ud.ac.ae}
\credit{Conceptualization; Methodology; Resources; Investigation; Writing original draft; Writing, review, and editing}

\author[3,4]{Ali Ismail Awad}
[orcid=0000-0002-3800-0757]
\ead{ali.awad@uaeu.ac.ae}
\credit{Conceptualization; Methodology; Resources; Visualization; Investigation; Writing original draft; Writing, review, and editing}
\address[1]{LSEA Laboratory, Electrical Engineering Department, University of Medea, 26000, Algeria}

\address[2]{College of Engineering and Information Technology, University of Dubai, Dubai, UAE}

\address[3]{College of Information Technology, United Arab Emirates University, Al Ain P.O. Box 15551, United Arab Emirates}

\address[4]{Faculty of Engineering, Al-Azhar University, Qena P.O. Box 83513, Egypt}

\tnotetext[1]{Corresponding author.}

\begin{abstract}
Globally, the external internet is increasingly being connected to industrial control systems. As a result, there is an immediate need to protect these networks from a variety of threats. The key infrastructure of industrial activity can be protected from harm using an intrusion detection system (IDS), a preventive mechanism that seeks to recognize new kinds of dangerous threats and hostile activities. This review examines the most recent artificial-intelligence techniques that are used to create IDSs in many kinds of industrial control networks, with a particular emphasis on IDS-based deep transfer learning (DTL). DTL can be seen as a type of information-fusion approach that merges and/or adapts knowledge from multiple domains to enhance the performance of a target task, particularly when labeled data in the target domain is scarce. Publications issued after 2015 were considered. These selected publications were divided into three categories: DTL-only and IDS-only works are examined in the introduction and background section, and DTL-based IDS papers are considered in the core section of this review. By reading this review paper, researchers will be able to gain a better grasp of the current state of DTL approaches used in IDSs in many different types of network. Other useful information, such as the datasets used, the type of DTL employed, the pre-trained network, IDS techniques, the evaluation metrics including accuracy/F-score and false-alarm rate, and the improvements gained, are also covered. The algorithms and methods used in several studies are presented, and the principles of DTL-based IDS subcategories are presented to the reader and illustrated deeply and clearly.
\end{abstract}

\begin{keywords}
Intrusion detection system \sep
Industrial control network \sep
Deep transfer learning \sep
Fine-tuning \sep
Domain adaptation \sep
Cybersecurity \sep


\end{keywords}

\maketitle

{
}

\begin{table*}[]
    \centering
{\small \section*{Abbreviations}}
\begin{multicols}{3}
\footnotesize
\begin{acronym}[AWGN] 
\acro{5G}{fifth generation }
\acro{6G}{sixth generation }
\acro{AA}{anomaly adapters}
\acro{ADR}{attack detection rate}
\acro{AE}{Auto-Encoder}
\acro{AI}{artificial intelligence}
\acro{ANN}{artificial neural network}
\acro{AP}{access point}
\acro{AUC}{area under the ROC curve}
\acro{APT}{advanced persistent threat}
\acro{ARTL}{adaptation regularization transfer learning}
\acro{AWID}{aegean wiFi intrusion dataset}
\acro{CAN}{control area network}
\acro{CAN}{controller area network}
\acro{CC}{cloud computing }
\acro{CNN}{convolutional neural network}
\acro{ConvLSTM}{convolutional long-short-term memory}
\acro{CORAL}{correlation alignmen}
\acro{CTF}{capture the flag}
\acro{DA}{domain adaptation}
\acro{DAE}{deep autoencoder}
\acro{DANN}{domain-adversarial neural network}
\acro{DBN}{deep belief network}
\acro{DL}{deep learning}
\acro{DM}{data mining}
\acro{DNN}{deep neural network}
\acro{DoS}{denial-of-service}
\acro{DDoS}{distributed denial of service}
\acro{DRL}{deep reinforcement learning}
\acro{DST-TL}{self-taught based transfer learning}
\acro{DTL}{deep transfer learning}
\acro{EEDTL}{extended equilibrium deep transfer learning}
\acro{EEO}{extended equilibrium optimizer}
\acro{EL}{ensemble learning}
\acro{eMBB}{enhanced mobile broadband}
\acro{ERP}{enterprise resource planning}
\acro{FAR}{false alarm rate}
\acro{FDIA}{false data injection attack}
\acro{FL}{federated learning}
\acro{FTP}{file transfer protocol}
\acro{FTL}{federated transfer learning}
\acro{FN}{false negative}
\acro{FP}{false positive}
\acro{GAN}{generative adversarial network}
\acro{GNN}{graph neural network}
\acro{HIDS}{host intrusion detection system}
\acro{HMM}{hidden markov model}
\acro{HPC}{high performance computing}
\acro{HTTP}{hypertext transfer protocol}
\acro{ICN}{industrial control network}
\acro{ICS}{industrial control system}
\acro{IDS}{intrusion detection system}
\acro{IGMM}{infinite Gaussian mixture model}
\acro{ILSVRC}{imageNet large-scale visual recognition challenge}
\acro{IoMT}{internet of medical things}
\acro{IoT}{internet of things}
\acro{IIoT}{industrial internet of things}
\acro{IoU}{intersection over union}
\acro{IoV}{internet-of-vehicles}
\acro{IP}{internet protocol}
\acro{KNN}{$k$-nearest neighbors}
\acro{LSTM}{long-short-term-memory}
\acro{MES}{manufacturing execution systems}
\acro{MITM}{man-in-the-middle}
\acro{ML}{machine learning}
\acro{MLP}{multilayer perceptrons}
\acro{MMD}{maximum mean discrepancy}
\acro{mMTC}{massive machine type communication}
\acro{MSA}{multi-stage network attacks}
\acro{MTL}{multi-task learning}
\acro{NFV}{ network function virtualization}
\acro{NIDS}{network intrusion detection system}
\acro{NLP}{natural language processing}
\acro{OS}{operating system}
\acro{PAD}{presentation attack detection}
\acro{PCA}{principal component analysis}
\acro{PN}{physical node}
\acro{PU}{primary user}
\acro{PUE}{primary user emulation}
\acro{QoS}{quality-of-service}
\acro{R2L}{root to local }
\acro{RFID}{radio frequency identification}
\acro{RNN}{recurrent neural network}
\acro{ROC}{receiver operating characteristics }
\acro{ROP}{return-oriented programming}
\acro{RR}{recognition rate}
\acro{RSS}{received signal strength}
\acro{RVAE}{recurrent variation autoencoder}
\acro{SCADA}{supervisory control and data acquisition}
\acro{SD}{source domain}
\acro{SDN}{software-defined network}
\acro{seq2seq}{sequence-to-sequence }
\acro{SM}{source model}
\acro{SOA}{sandpiper optimization algorithm}
\acro{SOTA}{state-of-the-art}
\acro{ST}{source task}
\acro{SU}{secondary user}
\acro{SVM}{support vector machine}
\acro{SWaT}{secure water treatment}
\acro{TCA}{transfer component analysis}
\acro{TD}{target domain}
\acro{TL}{transfer learning}
\acro{TM}{target model}
\acro{TN}{true negative}
\acro{TP}{true positive}
\acro{TT}{target task}
\acro{U2R}{user to root}
\acro{uRLLC}{ultra-reliable low latency communication}
\acro{WADI}{water distribution}
\acro{WLAN}{wireless local area network}
\acro{YOLO}{you only look once}
\end{acronym}

\end{multicols}
\end{table*}

\section{Introduction}
\label{sec1}
The potential of the \ac{IoT} to shape the future and make people's lives easier, safer, and more productive has led to numerous applications, hardware, and ecosystems being developed to support this technology \citep{himeur2022ai,sayed2022artificial}. The number of connected \ac{IoT} devices has become huge, and they have become more diverse to include various areas of industry. Making full use of this new technology requires a network infrastructure that takes several years to develop and even longer to implement \citep{himeur2022two}. The open-source community, \ac{IoT} provider companies, and high-technology government institutions have taken the lead in creating development infrastructure by producing software and hardware designs to help developers explore \ac{IoT} devices and applications \citep{atalla2023iot}. Furthermore, \ac{5G} mobile networks---which seek to provide several types of service, including massive machine-type communication, \ac{uRLLC}, and enhanced mobile broadband---make use of already-existing technologies like software-defined networks and network function virtualization, and they contribute significantly to spreading the use of \ac{IoT} devices \citep{adou2022modeling,kheddar2022all}. They thus provide a wide range of services and applications, including augmented reality \citep{zhang2022artificial}, e-healthcare \citep{zhao2022secure}, smart cities and transportation systems such as the \ac{IoV} \citep{yu2022construction}, and smart farming \citep{ullah2022optimization}.

The continuously increasing use of networks of all kinds has raised numerous security issues on internet and computer systems, and it can lead to breaching of the objectives of information security, such as, the \textit{\textbf{integrity}} of exchanged data, the \textit{\textbf{confidentiality}} of information sources, and the \textit{\textbf{availability}} of the services \citep{himeur2022latest}. The degree of threat varies from one network to another; for example, wireless-based networks are more vulnerable to attacks than wired networks, and differences in protocols and standards used on devices connected to the same network can create many access points for attack. Currently, introducing a small delay into \ac{uRLLC} is considered as an attack attempt, as investigated in a previous report \citep{gallenmuller20205g}. Therefore, cybercriminals are constantly looking for any weaknesses in a network to exploit \ac{IoT} devices; they consider these to be easy gateways to infiltrate information-system networks. Since \ac{IIoT} networks have expanded to include more industries and larger systems \citep{onate2023analysis}, malicious attacks have become increasingly sophisticated and prevalent \citep{himeur2022recent}.

An intrusion is an endeavor to bypass a computer's or a network's security measures. For example, an eavesdropper may attack a network to send malicious packets to a host system in an attempt to steal or modify sensitive data \citep{alshamrani2019survey,himeur2022blockchain}. A system for scrutinizing and evaluating network or computer-system events for indications of attacks is known as an \ac{IDS}.

\Acp{ICS}, with their diverse architectures \citep{kaiser2023review}, are used to control and monitor physical processes in industrial environments such as manufacturing plants that rely on \ac{IoT} devices and sensors, smart grids, and intelligent transportation systems such as \ac{IoV}. As these \acp{ICS} are increasingly connected to the internet, they have become more vulnerable to cyberattacks. Therefore, \acp{IDS} have become crucial for the protection of \acp{ICS} against cyber threats. It is worth mentioning that the \ac{IDS} deployed in an \ac{ICS} must be tailored to meet the specific security requirements and limitations of the industrial environment. Unlike a traditional \ac{IDS}, it should not disrupt the operational process of the \ac{ICS}, and it is critical for maintaining the security and integrity of industrial processes in the face of growing cybersecurity threats.

\ac{IDS} software has been frequently developed using \ac{AI} algorithms such as \ac{ML} and \ac{DM} techniques \citep{umer2022machine,jiang2019machine,liu2020web} because of their incredible ability to recognize intrusions. During the training phase of these techniques, an \ac{ML} model is trained using datasets. This includes two labeled sample classes: the normal and attacking network packets. However, because of the enormous size and complexity of the datasets, the majority of \ac{ML} and \ac{DM} techniques cannot be effectively used for intrusion detection. The tremendous processing time taken by these algorithms to identify intrusions makes their deployment in real-time situations more challenging. This is because network data contain gigantic feature vectors that must be assessed by an \ac{IDS}. For this reason, \ac{DL} algorithms have been employed by several researchers \citep{park2018anomaly,guizani2020network,elsayed2021novel,kao2022novel} so that the significant features of the vectors can be generated and selected automatically rather than being determined manually.

The most important and difficult problem encountered by \ac{DL} is data scarcity, and existing \ac{IDS} models are application specific. For example, when an intrusion technique is new or seldom occurs, an \ac{IDS} will not be sufficiently trained to detect this attack due to a lack of data. In other words, \ac{DL} models cannot perform well if (i)~small training datasets are used or (ii)~there is a discrepancy or data-distribution inconsistency not only \textit{between} the training and test data but also \textit{in} the training-data distribution.

From the above explanation, we can conclude that the quantity and quality of features are important for improving classification, since they will help a \ac{DL} model comprehend their significance and correlations. If relatively few features are used, the quality of classification will decrease, and the \textit{overfitting} phenomenon will occur; if too many features are used, generalization will be lost, and the \textit{underfitting} phenomenon will arise. To solve these problems, which are caused mainly by data scarcity and inconsistency, \ac{DTL} was invented. This is based on the principle of feeding knowledge about a \ac{TM} to a pre-trained \ac{SM}, so that the \ac{TM} starts with patterns gained by completing a task related to an \ac{SM} rather than starting anew. Multi-task learning, \ac{DA}, multiple and/or cross modalities, and using multiple datasets, are all \ac{DTL} techniques, and they can be viewed as ways of \textit{\textbf{fusing information}} from multiple sources to improve the overall performance of the model. By leveraging knowledge gained from one source to another, \ac{DTL} allows for more effective information fusion and can lead to better results than training models from scratch. The advantages offered by \ac{DTL} have drawn the attention of researchers starting to develop \ac{DTL}-based \ac{IDS} models to solve many problems in various applications, such as return-oriented programming, payload detection, \ac{WLAN} intrusion detection \citep{zhou2020indoor}, attacks on smart grids \citep{xu2021detecting}, and \ac{DoS} attacks.

In the context of networking, the use of an \ac{IDS} is paramount for automatically detecting and preventing malicious activities. These systems gather and analyze network traffic, security logs, and audit data from key nodes in the network to identify potential security breaches. As a crucial tool for ensuring the security of a network such as an \ac{ICN}, \acp{IDS} have attracted significant interest from industry and academia, particularly for \acp{ICS}. Consequently, numerous \acp{IDS} have been developed. This work aims to provide a comprehensive review of the most recent \ac{IDS} methods that use \ac{DTL} algorithms. \ac{DTL} is an innovative and efficient approach that can effectively improve the ability of \acp{IDS} to detect attacks and intrusions that traditional methods may miss. By examining the latest \ac{TL}-based \ac{IDS} methods, this review intends to contribute to the ongoing research efforts toward enhancing the security of telecommunication networks.



\subsection{Related reviews and our contributions}
There have been a few reviews on the use of \ac{ML} and \ac{DL} in \ac{IDS} applications in \acp{ICN} \citep{jiang2019machine,tama2022systematic,asharf2020review,hu2018survey}. These reviews, however, focus strongly either on traditional \ac{ML} strategies, such as ensemble learning, \ac{SVM}, and \ac{DRL}, or conventional \ac{DL}, such as recurrent neural networks, \acp{CNN}, deep \acp{AE}, and \acp{GAN}. As an alternative, recent studies have concentrated on the use of cutting-edge \ac{DL} algorithms---such as \ac{IDS}-based \ac{FL} \citep{al2020federated}, transformer-based approaches for \ac{IDS} \citep{wu2022rtids}, or graph-neural-network-based \acp{IDS} \citep{caville2022anomal}---to develop an efficient \ac{IDS} for an \ac{ICN} or any other type of communication network. There have also been some overview studies on \ac{DTL} approaches: one only presented a limited discussion of \ac{IDS}-based \ac{DTL} in a single subsection \citep{nguyen2022transfer}, and another considered their exclusive application to a wireless network \citep{vu2020deep}.

Notwithstanding the works noted above, to the best of our knowledge, there has not yet been a thorough survey of the applications of \ac{DTL}-based \acp{IDS}. Table~\ref{tab:1} presents a summary comparison of our review against the existing \ac{DTL}-based surveys. It is clear that this review addresses all of the relevant fields, including \ac{DTL} and \ac{ICS}/\ac{IDS} background, applications of \ac{DTL}-based \acp{IDS}, \ac{FTL}-based \acp{IDS}, pre-trained models, dataset descriptions, \ac{IDS} challenges, and future directions. In contrast, almost all of the other works have either left these fields unaddressed or have addressed them only partially. Therefore, this review can be considered comprehensive in its coverage of the various fields related to \ac{DTL} for \ac{IDS} applications.

\begin{table*}[t!]
\centering
\caption{Comparison of this review against other existing \ac{IDS} reviews and surveys. Tick marks (\cmark) indicate that a particular field has been considered, while cross marks (\xmark) indicate that a field has been left unaddressed. The symbol ($\sim$) indicates that the most critical concerns of a field have not been addressed.}
\label{tab:1}
\resizebox{\linewidth}{!}{%
\begin{tabular}{llccccccc}
\hline
{\scriptsize Ref.} & {\scriptsize Description} & {\scriptsize \ac{DTL}} &
{\scriptsize \ac{ICS} and \ac{IDS} } & {\scriptsize Applications of } & {\scriptsize \ac{FTL}-based } & {\scriptsize Pre-trained } & {\scriptsize Dataset }&{\scriptsize \ac{IDS} challenges and } \\
&  & {\scriptsize Background} & {\scriptsize Background} & {\scriptsize \ac{DTL}-based \ac{IDS} } & {\scriptsize \ac{IDS}} & {\scriptsize models } & {\scriptsize descriptions } & {\scriptsize
 future directions } \\ \hline
{\scriptsize \citep{lu2015transfer}} & {\scriptsize \ac{DTL} for intelligence comput.} & {\scriptsize \xmark} & {\scriptsize \xmark} & {\scriptsize \xmark} & {\scriptsize \xmark} & {\scriptsize \xmark} & {\scriptsize \xmark}& {\scriptsize \xmark}
  \\
  {\scriptsize \citep{agarwal2021transfer}} &{\scriptsize \ac{DTL} clustering and apps.}&{\scriptsize \xmark}&{\scriptsize \xmark}&{\scriptsize \xmark}&{\scriptsize \xmark}&{\scriptsize \xmark}&{\scriptsize \xmark}&{\scriptsize \xmark}\\
{\scriptsize \citep{vu2020deep}} & {\scriptsize \ac{DTL} for \ac{IoT} attacks} & {\scriptsize \xmark} & {\scriptsize \xmark} & {\scriptsize \xmark}
& {\scriptsize \xmark} & {\scriptsize \xmark} & {\scriptsize \xmark} &
{\scriptsize \xmark} \\
{\scriptsize \citep{himeur2023video}} & {\scriptsize \ac{DTL} for video surveillance} & {\scriptsize \cmark} & {\scriptsize \xmark} & {\scriptsize \xmark} & {\scriptsize \xmark} & {\scriptsize \cmark}
& {\scriptsize \xmark} & {\scriptsize $\sim$} \\
{\scriptsize \citep{nguyen2022transfer} } & {\scriptsize \ac{DTL} for wireless network} &
{\scriptsize \cmark} & {\scriptsize \xmark} & {\scriptsize $\sim$} &
{\scriptsize \xmark} & {\scriptsize \xmark} & {\scriptsize \xmark} &
{\scriptsize \xmark} \\
{\scriptsize Ours } & {\scriptsize \ac{DTL} for \ac{IDS} applications} & {\scriptsize \cmark} & {\scriptsize \cmark} & {\scriptsize \cmark} & {\scriptsize \cmark}
& {\scriptsize \cmark} & {\scriptsize \cmark} &
{\scriptsize \cmark} \\
\hline
\end{tabular}
}
\end{table*}

Considering the rapid growth of wireless communication technologies and the growing demands of mobile users for high quality of service, interoperability, robustness, privacy, and security, \ac{DTL}-based \acp{IDS} must act quickly to overcome the current limitations of traditional \acp{IDS}, particularly in \ac{5G} networks and beyond. As a consequence, it is anticipated that this review will fill a gap in the literature and make a substantial contribution to the advancement of future intelligent wired and wireless networks. This review stands out due to its diverse range of information, covering various aspects of \ac{DTL} and intrusion detection, including models, datasets, and attacks. Additionally, it provides comprehensive insights into the \ac{ICN} context, which is an area that requires further exploration and investigation. Overall, this review offers a unique and valuable perspective on \ac{DTL}-based \acp{IDS} applications in \acp{ICN}. The contributions of this survey can be summarized by the following points:
\addtolength{\leftmargini}{-0.5cm}
\begin{itemize}
    \item The review discusses the \ac{ICN} layers and covers their security needs. It also provides a thorough taxonomy of the different \ac{IDS} techniques used to secure \acp{ICN} in various fields.

    \item The review provides a comprehensive taxonomy of the different \ac{DTL} models that can be used for different \ac{IDS} techniques, including inductive, transductive, and adversarial \ac{DTL}.

    \item A comprehensive taxonomy of different \ac{IDS} techniques --- including signature-based, anomaly-based, and specific-ation-based \ac{IDS}--- is also provided, along with a review of \ac{IDS}-based \ac{DTL} applications, linked to the design decisions, pros, and cons of the suggested schemes, which are covered minutely.

    \item Finally, the review emphasizes the existing research challenges and introduces potential future paths for the \ac{IDS}-based \ac{DTL} field of study.
\end{itemize}

\begin{figure*}[!t]
    \centering
    \includegraphics[width=0.9\textwidth]{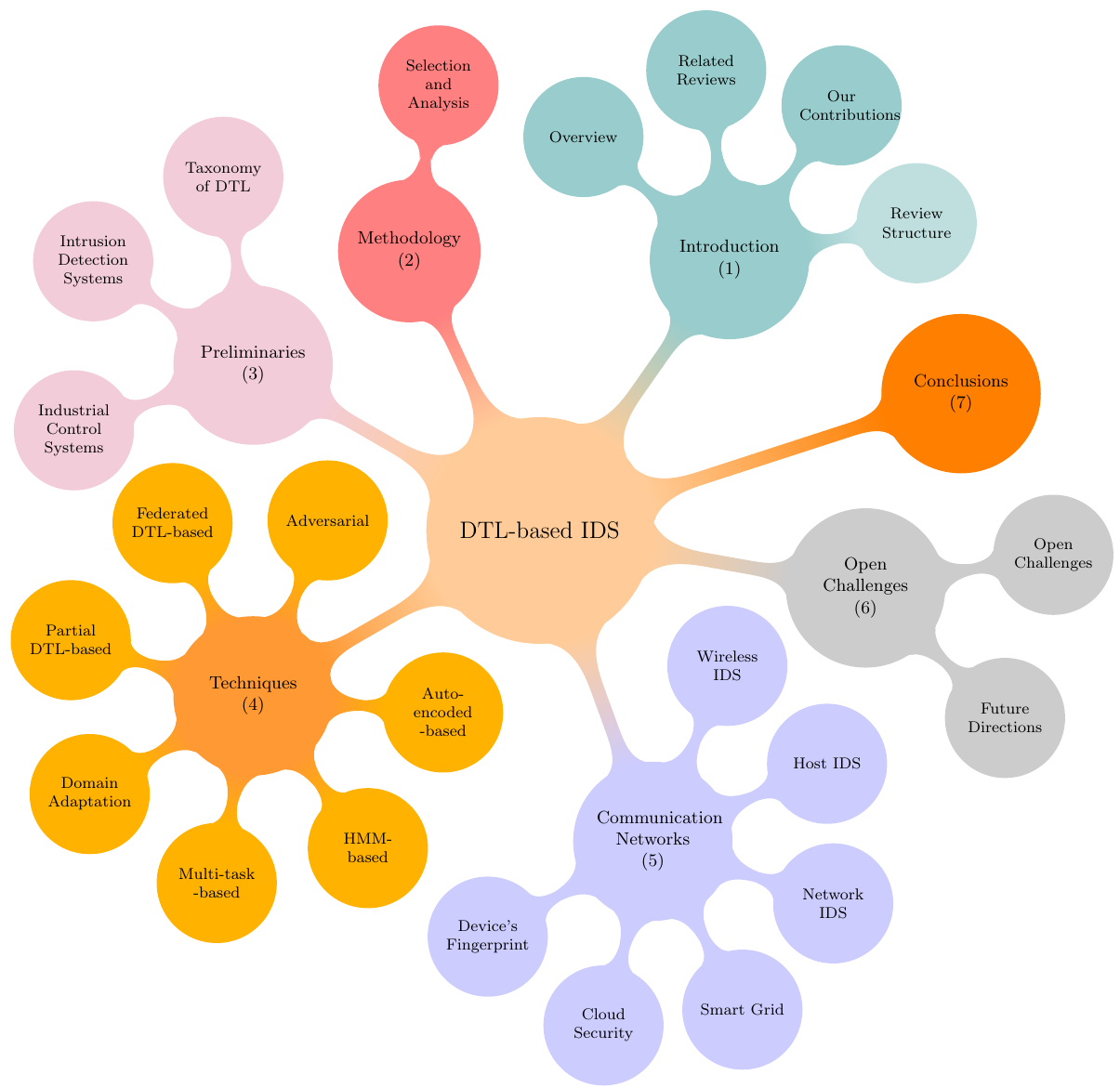}
    \caption{Mind map showing the main sections and key concepts covered in this review.}
    \label{mindMap}
\end{figure*}








\subsection{Paper structure}
The remainder of this review is organized as follows. Section~\ref{sec2} presents the methodology used for the review. Section~\ref{sec3} introduces a taxonomy of the existing \ac{IDS} techniques. Sections~\ref{sec4} and \ref{sec5} summarize the existing \ac{DTL} techniques and then present the categories of \ac{IDS} that are based on \ac{DTL} and their applications. Section~\ref{sec6} highlights the most important open challenges facing \ac{IDS}-based \ac{DTL} schemes and possible future contributions of \ac{IDS}-based \ac{DTL}. Section~\ref{sec7} presents the conclusions of this survey. Fig.~\ref{mindMap} provides a detailed mind map for this review and lists the key titles introduced in the study.

\begin{figure*}[!t]
\begin{center}
\includegraphics[scale=0.7]{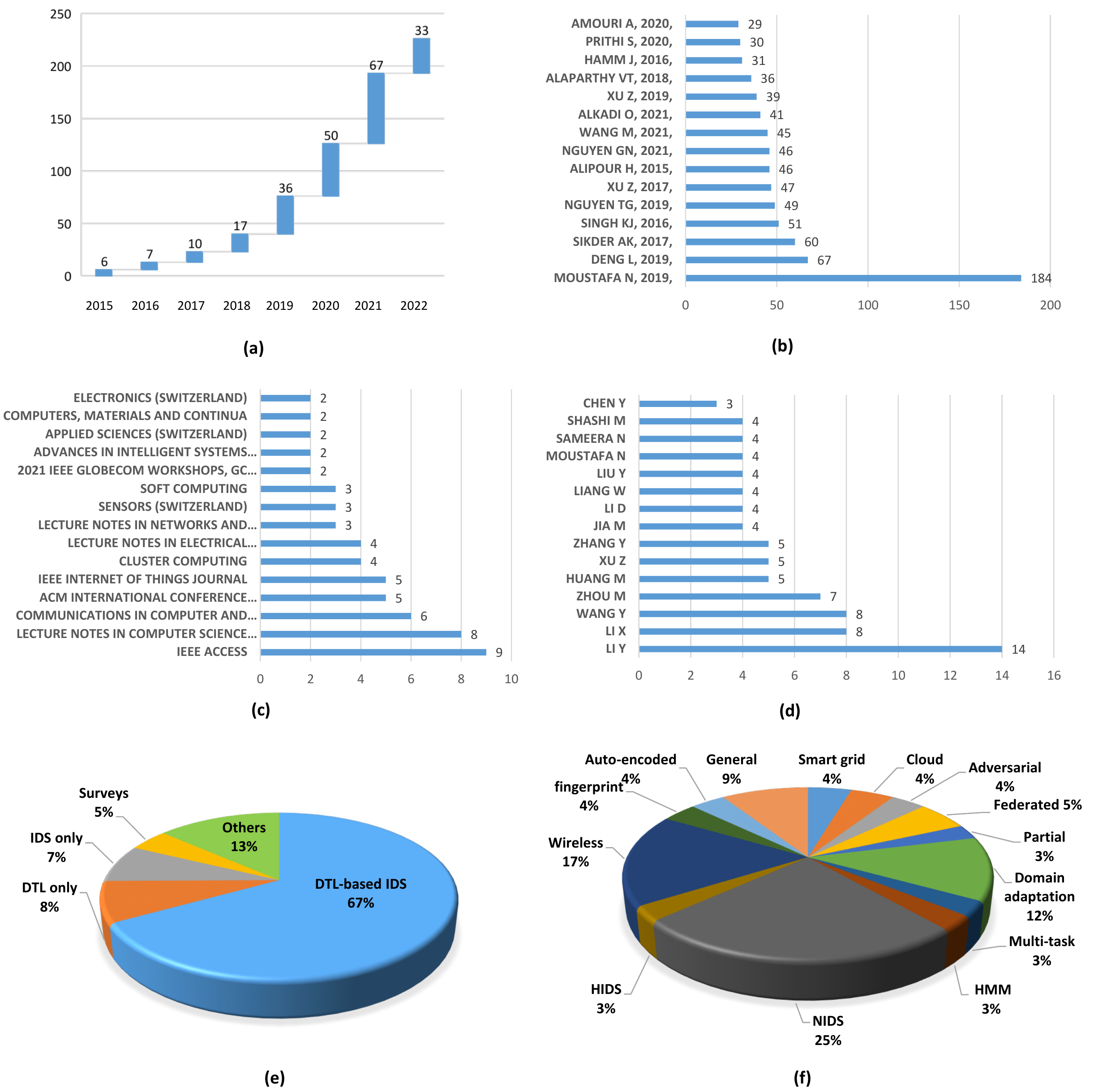}\\
\end{center}
\caption{Bibliometric analysis in terms of statistics on: (a)~Number of articles involved in this review from each year; (b)~the top-15 most-cited papers with their authors and publication years; (c)~the top-15 most-relevant papers' sources; (d)~the top-15 most active authors; (e)~the percentage of papers for \ac{DTL}-based \ac{IDS}, \ac{IDS} only, \ac{DTL} only, surveys, and others; and (f)~percentage of papers discussed in each field of \ac{DTL}-based \ac{IDS}.}
\label{statistics}
\end{figure*}

\section{Review methodology}
\label{sec2}
This section describes the process that was undertaken to carry out the review. The paper-selection strategy is covered first, a subsection on literature analysis follows, and the section finishes with a motivational explanation of the \ac{IDS}-based \ac{DTL} topic. Most of the papers in this review fall into three different categories. The first is related to different classes of \ac{IDS}, which are used to construct a thorough classification of \ac{IDS} techniques. The second category of papers focuses on \ac{DTL} techniques, and these are used to construct a comprehensive taxonomy of \ac{DTL}. The third category of papers concentrates on \ac{IDS} using \ac{TL} techniques, which makes up the majority of all papers considered in this review. Reviewing and summarizing these papers establishes the most significant and efficient techniques that empower \ac{IDS} in industry.

\subsection{Selection and analysis of the literature}
To compile a thorough list of study references examining \ac{IDS} based on \ac{DTL}, the renowned journal databases indexed in Scopus and Web of Science, including IEEE Xplore, ACM Digital Library, Science Direct, and SpringerLink, were searched.

These publications were determined based on the following four standards. (i)~Modularity: only the most typical methods and implementations were kept. (ii)~Coverage: reported a new (not repeated) or specific application domain. (iii)~Influence: published in high-quality journals with high impact factors or in book chapters or conference proceedings but with a high number of citations. (iv)~Modernity: only papers published from 2016 were included. The initial search was conducted using \ac{TL}-related keywords for intrusion detection. The obtained papers were divided into five groups using ``topic clustering'' based on the keywords associated with the application field: (a)~anomaly detection and transfer learning, (b)~intrusion detection and transfer learning, (c)~signature detection and transfer learning, (d)~industrial control system security and transfer learning, and (e)~application of transfer learning. Fig.~\ref{statistics} presents the statistics obtained after applying our selection criteria.

\section{Preliminaries}
\label{sec3}
\subsection{Industrial control systems}
\begin{figure*}[!t]
    \centering
    \includegraphics[scale=1]{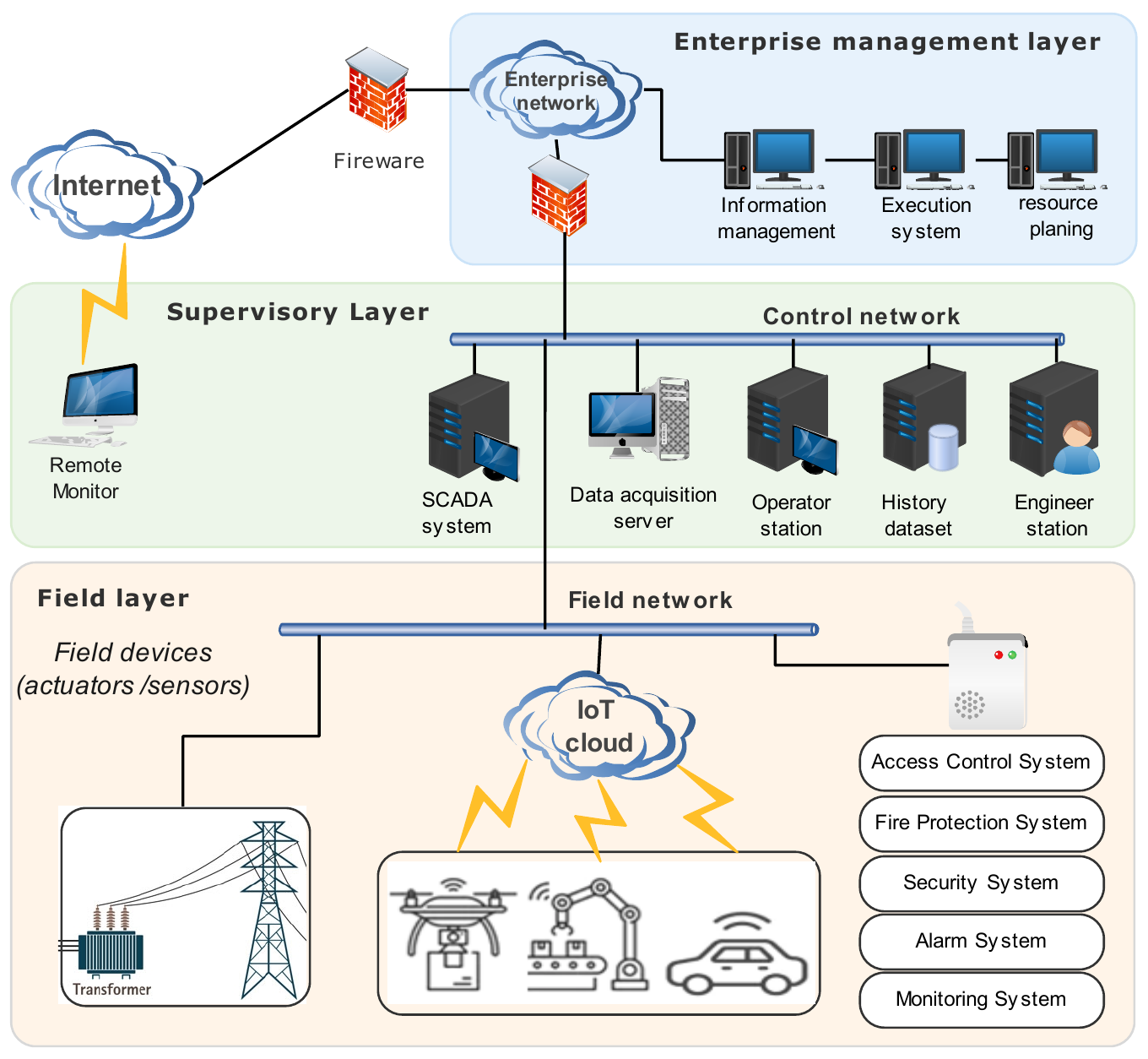}
    \caption{General layout of an \ac{ICS}.}
    \label{fig:icn}
\end{figure*}

A traditional \ac{ICS} is a kind of closed system; only the internal network is used for device manipulation and administration. The physical separation of the internal network from external networks can guarantee an \ac{ICS}'s security. Fundamental connections to the real world are what distinguishes \acp{ICS} from conventional information systems \citep{alanazi2022scada}. The enterprise management layer, supervisory layer, and field layer make up the majority of the architecture of a contemporary \ac{ICS} \citep{onate2023analysis,williams1994purdue,xu2022improved}, as depicted in Fig.~\ref{fig:icn}. These layers have other names, and their functions can be summarized as follows:
\addtolength{\leftmargini}{0cm}
\begin{itemize}
    \item \textbf{Application layer} (also called the \textit{enterprise administration/management layer}). The \ac{ERP} system, \ac{MES}, and other application systems are the key components of this layer. To provide real-time monitoring and administration of industrial processes and support enterprise-level intelligent decision-making, this layer connects to the internet via network-communication technologies.

    \item \textbf{Network layer} (also called the \textit{supervisory layer}). The historical and real-time databases, a number of operator and engineer stations, and process-monitoring devices are the key functions of this layer. This layer is in charge of gathering and transmitting data between the enterprise management layer and the field layer, as well as managing field equipment in accordance with predetermined control logic. \Ac{SCADA} is an example of an \ac{ICN} commonly used in various industries such as manufacturing \citep{alanazi2022scada}.

    \item \textbf{Physical layer} (also called the \textit{field layer}). Sensors, actuators, transmitters, and I/O devices of various kinds are all included in this layer. The primary functions of this layer are field-device manipulation, field-information perception, and digital or analog data exchange through the field bus between various field devices.
\end{itemize}

The lack of authentication and encryption mechanisms in some \ac{ICS} protocols makes \ac{MITM} attacks and tampering with network packets simple. An \ac{ICN} can also be exposed to several risks and vulnerabilities due to the extensive use of general-purpose hardware and software. The following is a summary of the  security needs of an \ac{ICS}:
\addtolength{\leftmargini}{0cm}
\begin{itemize}
    \item \textbf{Real-time operation:} each piece of hardware equipment has a precisely regulated running time. A small delay might cause physical equipment damage and result in catastrophic industrial disasters. Devices such as sensors and actuators can have limited available energy, computing, or storage resources, hence making it challenging for an \ac{ICS} to support the operation of security solutions.

    \item \textbf{Fixed business logic:} to accomplish specified task objectives, an \ac{ICS} should adhere to certain business logic. Serious events are expected to occur when business logic is breached.

    \item \textbf{Legacy systems and updates:} the software in an \ac{ICS}'s subsystems needs to be continuously updated to detect security threats, and this poses great challenges to existing \ac{IDS} solutions. It is difficult to make upgrades when legacy systems are present in an \ac{ICS}. Furthermore, it is very risky to reboot components of an \ac{ICS} for software upgrades or bug fixing.

    \item \textbf{Security of industrial protocols:} to enable remote process monitoring and widespread information interchange, an \ac{ICS} may use completely standard protocols, hardware resources, and software on the internet. As a result, industrial protocols that were previously safe in closed contexts are now susceptible to cyberattacks in open environments, raising the possibility that sensitive and potentially critical process data will be revealed to attackers.
\end{itemize}

Because of the aforementioned security needs, \ac{ICS} security has become a major problem for both academics and businesses, and pattern-recognition detection in \acp{ICN} is becoming increasingly popular. \acp{IDS} have several benefits for organizations, as follows. (i)~Improved security: an \ac{IDS} helps to detect and prevent unauthorized access, misuse, and theft of sensitive information. (ii)~Early detection: an \ac{IDS} can detect security threats and breaches in real time, allowing organizations to respond quickly and effectively. (iii)~Customization: an \ac{IDS} can be configured to match the specific security needs of an organization, providing tailored protection. (iv)~Cost savings: by detecting and preventing security breaches, an \ac{IDS} can save organizations money that would otherwise be spent on recovery and cleanup efforts. Overall, \acp{IDS} play a crucial role in helping organizations maintain the security and integrity of their networks and data.

\subsection{Intrusion detection systems}
Many \acp{IDS} share the same common structure, which includes \citep{asharf2020review}:
\addtolength{\leftmargini}{0cm}
\begin{itemize}
    \item \textbf{A data-collection module.} This gathers data that may include indications of an intrusion. Each component of an \ac{ICN} system's input data can be collected and analyzed to identify typical interaction patterns, allowing for early detection of malicious activities. These data might be host-based, network-based, or hybrid-based inputs. The data types include: (i)~audit trails on a host, including system logs and system instructions; (ii)~network packets or connections; (iii)~wireless network traffic; and (iv)~application logs.

    \item \textbf{An analysis module.} This processes the data and searches for breaches, and it is equipped with an attack-reporting algorithm. The analysis module may use a variety of approaches and methodologies, although \ac{ML}- and \ac{DL}-based methods are prevalent, and these approaches are well suited to data investigation for understanding benign and anomalous behavior based on how \ac{IoT} devices and systems communicate with one another. Additionally, \ac{ML}/\ac{DL} algorithms may intelligently predict upcoming unknown assaults by learning from current valid samples. This allows them to identify new attacks, which are frequently distinct from previous attacks.

    \item \textbf{Algorithms for detection.} Detection algorithms can use anomaly-based, signature-based, or hybrid detection methods. The detection mode could be online (real time) or offline (not real time). Additionally, time specificity includes ``continuous,'' ``periodic,'' or ``batch'' processing of signals to detect intrusions. In addition, there are two sorts of detection responses to intrusions: ``passive'' responses, where an \ac{IDS} generates alerts and does not take any corrective or preventative action, and ``active'' responses, when an \ac{IDS} takes some action to respond to an intrusion.

    \item \textbf{Architecture.} Architectures fall within the following categories. (i)~``Distributed'' architecture gathers information from several systems under monitoring to identify thorough, distributed, and coordinated cyberattacks; this could be cloud based, parallelized, or grid based. (ii)~``Centralized'' architecture gathers and examines data from a single monitored system. (iii)~Finally, ``hybrid'' architecture is a combination of both distributed and centralized architectures.
    \item \textbf{Environment.} These networks can be ad hoc, wired, wireless, structured, or mixed. In particular, wireless \acp{IDS} have numerous and diverse needs that can be set up in standalone, collaborative, or hierarchical environments.

    \item \textbf{Detection approach.} The detection approach can either be ``transition based'' (insecure to secure and vice versa) or ``state based'' (insecure or secure), and both can include a stimulating or unassuming evaluation.
\end{itemize}

\begin{figure*}[!t]
    \centering
    \includegraphics[width=\textwidth]{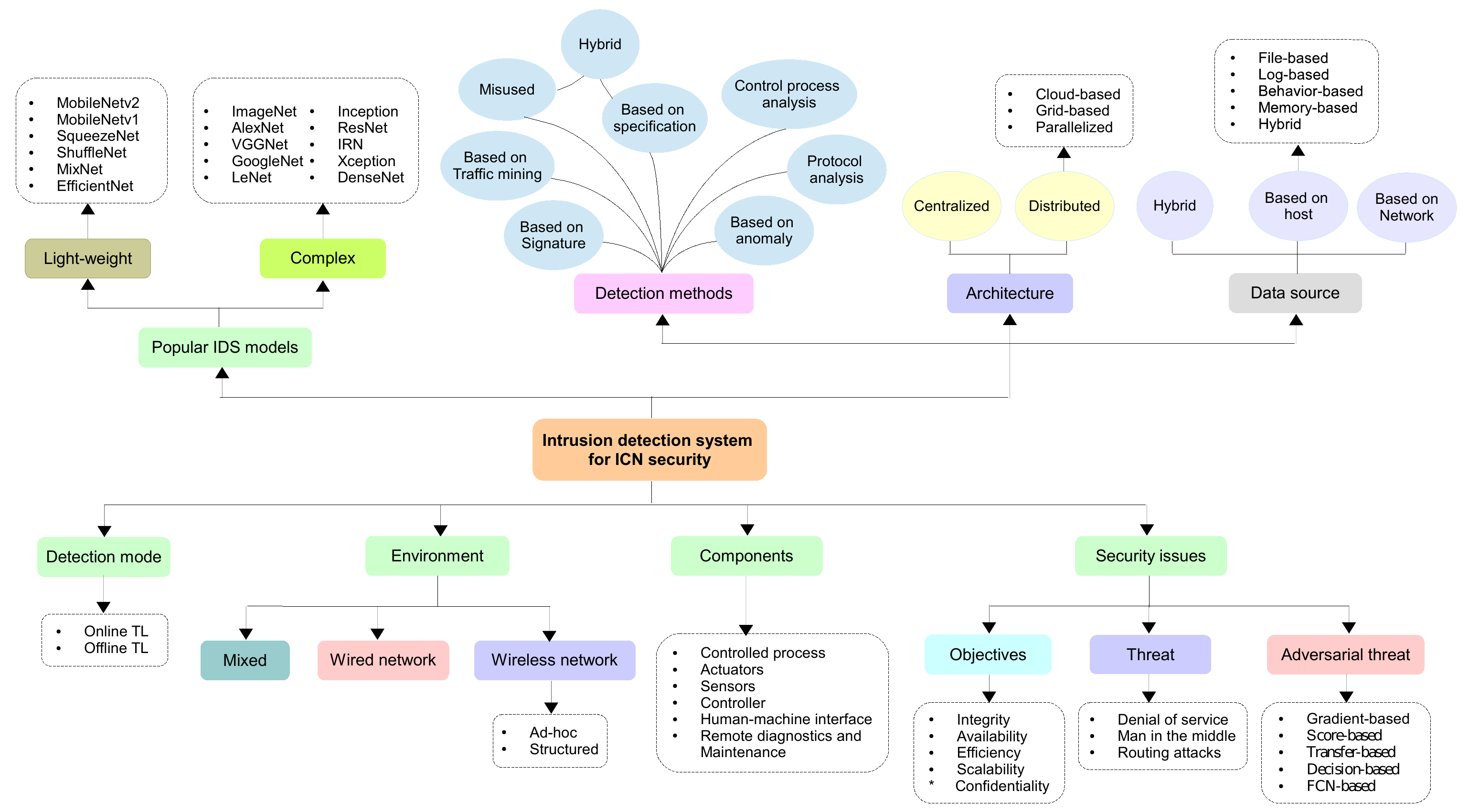}
    \caption{Summary of the \ac{IDS} taxonomy reported in this review.}
    \label{Taxonomyids}
\end{figure*}

Fig.~\ref{Taxonomyids} presents a taxonomy of \acp{IDS}. It also shows that the aforementioned elements are the key determinants for designing different \ac{IDS} schemes.

\subsubsection{Detection methods} \label{sec321}

In the following, classifications of the detection methodologies --- namely, \textit{signature-based}, \textit{anomaly-based}, \textit{specifi-cation-based}, \textit{misused-based}, and \textit{hybrid} approaches --- are detailed \citep{sicato2020comprehensive,asharf2020review}.

\vskip2mm
\noindent\textbf{(a)~Signature-based \acp{IDS}.}
Signature-based detection approaches involves comparing network or system activities to a library of threat signatures. A detection warning is then triggered as soon as a match is detected. Although this approach is adequate against known attacks for which signatures are present in the repository, it is unable to identify zero-day (new) intrusions, even though it is useless against variations of an already existing assault \citep{asharf2020review,keshk2019integrated,keshk2019privacy,masdari2020survey}. Some studies have employed \ac{HTTP} payload signatures to explicitly exploit high-speed streams carrying \ac{HTTP} data for signature-based intrusion detection \citep{erlacher2018fixids}. Another report \citep{li2019designing} proposed a generic framework for cooperative blockchain signature-based \acp{IDS}. This can incrementally create and update a trusted signature database in a distributed, cooperative \ac{IoT} context without the need for a trusted intermediary.

\vskip2mm
\noindent\textbf{(b)~Anomaly-based \acp{IDS}.}
As detailed in previous reviews \citep{asharf2020review,aldweesh2020deep}, a baseline profile of typical behavior for a monitored environment is used as the foundation for anomaly-based detection algorithms \citep{keshk2019privacy}. Then, system behaviors at any particular time are compared to this usual baseline. Any deviations from permitted thresholds result in the generation of an alert without classifying the sort of assault that was discovered. Building normal profiles is preferable to learning normal behavior vs attack-event behaviors, as this approach cannot incorporate new attack events in real-world networks. There have been attempts to use \ac{ML} models to learn normal and attack events to create behavioral detection models. Anomaly-based detection approaches are more successful in finding novel threats than signature-based detection techniques. Anomaly-based approaches can generally be divided into two groups \citep{kaouk2019review}:
\addtolength{\leftmargini}{0cm}
\begin{itemize}
\item \textbf{Statistical-based approaches.} These methods involve processing events or network traffic using statistical algorithms (such as time-series analysis, non-parametric and parametric-based methodologies and analysis, and Markov chains) to ascertain whether a particular piece of data complies with a specific statistical model to establish the presence of intrusions. Statistical profiles that were previously used to train the model are compared with the currently observed profiles as part of the anomaly-detection process, and the result is an anomaly score. The anomaly score describes how unusual the particular occurrence is, and the \ac{IDS} will issue an alert if the anomaly score is greater than a certain threshold \citep{kumar2021research}. Following the same principle, one study \citep{malek2019user} used the approach of extracting a number of internet traffic metrics from a user log, including the number of device accesses and resource usage. The authors' suggested approach focuses on applying statistical methods such as aggregation measures and logistic-regression techniques to determine whether user behavior on host-based systems is normal or irregular.

\item \textbf{\ac{ML}-based approaches.} These are founded on creating mathematical models that permit categorization of the events that are being investigated. \ac{ML} approaches can be supervised or unsupervised depending on the learning methodology used. The training data in supervised learning is labeled. As a result, every piece of input data in the training set is classified as either indicating conventional or unconventional behavior. Conversely, in unsupervised learning, the data is not labeled, and the computer generates the classifier by evaluating the properties of the data. \acp{ANN}, for example \acp{SVM} \citep{mohammadi2021comprehensive}, Bayesian networks \citep{vaddi2020dynamic}, fuzzy logic \citep{ap2019secure}, and \ac{DL} \citep{lansky2021deep}, are examples of \ac{ML} algorithms.
\end{itemize}

\vskip2mm
\noindent\textbf{(c) Specification-based \acp{IDS}.}
The fundamental idea behind both specification- and anomaly-based detection strategies is the same. To identify out-of-range deviations, a system's typical behavior is profiled using some method, and this is contrasted with the actions currently being taken. While specification-based strategies need manual specification of normal behavior using a repository of rules and related ranges of deviations by a human expert, anomaly-based techniques use \ac{ML} to discover normal behavior. A specification-based \ac{IDS} enables a reduction in \ac{FP} rates. This strategy has the benefit of not requiring a learning phase after defining a rule set, but it has a limited ability to adapt to different settings and is susceptible to specification errors. It frequently makes use of formal approaches, state diagrams, finite automata, etc. \citep{asharf2020review,kaouk2019review}. For example, the authors of one study built the so-called SAIDuCANT scheme \citep{olufowobi2019saiducant}, a specification-based \ac{IDS} employing anomaly-based supervised learning. This offers a unique approach to extracting the real-time model parameters of a \ac{CAN} bus.

\vskip2mm
\noindent\textbf{(d) Misuse-based \acp{IDS}.}
In misuse-based \acp{IDS}, to efficiently identify known intrusions, it is primarily necessary to compare the system data obtained with known signatures in a misuse-pattern database. The high rate of known threats being detected is a benefit of misuse-based \acp{IDS}. However, they are unable to recognize zero-day (unknown) threats \citep{sicato2020comprehensive}. Many proposed \acp{IDS} based on misuse patterns have been previously surveyed \citep{le2021machine}.

\vskip2mm
\noindent\textbf{(e) Hybrid \acp{IDS}.}
To overcome the drawbacks and maximize the benefits of detecting both old and new intrusions, hybrid-based \acp{IDS} combines a number of the aforementioned strategies \citep{asharf2020review}. One such suggested scheme incorporates a signature-based \ac{IDS} and an anomaly-based \ac{IDS} to detect both known and unknown attacks on vehicular networks \citep{yang2021mth}. Another proposed hybrid \ac{IDS} employs different detection methods \citep{yao2019hybrid}, combining the advantages of a distributed \ac{IDS}, which operates on each physical device in the network, with a centralized \ac{IDS}, which runs on several central devices such as servers.

\subsubsection{IDS datasets}
Datasets are crucial to \ac{ML}/\ac{DL} since they influence the results of the training process. Any \ac{IDS} must be evaluated using a current, precise, and reliable dataset that includes both regular and anomalous behaviors that are currently occurring. Due to the sensitive nature of \acp{ICS}, there are not many publicly accessible datasets that allow researchers to assess how well a suggested solution works, especially in real time; instead, the majority of these datasets, such as KDDCup99 and NSL-KDD \citep{asharf2020review}, are produced using simulation platforms or testbeds. However, the generated datasets appear to have a detrimental impact on the \ac{IDS} results \citep{koroniotis2017towards}. Several research projects have been carried out to address the shortcomings of these datasets. The most popular datasets for assessing and benchmarking \acp{IDS} are briefly described in the following list \citep{asharf2020review,soni2021systematic}:

\addtolength{\leftmargini}{0cm}
\begin{itemize}

    \item \textbf{KDDCup99.}
When it comes to \acp{IDS}, this is the most-often-used dataset, with 4,900,000 attack records available. This dataset was obtained by processing the tcpdump dataset, which was first made available in the 1998 DARPA ID challenge. There are 41 features for each record in the dataset.

\item \textbf{NSL-KDD.} This is an upgraded version of the dataset used in KDDCup99. In 2009, redundant records were deleted, and the full dataset was reorganized to create the NSL-KDD dataset. For the purpose of training the models, it was split into five classes: \ac{DoS}, normal, probe, \ac{U2R}, and \ac{R2L} attacks. However, the dataset still has drawbacks, such as its lack of representation of low-footprint threats.

\item \textbf{UNSW-NB15.} This dataset, which was developed in 2015 ; it is very recent and thus addresses contemporary minimal-footprint intrusions. It has 49 features, and these are broken down into seven classes: basic, time, labeled, connection, flow, content, and general.

\item \textbf{DEFCON-10.} Using faulty packets, the \ac{FTP} via telnet protocol, administrator privileges, port scans, and sweep assaults, the DEFCON-10 dataset was created in 2002 \citep{sharafaldin2018towards}. A ``capture the flag'' competition generates traffic that is different from network traffic in the real world, since it mainly consists of attack traffic rather than regular background traffic, which limits its relevance for testing \acp{IDS}. The dataset is primarily employed to evaluate alert-correlation methods.

\item \textbf{CSE-CIC-IDS2018.} This contains seven distinct attack scenarios: brute-force, Heartbleed, botnet, \ac{DoS}, \ac{DDoS}, web assaults, and network infiltration. The victim organization comprises five departments, 420 computers, and 30 servers, in addition to the 50 machines that make up the attacking infrastructure. Each machine's system logs and network traffic are captured in the dataset, along with 80 characteristics \citep{CICIDS2018}.

\item \textbf{CIC-DDoS2019.} This dataset covers the most recent, harmless \ac{DDoS} assaults, closely representing actual real-world data. It also contains the outcomes of a network-traffic analysis, with flows labeled according to the timestamp, ports, protocols, source and destination \ac{IP} addresses, and attack. A total of 25 users' abstract behavior based on the \ac{HTTP}, hypertext transfer protocol secure (HTTPS), \ac{FTP}, secure shell (SSH), and email protocols were used to create the dataset \citep{sharafaldin2019developing}.

\item \textbf{Edge-IIoTset.} This dataset was created using a set of devices, sensors, protocols, and cloud/edge settings. The data were produced by various \ac{IoT} devices that are used to sense variables such as temperature and humidity. Additionally, it can be used to identify 14 kinds of attack against \ac{IoT} communication protocols. These are broken down into five types of threats: \ac{DoS} and \ac{DDoS} attacks, information-gathering attacks, \ac{MITM} attacks, infiltration attacks, and malware assaults. Furthermore, it provides features from various sources, such as warnings, system resources, logs, and network traffic. From 1176 previously discovered features with significant correlations, an additional 61 can be used \citep{ferrag2022edge}.

\item \textbf{BoT-IoT.}
A realistic network environment was established to develop the BoT-IoT dataset. The environment includes both botnet traffic and regular traffic. To make labeling easier, the files were divided depending on threat type and subcategory. The collection includes keylogging, \ac{DoS}, \ac{DDoS}, and operating-system and service-scan attacks. The \ac{DDoS} and \ac{DoS} intrusions are further grouped based on the protocol being used.

\item\textbf{SWaT.} The \ac{SWaT} dataset is a publicly available dataset used for research in the field of \ac{ICS} security. It was collected from a full-scale water treatment plant and includes data from various sensors and actuators, including flow meters, pumps, valves, and pH sensors. \Ac{SWaT} includes both normal and attack scenarios, making it suitable for use in research on intrusion detection and cybersecurity for \acp{ICS} \citep{goh2017dataset}.

\item\textbf{WADI.} The \ac{WADI} dataset is an extension of the \ac{SWaT} dataset, and it similarly contains data relating to chemical-dosing systems, booster pumps and valves, and instrumentation and analyzers. A total of 16~days of continuous operation are included, with 14~days under normal operation and two~days with attack scenarios, during which 15 attacks were launched \citep{ahmed2017wadi}.

\item \textbf{ADFA-LD.} This dataset was created for the analysis of \acp{HIDS}. It contains three subsets of raw system-call traces, including 746 attack traces and 4372 validation traces, in addition to 833 regular data traces. The operating system's regular operations were used to collect training and validation traces. The attack traces were acquired under the context of a cyberattack. Some of the attack techniques it includes are ``brute-force password guessing,'' ``create new superuser,'' ``Java Meterpreter payload,'' ``Linux Meterpreter payload,'' and ``C100 Webshell'' \citep{creech2013semantic}.
\end{itemize}

\subsubsection{Evaluation metrics}
Understanding the various assessment measures that are often employed in the community to evaluate the effectiveness of the proposed \ac{IDS} models is crucial \citep{soni2021systematic}. Table~\ref{tab:metrics} summarizes the evaluation metrics used in \ac{IDS}-based \ac{DTL}. The \ac{ROC} is the area under the curve of the \ac{TP} rate as a function of the \ac{FP} rate.

\begin{table*}[t!]
\caption{Summary of the evaluation metrics used in \ac{IDS}-based \ac{DTL}: TP is true positive, TN is true negative, FP is false positive, and FN is false negative.}
\label{tab:metrics}
\begin{tabular}{p{4cm}cp{9cm}}
\hline
 Metric & Math. description & Notes \\
 \hline
\multirow{4}{*}{\begin{tabular}[c]{@{}l@{}}Confusion\\ matrix\end{tabular}} &$\mathrm{TP}$&  Abnormal sample categorized correctly as an attack. \\ [3pt]
 &$\mathrm{FP}$&   Normal sample categorized incorrectly as an attack.\\[3pt]
 &$\mathrm{TN}$&   Normal sample categorized correctly as not being an attack. \\[3pt]
 &$\mathrm{FN}$ &   Abnormal sample categorized incorrectly as not being an attack. \\[3pt]
\Ac{ADR} & \(\displaystyle \mathrm{\frac{TP}{TP+FN}}\) & Represents the \ac{TP} rate. \\[10pt]
 \Ac{FAR} & \(\displaystyle \mathrm{\frac{FP}{FP+TN}}\) & Represents the \ac{FP} rate. \\[10pt]
 Accuracy & \(\displaystyle \mathrm{\frac{TP+TN}{TN+FN+TP+FP}}\) &  Defined as the percentage of accurate predictions among all predictions produced by a classifier. \\[7pt]
 Precision & \(\displaystyle \mathrm{\frac{TP}{TP+FP}}\) & Defined as the percentage of accurate predictions among all positive predictions produced by a classifier. \\[7pt]
Recall &  \(\displaystyle \mathrm{\frac{TP}{TP+FN}}\) & Defined as the percentage of positive labels that the classifier correctly predicts to be positive. \\[7pt]
F-score & \(\displaystyle 2 \times \mathrm{\frac{Precision \times Recall}{Precision+Recall}} \) & Defined as the harmonic mean of precision and recall. \\[10pt]
\Ac{IoU}&\(\displaystyle \mathrm{\frac{B_1 \cap B_2}{B_1 \cup B_2 }} \)& Used for object detection; described as the area of overlap $\mathrm{B_1 \cap B_2}$ divided by the area of union $\mathrm{B_1 \cup B_2}$  between the two shapes $\mathrm{B_1}$ and $\mathrm{B_2}$. \\ [5pt]
 \hline
\end{tabular}
\end{table*}

\subsection{Taxonomy of \ac{DTL}}
\ac{DTL} often entails training a \ac{DL} model on a particular domain (or task) and then adapting the learned features to another related or unrelated domain (or task). Three definitions that are crucial to comprehending the principles of \ac{IDS}-based \ac{DTL} are presented in the subsequent paragraphs.

\vskip2mm
\noindent \textbf{\textit{Def.~1: Task.}} Let us consider $\chi$ and $\gamma$ as the feature and label spaces, respectively. The dataset can be denoted as $X=\allowbreak \left\{ x_{1} , \cdots , x_{n}\right\}|_{X \in \chi}$, labeled with $Y=\left\{ y_1,\cdots ,y_n\right\}|_{Y \in \gamma}$. A task is a conditional distribution $P(Y|X)$, and we can define $\mathbb{T}=\left\{Y,F(X)\right\}$, where $F$ denotes the learning objective predictive function \citep{ramirez2019learning}. In \ac{DTL}, the tasks that belong to the initial knowledge and unknown knowledge to be learned are the \ac{ST} (denoted by $\mathbb{T}_{\mathrm{S}}$) and \ac{TT} (denoted by $\mathbb{T}_{\mathrm{T}}$), respectively.

\vskip2mm
\noindent \textbf{\textit{Def.~2: Domain.}} Let us consider $P(X)$ as the marginal probability distribution of a specific dataset $X=\left\{x_{1},\cdots ,x_{n}\right\}|_{X \in \chi}$. A domain is defined as $\mathbb{D}=\left\{ X,~P(X)\right\}$. In \ac{DTL}, there are two domains: the \ac{SD} (denoted by $\mathbb{D}_{\mathrm{S}}$) and the \ac{TD} (denoted by $\mathbb{D}_{\mathrm{T}}$)~\citep{lu2021general}. When $\mathbb{D}_{\mathrm{S}} \neq \mathbb{D}_{\mathrm{T}}$, \ac{DA} is the process that attempts to understand the prediction function $\mathbb{F}_{\mathrm{S}}$ of the \ac{SM} that can be employed to enhance $\mathbb{F}_{\mathrm{T}}$ of the \ac{TM} using the knowledge acquired from $\mathbb{D}_{\mathrm{S}}$ and $\mathbb{T}_{\mathrm{S}}$ \citep{Kheddar023ASR}.

\vskip2mm
\noindent \textbf{\textit{Def.~3: \ac{DTL}.}} \Ac{DTL} attempts to enhance the performance of $\mathbb{F}_{\mathrm{T}}$ by transferring the knowledge of $\mathbb{F}_{\mathrm{S}}$. In particular, \ac{DTL} tries to find $\mathbb{F}_{\mathrm{T}}$ for $\mathbb{T}_{\mathrm{T}}$ and the $\mathbb{D}_{\mathrm{T}}$. The major difficulty that \ac{DTL} techniques can face is when $\mathbb{D}_{\mathrm{S}} \neq \mathbb{D}_{\mathrm{T}}$ or $\mathbb{T}_{\mathrm{S}} \neq \mathbb{T}_{\mathrm{T}}$. \Ac{DTL} can be expressed as follows \citep{lu2015transfer}:
\begin{equation}
\label{equaFT}
\begin{split}
\mathbb{D}_{\mathrm{S}}=\{X_{\mathrm{S}},P(X_{\mathrm{S}})\}, \quad \mathbb{T}_{\mathrm{S}}=\{Y_{\mathrm{S}},P(Y_{\mathrm{S}}/X_{\mathrm{S}})\} \\ \rightarrow \mathbb{D}_{\mathrm{T}}=\{X_{\mathrm{T}},P(X_{\mathrm{T}})\}, \quad \mathbb{T}_{\mathrm{T}}=\{Y_{\mathrm{T}},P(Y_{\mathrm{T}}/ X_{\mathrm{T}})\}
\end{split}
\end{equation}

\begin{figure*}[t!]
\begin{center}
\includegraphics[width=1.0\textwidth]{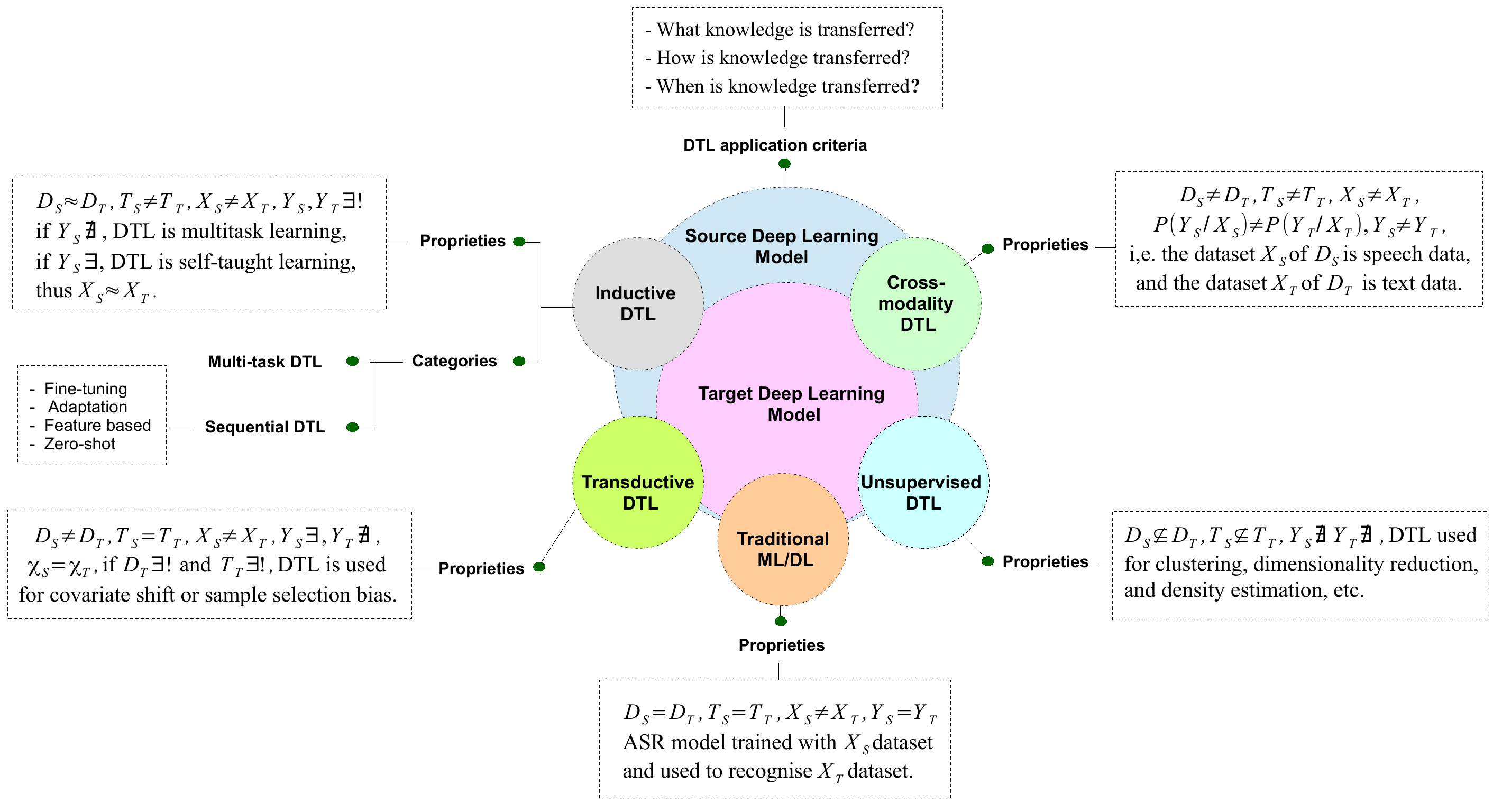}\\
\end{center}
\caption{\ac{DTL} possibilities. The symbol $\nsubseteq$ indicates that the domains/tasks are different but related, $\exists!$ indicates that there is one and only one domain/task, and $\approx$ indicates that domains, tasks, or spaces are not always equal.}
\label{fig:dtl}
\end{figure*}

Until now, there has not been a universally accepted and thorough methodology for categorizing \ac{DTL} into various groups. Nonetheless, \ac{DTL} algorithms may be grouped into different kinds based on the nature of knowledge transfer regarding ``what,'' ``when,'' and ``how.'' Additionally, certain research has endeavored to establish a classification system for \ac{DTL} techniques. An example of this can be seen in the study conducted by Niu et~al. \citep{niu2020decade}, in which they categorize \ac{DTL} methods into two levels. The first level is further divided into four sub-groups based on (i)~the availability of labeled data and (ii)~the data modality in the \ac{SD} and \ac{TD}. This typically results in the classification of \ac{DTL} techniques as inductive, transductive, cross-modality, or unsupervised \ac{DTL}. Fig.~\ref{fig:dtl} summarizes these possibilities. To delve deeper, each sub-group of the initial level can be broken down into four distinct types of learning: instance-, feature-, parameter-, and relation-based learning. In the next subsections, the taxonomy of \ac{DTL} techniques, namely, \textit{inductive \ac{DTL}}, \textit{transductive \ac{DTL}}, and \textit{adversarial \ac{DTL}}, along with the pre-trained models used, are detailed.

\subsubsection{Inductive \ac{DTL}}
When compared to traditional \ac{ML}, which can be used as a benchmark for \ac{DTL}, the objective of inductive \ac{DTL} is to improve the target prediction function $\mathbb{F}_{\mathrm{T}}$ in the \ac{TD} (as noted in Eq.~(\ref{equaFT})), given that the target tasks $\mathbb{T}_{\mathrm{T}}$ differ from the source tasks $\mathbb{T}_{\mathrm{S}}$. However, it is important to note that the \ac{SD} $\mathbb{D}_{\mathrm{S}}$ and \ac{TD} $\mathbb{D}_{\mathrm{T}}$ may not necessarily be identical (as depicted in Fig.~\ref{fig:dtl}). Depending on whether labeled or unlabeled data are available, inductive \ac{DTL} can be described as falling into one of the following two cases:

\vskip2mm
\noindent \textbf{(a)~Multi-task \ac{DTL}.} In multi-task \ac{DTL}, the \ac{SD} possesses a vast labeled dataset ($X_{\mathrm{S}}$ labeled with $Y_{\mathrm{S}}$), which is a unique type of multi-task learning. Nonetheless, it is unlike multi-task approaches, which learn several tasks $(T_1, T_2,\dots, T_n)$ simultaneously (in parallel), encompassing both source and target activities.

\begin{figure*}[t!]
\begin{center}
\includegraphics[width=1.0\textwidth]{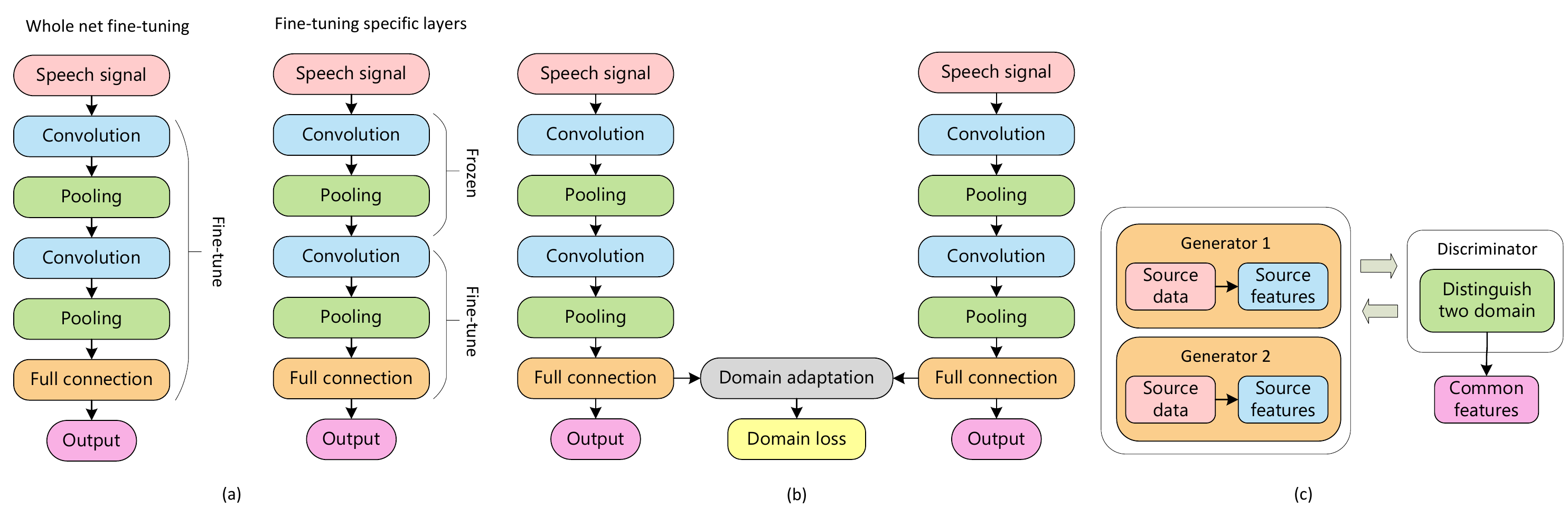}\\
\end{center}
\caption{Structures of: (a)~fine-tuning, (b)~\ac{DA}, and (c)~\ac{GAN}-based \acp{DTL}.}
\label{fig:4}
\end{figure*}

\vskip2mm
\noindent \textbf{(b)~Sequential \ac{DTL}.} Sequential \ac{DTL} is also referred to as ``self-taught learning.'' The \ac{SD} lacks labeled data ($X_{\mathrm{S}}$ is not labeled with $Y_{\mathrm{S}}$), but the labels are available in the \ac{TD} ($X_{\mathrm{T}}$ is labeled with $Y_{\mathrm{T}}$). Sequential learning is a type of \ac{DL} system used for classification purposes, and it involves two steps. The first step is the transfer of feature representation, which is learned from a vast collection of unlabeled datasets. The second step involves the application of the learned representation to labeled data to carry out classification tasks. Therefore, sequential \ac{DTL} is a method of sequentially learning several activities (tasks). There may be differences between the \ac{SD} and \ac{TD}. For example, suppose we have a pre-trained model $M$ and aim to apply \ac{DTL} to several tasks $(T_1, T_2,\dots, T_n)$. In this case, we learn a specific task $\mathbb{T}_{\mathrm{T}}$ at each time step $t$, which is slower than multi-task learning. However, when not all tasks are available during training, sequential \ac{DTL} might be advantageous. Sequential \ac{DTL} can also be categorized into several types \citep{alyafeai2020survey}:
\begin{enumerate}[leftmargin=0.5cm]

\item \textbf{Fine-tuning.} The main idea of fine-tuning is to use a pre-trained model $M_{\mathrm{S}}$ with weights $W_{\mathrm{S}}$ and a target task $\mathbb{T}_{\mathrm{T}}$ with weights $W_{\mathrm{T}}$ to learn a new function $\mathbb{F}_{\mathrm{T}}$ that maps the parameters $\mathbb{F}_{\mathrm{T}}(W_{\mathrm{S}}) = W_{\mathrm{T}}$. This can be done by adjusting the settings across all layers or just some of them (as shown in Fig.~\ref{fig:4}(a)). The learning rate can also be different for each layer, and this is referred to as discriminative fine-tuning. Additionally, a new set of parameters $K$ can be added to most tasks so that $\mathbb{F}_{\mathrm{T}}(W_{\mathrm{T}}, K) = W_{\mathrm{S}}\circ K$.

\item \textbf{Adapter modules.} The adapter module is designed to work with a pre-trained model $M_{\mathrm{S}}$ that outputs $W_{\mathrm{S}}$ and is used for a target task $\mathbb{T}_{\mathrm{T}}$. Its objective is to learn a new set of parameters $K$ that is much smaller than $W_{\mathrm{S}}$, i.e., $K\ll W_{\mathrm{S}}$. $W_{\mathrm{S}}$ and $K$ can be broken down into more compact modules, such that $W_{\mathrm{S}}={w}_n$ and $K={k}_n$. The adapter module facilitates the learning of a new function $\mathbb{F}_{\mathrm{T}}$, which can be expressed as:
\begin{equation}
\label{adapt}
 \mathbb{F}_{\mathrm{T}}(K, W_{\mathrm{S}})= k_{1}'\circ w_{1}\circ \dots k_n'\circ w_n.
\end{equation}
The adaptation process described in Eq.~(\ref{adapt}) does not modify the set of original weights $W_{\mathrm{S}}={w}_n$ but rather modifies the set of weights $K$ to $K'=k_{n}'$. This principle is demonstrated in Fig.~\ref{fig:4}(b).

\item \textbf{Feature-based \ac{DTL}.} The focus of feature-based \ac{DTL} is on learning different concepts and representations at various levels, including word, character, phrase, or paragraph embedding $E$, without regard to the specific task. The model $M$ is used to create a collection of $E$ that remains unchanged, and the fine-tuning of $W'$ is carried out so that $\mathbb{F}_{\mathrm{T}}(W_{\mathrm{S}}, E) = E \circ W'$. For example, one approach involves using the \ac{GAN} concept in \ac{DTL}, in which generators send features from both the \ac{SD} and \ac{TD} to a discriminator, which determines the source of the features and sends the result back to the generators until they become indistinguishable. Through this process, a \ac{GAN} is able to identify common features between the two domains, as depicted in Fig.~\ref{fig:4}(c).

\item \textbf{Zero-shot \ac{DTL}.} This is the most straightforward compared to other methods. It assumes that the pre-trained model's parameters $W_{\mathrm{S}}$ cannot be changed or extended with new parameters $K$. Essentially, in zero-shot \ac{DTL}, there is no training process to learn or optimize new parameters.
\end{enumerate}

\subsubsection{Transductive \ac{DTL}}\label{TDTL}
In the context of \ac{DTL}, which is compared to traditional \ac{ML}, the \ac{SD} and the \ac{TD} have distinct datasets $\mathbb{D}_{\mathrm{S}}$ and $\mathbb{D}_{\mathrm{T}}$, respectively. The \ac{SD} has a labeled dataset $X_{\mathrm{S}}$ with corresponding labels $Y_{\mathrm{S}}$, while the \ac{TD} has no labeled dataset. The source and target tasks are the same (as shown in Fig.~\ref{fig:dtl}). Transductive \ac{DTL} aims to construct the target prediction function $\mathbb{F}_{\mathrm{T}}$ for $\mathbb{D}_{\mathrm{T}}$ based on knowledge of $\mathbb{D}_{\mathrm{T}}$ and $\mathbb{T}_{\mathrm{T}}$. Transductive \ac{DTL} can be divided into two categories based on the conditions between the \ac{SD} and \ac{TD}:
\vskip2mm
\noindent \textbf{(a)~Unsupervised \ac{DTL}.} The objective of \ac{DTL} is to improve the learning of the target prediction function $\mathbb{F}_{\mathrm{T}}$ in $\mathbb{D}_{\mathrm{T}}$ by using the information available in $\mathbb{D}_{\mathrm{S}}$ and $\mathbb{T}_{\mathrm{S}}$, where $\mathbb{T}_{\mathrm{S}}$ is a task that is different but related to $\mathbb{T}_{\mathrm{T}}$, and $Y_{\mathrm{S}}$ and $Y_{\mathrm{T}}$ are not observable. This is achieved by considering an \ac{SD} $\mathbb{D}_{\mathrm{S}}$ with a learning task $\mathbb{T}_{\mathrm{S}}$, a \ac{TD} $\mathbb{D}_{\mathrm{T}}$, and a matching learning task $\mathbb{T}_{\mathrm{T}}$, where $\mathbb{D}_{\mathrm{S}}$ and $\mathbb{D}_{\mathrm{T}}$ are different but related.

\vskip2mm
\noindent \textbf{(b)~\Ac{DA} \ac{DTL}.} The feature spaces $\chi_{\mathrm{S}}$ and $\chi_{\mathrm{T}}$ are the same in different domains, but the probability distributions of the input dataset $P(Y_{\mathrm{S}}/ X_{\mathrm{S}})$ and $P(Y_{\mathrm{T}}/ X_{\mathrm{T}})$ are different. For instance, if an evaluation is performed on a resort topic in $\mathbb{D}_{\mathrm{S}}$, it can be used to train a model for restaurants in $\mathbb{D}_{\mathrm{T}}$. \ac{DA} is highly useful when $\mathbb{T}_{\mathrm{T}}$ has a unique distribution or there is a lack of labeled data. To measure the adaptability level between $\mathbb{D}_{\mathrm{S}}$ and $\mathbb{D}_{\mathrm{T}}$, many advanced techniques use \ac{MMD} as a metric, which is calculated:
\begin{equation}
    \mathrm{MMD}(\mathbb{D}_{\mathrm{S}}, \mathbb{D}_{\mathrm{T}})=\norm[\bigg]{ \frac{1}{n_{\mathrm{S}}} \sum_{i=1}^{n_{\mathrm{S}}} x_{\mathrm{S}}^i \cdot \beta_{\mathrm{S}}-\frac{1}{n_{\mathrm{T}}} \sum_{i=1}^{n_{\mathrm{T}}} x_{\mathrm{T}}^i \cdot \beta_{\mathrm{T}}}_\mathcal{H},
    \label{mmd}
\end{equation}
where: $n_{\mathrm{S}}$ and $n_{\mathrm{T}}$ are the numbers of samples of the \ac{SD} and \ac{TD}, respectively; $\beta_{\mathrm{S}}$ and $\beta_{\mathrm{T}}$ denote the representations of the source and target datasets (i.e., $x_{\mathrm{S}}^i$ and $x_{\mathrm{T}}^i$), respectively; and $\Vert \cdot \Vert_\mathcal{H}$ represents the 2-norm operation in reproducing kernel Hilbert space \citep{vu2020deep}.

\vskip2mm
\noindent \textbf{(c)~Cross-modality \ac{DTL}.} In the field of spoken language, cross-lingual \ac{DTL} is also referred to as a \ac{DTL} method. However, it is one of the most challenging issues in \ac{DTL} because it differs from other \ac{DTL} approaches in requiring a connection between the feature or label spaces of $\mathbb{D}_{\mathrm{S}}$ and $\mathbb{D}_{\mathrm{T}}$. In other words, \ac{DTL} is only possible when the source and destination data are in the same modality, such as text, speech, or video. Cross-lingual \ac{DTL} assumes that the feature spaces of the source and destination domains are completely distinct ($\chi_{\mathrm{S}} \neq \chi_{\mathrm{T}}$), such as speech-to-image, image-to-text, or text-to-speech, and the label spaces of the source $Y_{\mathrm{S}}$ and destination $Y_{\mathrm{S}}$ domains may differ ($Y_{\mathrm{S}} \neq Y_{\mathrm{T}}$).

\subsubsection{Adversarial \ac{DTL}}
Compared to the \ac{DTL} methods described above, adversarial learning \citep{wang2020transfer} aims to learn more transferable and distinguishable representations. The \ac{DANN} \citep{ganin2016domain} was introduced as the first approach that used a domain-adversarial loss in the network instead of a predefined distance function like \ac{MMD}. This technique greatly improves the network's ability to learn more discriminative data. Several studies have used domain-adversarial training following the idea of \ac{DANN} \citep{bousmalis2016domain,chen2019joint,long2017deep,zhang2018collaborative}. In these works, the different impacts of marginal and conditional distributions in adversarial \ac{TL} are disregarded. However, the proposed dynamic distribution alignment method in one previous study \citep{wang2020transfer} can dynamically evaluate the significance of each distribution.

\subsubsection{Pre-trained models}
The \ac{IDS} research community has come to the conclusion that using pre-trained models as the core for target tasks is preferable to building models from scratch. We now briefly introduce the most widely used pre-trained models in \ac{DTL}-based \acp{IDS}:
\addtolength{\leftmargini}{0cm}
\begin{itemize}
    \item \textbf{ResNet.} ResNet was introduced in 2015 by Microsoft Research. The model is designed to handle the vanishing-gradient problem in deep networks, which can lead to underfitting during training. The pre-trained ResNet model is used as a meaningful extractor of residuals, instead of features, from the raw data using identity shortcut connections \citep{he2016deep}. Refined variants of ResNet, such as ResNet-34, ResNet-50, and ResNet-101, have been created using various combinations of layers. ResNet has already been employed for \ac{IDS}-based \ac{DTL} \citep{li2017intrusion,zhang2021dual}.

    \item \textbf{GoogLeNet.} Also known as Inception-v1, GoogLeNet is a deep \ac{CNN} architecture developed by Google for the \ac{ILSVRC}. It is widely used for image-classification tasks and has achieved \ac{SOTA} results on a variety of benchmarks \citep{szegedy2015going}. GoogLeNet has been employed as a \ac{SM} by retraining the last few layers of the network to adapt to the new task for \ac{DTL}-based \ac{IDS} \citep{li2017intrusion}.

    \item \textbf{VGG-16.} VGG-16 is a 16-layered \ac{CNN} architecture that was developed by the Visual Geometry Group at the University of Oxford. It is designed for image-classification tasks, and it has been trained on a large dataset, ImageNet, to recognize a variety of objects and scenes \citep{simonyan2014very}. \ac{IDS} schemes with datasets that can be converted to images (matrices) could employ VGG-16 as a backbone model. VGG-16 has also been used as an \ac{SM} to build an efficient \ac{DTL}-based \ac{IDS} \citep{masum2020tl}.

    \item \textbf{LeNet.} LeNet is a shallow \ac{CNN} architecture that was developed by Yann LeCun and his colleagues in the late 1990s. It was one of the first \ac{DL} models applied to the task of image recognition, and it was used for recognition of handwritten digits \citep{lecun1998gradient}. The main reason behind the popularity of this model was its simple and straightforward \ac{CNN} architecture. LeNet and its variant LeNet-5 have been used as \acp{SM} to transfer knowledge to a target \ac{IDS} model \citep{lin2018using,mehedi2021deep}.

    \item \textbf{AlexNet.} AlexNet is a deep CNN architecture that was introduced by Alex Krizhevsky et~al. in 2012 \citep{krizhevsky2017imagenet}. It comprises three fully connected layers, five \ac{CNN} layers, and max pooling to minimize the dimensionality of the data. It was one of the first \ac{DL} models applied to the task of image recognition, and it achieved \ac{SOTA} results in the \ac{ILSVRC} at the time. AlexNet has been used in many \ac{DTL}-based \acp{IDS} \citep{sreelatha2022improved,zhang2021dual}.

    \item \textbf{DenseNet.} DenseNet is a \ac{CNN} architecture that was introduced by Gao Huang and his colleagues in 2016. It is known for its dense connectivity pattern, in which each layer is connected to all previous layers rather than just the previous few layers, as in a traditional \acp{CNN}. This reduces the size of the feature maps, making the final \ac{CNN} smaller and more compact \citep{huang2017densely}. DenseNet has been used to build an efficient \ac{IDS} scheme based on \ac{DTL} \citep{zhang2021dual}.
\end{itemize}

\section{Techniques of \ac{DTL}-based \acp{IDS}}\label{sec4}

\ac{DL} approaches have been regarded as one of the most successful strategies for anomaly-based intrusion detection because of their capacity to automatically identify behaviors. However, the distributions of the training data and the real data must match for \ac{ML} approaches to be effective, which is exceedingly improbable in the case of zero-day cyberattacks \citep{vu2020deep}. \Ac{DL} models must be trained on a massive amount of labeled data to address this issue. However, in reality, this is sometimes impossible to accomplish, which limits the use of \ac{DL} for intrusion detection. Many studies suggest \ac{DTL} strategies for intrusion detection as a solution to this issue. Furthermore, the hyperparameter-tuning process of \ac{DTL} takes advantage of a model pre-trained on a large-scale dataset, which tends to extract useful features from the input and therefore requires dramatically less training target data (for the task of interest) to converge without overfitting.

Some works have adopted the principle of converting the \ac{IDS} dataset into image representations \citep{wen2019time,xu2020intrusion}. Wen and Keyes \citep{wen2019time} created a \ac{CNN} to identify unusual time-series segments in exactly the same way that desired items are found in images. The authors also suggest a \ac{TL} technique that applies the top-12 layers of a pre-trained \ac{CNN} to the \ac{TM} and then uses target data to refine (fine-tune) the \ac{TM}. The earliest layers often capture the features shared by the tasks, and the latter layers typically catch the more task-specific properties; hence, this \ac{TL} method is expected to be more effective than just transferring the entire model. Results from experiments indicate that the \ac{TL} technique can raise the \ac{CNN}'s \ac{IoU} score by up to 21\%. Similarly, Xu et~al. \citep{xu2020intrusion} employed the LeNet pre-trained model as a source of knowledge to construct an efficient \ac{IDS} \ac{TM}, the so-called \ac{CNN}-\ac{IDS}. Tests were conducted for many kinds of attacks, and these resulted in the best performance compared with other \ac{SOTA} methods.

In the field of biometrics, \ac{DTL} can improve the performance of an \ac{IDS} in identifying relevant attacks. Raghavendra et~al. \citep{raghavendra2017transferable} suggested a framework that feeds the finger-vein \ac{PAD} scheme by exploiting the \ac{DTL} ability of the AlexNet pre-trained model. Other schemes have been proposed to solve many intrusion attacks in the field of video surveillance \citep{xia2022pedestrian,li2020efficient, nayak2019video}. Some intrusion-attack detection systems can be beneficial for users' security by automatically making decisions to interrupt or allow dangerous activities. Some research works have discussed protecting pedestrians in hazardous areas using video surveillance and \acp{ICS} to determine when to launch intrusions. Hence, automatic launch decisions for object detection and intrusion are needed in these situations. For example, an enhanced mask regional \ac{CNN}-based target-identification approach has been suggested using the ResNet50-FPN structure as a \ac{SM} to address the issue of pedestrian-infiltration detection in hazardous zones of electric power construction sites. In this system, the multi-scale transformation approach is used to address the issue of misdetection of smaller targets. The electrical-scene identification model is generated by the \ac{DTL} training dataset.

Nayak et~al. \citep{nayak2019video} suggested the development of a modern \ac{DL}-based \ac{IDS} for video. They employed a \ac{YOLO} pre-trained model to identify objects, and the proposed \ac{TM} determines intrusion based on the shifting center of mass of the observed objects. Tariq et~al. \citep{tariq2020cantransfer} presented a \ac{DTL}-based \ac{IDS} for a \ac{CAN}. They proposed using a \ac{ConvLSTM} network as a source to build CANTransfer. By using a \ac{DTL} strategy, CANTransfer can detect both known and unknown attacks in a \ac{CAN} bus.

In the next subsections, other \ac{DTL}-\ac{IDS} strategies are discussed and grouped according to their \ac{DTL} sub-field and domain of application. 

\subsection{Auto-encoded \ac{DTL}-based \ac{IDS}}
\begin{figure}[!t]
    \centering
    \includegraphics[scale=0.7]{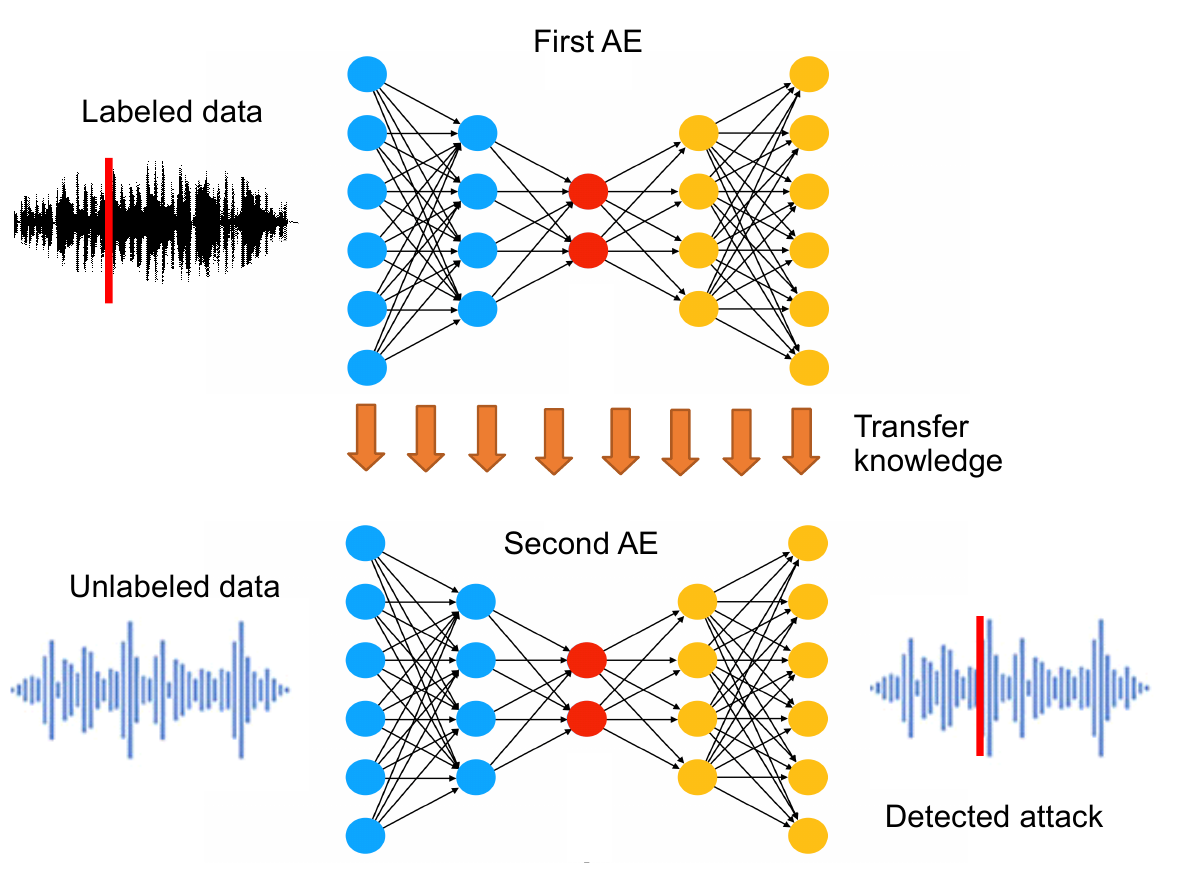}
    \caption{Example of a \ac{TL}-based \ac{IDS} approach using an \ac{AE}.}
    \label{fig:AETL}
\end{figure}

An \ac{AE} is a type of \ac{AI} network used for unsupervised learning. It consists of an encoder, which compresses the input data into a lower-dimensional representation, often called the bottleneck, and a decoder that then takes the bottleneck representation and uses it to reconstruct the original input. \acp{AE} can be used for \acp{IDS}, and they can be trained on normal network behavior and used to detect anomalies in network traffic. If an \ac{AE} is trained well, it should be able to reconstruct normal network traffic accurately, but it will struggle to reconstruct anomalous traffic, indicating that it may be malicious. For example, the approach suggested by Vu et~al. \citep{vu2020deep} can use labeled and/or unlabeled data to identify intrusions such as \acp{DDoS}, UDP flooding, TCP flooding, and data spamming. Specifically, the proposed model consists of two \acp{AE}, as illustrated in Fig.~\ref{fig:AETL}. The first \ac{AE} conducts supervised learning of the source data, whereas the second \ac{AE} is trained with the unlabeled target data. The knowledge gained from the \ac{SD} will be transferred to the \ac{TD} (transductive \ac{TL}) by minimizing the \ac{MMD} (expressed by Eq.~(\ref{mmd})) distances between each layer of the first \ac{AE} and its corresponding layer of the second \ac{AE}.

Kim et~al. \citep{kim2020transfer} suggested an alternative approach to botnet detection by incorporating \ac{DTL}. Their proposed method leverages well-curated source data and transfers the knowledge to new, unseen problem domains; it involves training a \ac{RVAE} structured neural network on the source data and also using it to calculate anomaly scores for records from the \ac{TD}. The approach is effective even when the \ac{TD} data is unlabeled. The difficulty of obtaining enough labeled training data for efficient intrusion detection using \ac{ML} is addressed by dos~Santos et~al. \citep{dos2021reminiscent}. The suggested approach uses a two-stage process in which features are extracted using an \ac{AE} and a pattern-recognition classifier is constructed using the encoder's output. The classifier takes into consideration the output features of the encoder layers, which describe the network's history behavior, making model updates simpler. Each time a model is updated, \ac{DTL} is also used on the older \ac{AE} to minimize the number of labeled events and the training costs.

In the technique suggested by Qureshi et~al. \citep{qureshi2020intrusion}, features from the NSL-KDD dataset are extracted using a network pre-trained on regression-related tasks. The pre-trained network's extracted features, as well as the original features, are then delivered as input to the sparse \ac{AE}. The performance of the sparse \ac{AE} is improved when self-taught learning-based extracted features are combined with the original features of the NSL-KDD dataset. Ten separate runs were used to demonstrate the effectiveness of this self-taught learning-based strategy over other methods.

\subsection{\Ac{DTL} with \ac{HMM}-based \ac{IDS}}
\begin{figure*}[!t]
    \centering
    \includegraphics[scale=1.8]{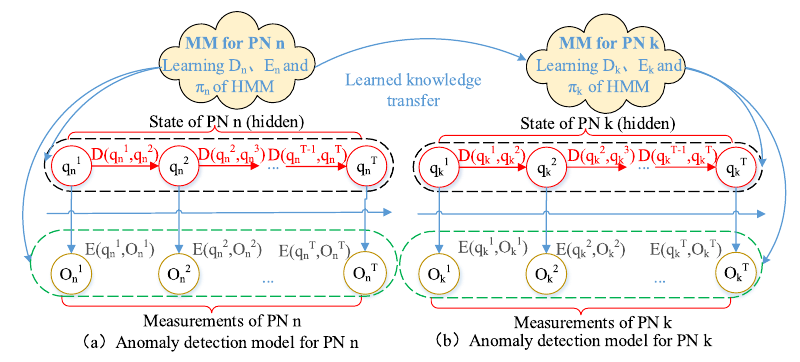}
    \caption{Coordinated anomaly detection using \ac{TL}-based \ac{HMM}.}
    \label{fig:hmm}
\end{figure*}

The following studies emphasize the need for apply serious efforts to updating \ac{ML} methodologies. It is therefore expected that \ac{HMM} learning will be able to go beyond present cutting-edge algorithms such as \ac{TL}-based Baum-Welsh (BW). \ac{HMM} is a successful way to deduce a system's unknown data from its observable outputs. Based on observations from the virtual nodes, Wang et~al. \citep{wang2019cooperative} suggested employing \ac{TL} in \ac{HMM} for capturing anomalous states of the \acp{PN} substrate in a network-slicing scenario. To communicate information between two physical nodes, the BW algorithm was updated. Fig.~\ref{fig:hmm} illustrates the principle of anomaly detection using \ac{TL}-based \ac{HMM}. Another initiative that has been designed to handle multi-stage network attacks and advanced persistent threats with relation to research work in traditional \acp{HMM} was suggested by Chadza et~al. \citep{chadza2020learning}. To detect attack stages in network traffic as well as predict the next most likely attack stage and how it will manifest, five unsupervised \ac{HMM} techniques---BW, Viterbi training, gradient descent, differential evolution, and simulated annealing---were developed using a \ac{TL} approach, and these were compared to traditional \ac{ML} approaches. The maximum accuracy increase was found to occur with differential evolution-based \ac{TL}, reaching up to 48.3\%. Table~\ref{tab:3} summarizes and compares most of the aforementioned methods.

\subsection{Multi-task \ac{DTL}-based \ac{IDS}}
In this \ac{DTL} category, a shared model is used to learn multiple tasks at the same time. This is known as \ac{MTL}, which is a subset of inductive \ac{DTL} and is often used in computer-vision applications. The work of Huang et~al. \citep{huang2018automatic} showed the potential of this approach in the field of network security. They created multiple \acp{IDS} using a \ac{CNN} to detect malware, virtual private network encapsulation, and trojans all at once. The data were transformed into a three-dimensional grayscale array to fit the design of the \ac{CNN}; however, this approach has the drawback of taking longer to execute due to the size of the three-dimensional array. Demertzis et~al. \citep{demertzis2019cyber} introduced the Cyber-Typhon framework, which enhances passive infrastructure using advanced computational intelligence algorithms. This approach employs a series of complex anomaly-detection equations using \ac{MTL} and combining online sequential extreme learning machines with restricted Boltzmann machines. Although this approach has been successful, it requires a large amount of training data to achieve high performance. Similarly, Albelwi \citep{albelwi2022intrusion} proposed an \ac{MTL} model for a \ac{DTL}-based \ac{IDS}. To do this, each sample in the UNSW-NB15 and CICID2017 datasets was combined into a single feature vector, resulting in some data points in both the training and testing sets containing two types of threat at the same time.

\subsection{\ac{DA}-based \acp{IDS}}
\begin{figure}[!t]
    \centering
    \includegraphics[scale=1]{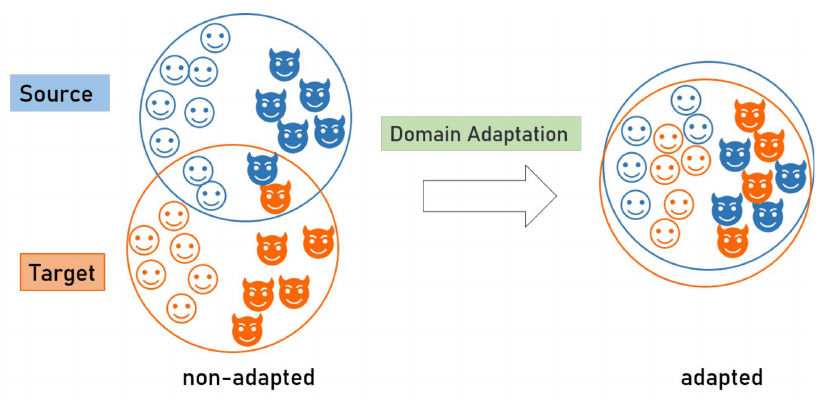}
    \caption{Feature distributions in the \ac{SD} and \ac{TD} before and after \ac{DA} in the \ac{IDS} field. The devils symbolize malicious network traffic, while the smiling faces represents regular network traffic. The feature distribution of the \ac{SD} is shown in blue, while the feature distribution of the \ac{TD} is shown in orange.}
    \label{fig:DA}
\end{figure}

\Ac{TL} has a sub-field called deep \ac{DA} (cf. Section~\ref{TDTL}). The feature distributions in the \ac{SD} and \ac{TD} before and after \ac{DA} are shown in Fig.~\ref{fig:DA}. \ac{DA} is an excellent option for \acp{IDS} because it can transmit information and provide strong learning in the \ac{TD}. \ac{DA} could be introduced by the ``IoTDefender'' algorithm to address the problem of disparate distribution between regular networks and \ac{IoT} \citep{wang2018deep}. Juan et~al. \citep{zhao2017feature} introduced a feature-based \ac{TL} approach, showing its potential use for cybersecurity research by improving classification results, compared to the HeMap \citep{shi2010transfer} and \ac{CORAL} methods \citep{sun2016return}, on the popular NSL-KDD dataset of TCP traffic, demonstrating its potential use in cybersecurity assessment.

Analogously, the source and target datasets, which are regarded as two manifolds in the frameworks suggested by Taghiyarrenani et~al. \citep{taghiyarrenani2018transfer} and Sameera and Shashi \citep{sameera2020deep}, are mapped onto a shared feature space to avoid the issue of dissimilar feature spaces and marginal probability across the domains. Additionally, in Sameera and Shashi's work \citep{sameera2020deep}, the use of cluster correspondence processes is suggested as a way to generate target soft labels to overcompensate for the absence of labeled target zero-day attacks. By minimizing the greatest mean discrepancy, Vu et~al. \citep{vu2020deep} were able to bridge the gaps between the bottleneck layers of these two \acp{AE} for a label-rich and a label-sparse \ac{IoT} domain (using \ac{MMD}). A deep sub-\ac{DA} network with an attention mechanism was suggested by Hu et~al. \citep{hu2022deep} that used the local \ac{MMD} to improve prediction accuracy and an attention mechanism to avoid having too long of a convergence period; the proposed method outperforms the data-source attention network method \citep{zhu2020deep} in terms of average accuracy, but it increases the specific time consumption. Ning et~al. \citep{ning2021malware} suggested the ConvLaddernet, which is based on knowledge transfer and operates in semi-supervised scenarios, i.e., transfer of knowledge from a small-scale \ac{SD} to assist intrusion detection of the \ac{TD} to get around the labor-intensive dataset-gathering procedure.

Although prior \ac{DA}-based approaches have been used to conduct intrusion detection, their performance may be hampered by a failure to simultaneously incorporate the implicit categorical and explicit distance semantics during knowledge transfer. Additionally, in working on these \ac{DA}-based approaches, the researchers were unaware that using a network-intrusion domain in conjunction with a small-scale \ac{IoT} intrusion domain might improve the intrusion-detection performance of a large-scale poorly tagged target \ac{IoT} domain. As a result, they missed the possibility of scenario semantics \citep{wu2022joint}. The latter method is achieved by the use of a confused domain discriminator and categorical-distribution knowledge preservation to enable the knowledge-transfer method. Hence, it lessens the source--target disparity to achieve domain invariance for the shared feature space. A 10.3\% accuracy boost was achieved compared to the \ac{SOTA} methods \citep{yao2021multisource,yao2019heterogeneous,yao2020discriminative}.

\begin{figure*}[!t]
    \centering
    \includegraphics[scale=1.2]{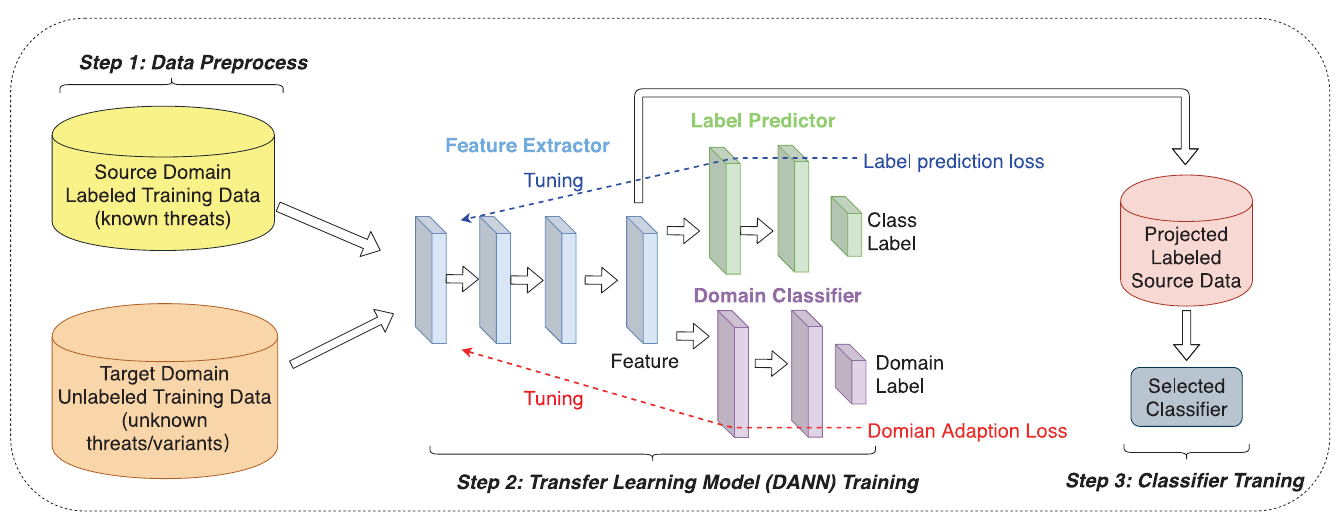}
    \caption{Domain-adversarial \ac{TL} framework.}
    \label{fig:AdvDA}
\end{figure*}

By applying an adversarial learning framework, a family of techniques known as adversarial \ac{DA} uses \acp{GAN} to develop a common domain-invariant mapping between the source and target datasets. The majority of adversarial \ac{DA} techniques use some variation of a generic \ac{GAN} architecture, with differences in the underlying models' generativeness or discriminativity, the loss functions used, and the distribution of weights between source \acp{GAN} (G) and target discriminators (D). Singla et~al. \citep{singla2020preparing} supposed that there are not many labeled examples available in the \ac{TD}; they created a hybrid version of the adversarial discriminative \ac{DA} and \ac{DANN} architectures (Fig.~\ref{fig:AdvDA}) to suit their particular case.

In contrast to the \ac{DANN} technique, which employs a single feature extractor and two classifiers---one for label prediction and one for domain prediction---the ADDA strategy completely decouples the source and target G weights while maintaining a single D. Experiments show that when both the source and the target datasets have similar features (homogeneous \ac{DA}) and when the two datasets have distinct features (heterogeneous \ac{DA}), the adversarial \ac{DA} technique performs better than alternative approaches such as \ac{TL} using fine-tuning.

To provide reliable intrusion detection against smart-grid threats, Zhang and Yan \citep{zhang2019domain} suggested a domain-adversarial \ac{TL} system. This approach adds domain-adversarial training to establish a mapping between the labeled \ac{SD} and the unlabeled \ac{TD}, using DAAN architecture, as depicted in Fig.~\ref{fig:AdvDA}, so that the classifiers can acquire knowledge in a new feature space while protecting themselves against unidentified threats. Using a real-world hardware-in-the-loop security testbed, a smart-grid cyberattack dataset was gathered to assess the proposed framework with various baseline classifiers. The findings showed that trained classifiers perform better against various kinds and locations of invisible threats, with improvement from 7\% up to 36.8\%. 
Table~\ref{tab:3} presents a comprehensive comparison of \ac{DTL}-based \acp{NIDS}, \acp{HIDS}, and \ac{DA} schemes, each providing different performance improvements, methodologies, and considerations. The selection of \acp{SM}, datasets, performance metrics, and \ac{DTL} techniques used in each scheme provides diverse insights into the implementation and effectiveness of these \ac{IDS} methods. For instance, the selection of ResNet50 and GoogLeNet as \acp{SM} by Li et~al. \citep{li2017intrusion} achieved 15.41\% and 15.68\% improvements, respectively, with their \ac{CNN} outperforming standard classifiers. This was achieved by converting the NSL-KDD dataset to image format, making it a \ac{DTL}-based \ac{NIDS}. Masum and Shahriar \citep{masum2020tl} also used KDDTest+ as a target dataset with VGG-16 and achieved an 8.65\% improvement, demonstrating the potential effectiveness of \ac{DTL}-based \acp{NIDS}. AlexNet was used by Sreelatha et~al. \citep{sreelatha2022improved} and attained impressive performance rates on the NSL-KDD and UNSW-NB15 datasets, suggesting the potential benefit of \ac{EEO} for updating \ac{DTL} weights and the use of a \ac{SOTA} optimizer for significant feature selection. The application of ROBERTa by Ünal et~al. \citep{unal2022anomalyadapters} led to a 15.5\% improvement, exemplifying a \ac{DTL}-based \ac{HIDS} for multi-anomaly task detection, which further points to the versatility of \ac{DTL}. \Ac{LSTM}- and attention-based models also exhibited improvements in the work of Ajayi et~al. \citep{ajayi2021dahid} using an \ac{AUC} metric for a host-based \ac{IDS} scheme and the work of Nedelkoski et~al. \citep{nedelkoski2020self} using \ac{PCA} for vector improvements. Interestingly, Wu et~al. (2019) employed a two-stage learning process with ConvNet, resulting in a significant 22.02\% improvement on the KDDTest-21 dataset. The traditional \ac{ML} approach of Niu et~al. \citep{niu2019abnormal} and the hyperparameter models of Zhao et~al. \citep{zhao2019transfer} achieved 75\% and 16\% improvements, respectively, both using unsupervised \ac{DTL}-based \acp{NIDS}. Notably, the latter used a novel clustering-enhanced hierarchical \ac{DTL}. Li et~al. \citep{li2021transfer} used a \ac{seq2seq} model and employed IP2Vect to convert string fields into vectors for visual clustering. Chadza et~al. \citep{chadza2020learning} applied \acp{HMM} to detect sequential network attacks, achieving a 59.95\% improvement, and Fan et~al. \citep{fan2020iotdefender} used IoTDefender, a federated \ac{DTL}-based \ac{IDS}, showing an improvement of 3.13\%. Lastly, Singh et~al. \citep{singh2021novel} used the WideDeep model, achieving a notable 19.91\% improvement, and Phan et~al. \citep{phan2020q} employed an \ac{MLP} that achieved a 43.58\% improvement on their \ac{DDoS} study.

\begin{table*}
\caption{ Summary of \ac{DTL} schemes: \ac{NIDS}, \ac{HIDS}, and \ac{DA}. $P$ is the performance of each method; $\Delta P$ represents the improvement obtained between \ac{IDS}-based \ac{DTL} and the same scheme without \ac{DTL}, if any, or with the best \ac{IDS} compared to it; HPC is high-performance computing.} \label{tab:3} \scriptsize 
\begin{tabular}[!t]{m{0.5cm} m{2.5cm} m{2cm} m{1cm} m{1cm} m{8cm}}
\hline
Ref.& Source model &  Dataset & $P$ (\%)  &   $\Delta P$ (\%)   &   Comments \\
\hline

\citep{li2017intrusion}
&   ResNet50 \newline GoogLeNet &  KDDTest-21 &   81.57 \newline 81.84    & 15.41 \newline 15.68 &  Metric is accuracy. \ac{CNN} outperforms standard classifiers. NSL-KDD dataset is converted to image format. \ac{DTL}-based \ac{NIDS}. \\

\citep{masum2020tl} &VGG-16 & ImageNet &89.30&  08.65&  KDDTest+ is a target dataset. Metric is accuracy. \ac{DTL}-based \ac{NIDS}. $\mathrm{FAR}=7.9\%$.\\

\citep{sreelatha2022improved} & AlexNet&NSL-KDD, UNSW-NB15 &99.98 \newline 99.91 &  0.12 \newline 5.6& \ac{EEO} is used to update \ac{DTL} weights. \ac{DTL}-based cloud \ac{IDS}. \ac{SOA} optimizer is used to find the significant features. Metric is accuracy.\\

\citep{unal2022anomalyadapters} &ROBERTa &Firewall log, HDFS &94.5&  15.5 &\ac{DTL}-based \ac{HIDS} , multi-anomaly task detection model based on log semantic-based anomaly detection models. Metric is F1 score. $\Delta P$ with \ac{PCA}.\\

\citep{ajayi2021dahid} &\ac{LSTM}-based & ADFA-WD:SAA &91&  8& Host-based \ac{IDS} scheme, fine-tuning uses the \ac{DTL} technique. Uses \ac{AUC} metric.\\

\citep{nedelkoski2020self}&Attention-based &HPC log &99&   25&Host-based \ac{IDS} scheme, when adopting \ac{PCA} with Logsy, the improvement reaches 28\%. Uses F1-score metric.\\

\citep{wu2019transfer} & Learning stage 1 (\mbox{ConvNet})&  KDDTest+ \newline KDDTest-21& 87.30 \newline 81.94  &  2.68 \newline 22.02  &  \ac{TL}-ConvNet outperforms other \ac{SOTA} methods. \ac{DTL}-based \ac{NIDS}. Metric is accuracy. \\

\citep{niu2019abnormal}&Traditional \ac{ML} & CTU, KDD99, CICIDS2017 &71.09&  75& Uses \ac{TCA} mapping method. Unsupervised \ac{DTL}-based \ac{NIDS}. Metric is accuracy.\\

\citep{zhao2019transfer}&Hyper\-parameters&NSL-KDD&90&  16& Uses proposed clustering enhanced hierarchical \ac{DTL}. Unsupervised \ac{DTL}-based \ac{NIDS}. Metric is accuracy.\\

\citep{li2021transfer}&seq2seq& CIDDS02 &--&  --& Uses CIDDS01 as target dataset and IP2Vect to convert string fields, such as \ac{IP} addresses, into vectors; visual clustering metric. \\

\citep{wang2021network}&BoTNet&UNSW-NB15&94.5&
 4.08& Self-supervised \ac{DTL}-based \ac{IDS}. A new data-augmentation technique is used to increase the feature representation. Metric is accuracy.\\

\citep{li2020cross}& LSTM & Many &13& 8&Cross-domain \ac{DTL}-based \ac{NIDS}. Uses TrAdaBoost-based LSTM. Metric is error rate ratio.\\

\citep{singla2019overcoming} & DNN & USNW-NB15 & $\approx$98&   26.4& Metric is accuracy. $\Delta P$ of the reconnaissance-attack type. \ac{DTL}-based \ac{NIDS}\\

\citep{sun2018network} & Maxent & Cambridge's Nprobe &98.7 &   13.3& Uses Maxent-based TrAdaBoost model and accuracy metric. \ac{DTL}-based \ac{NIDS}. \\

\citep{dhillon2020towards}&CNN-LSTM &UNSW-15&98.30&  98.43&Accuracy metric, \ac{DTL}-based \ac{NIDS}, classification speed is boosted. \\
\citep{andresini2021network}&Str-MINDFUL&CICIDS2017&99.49&  11.48&Uses accuracy metric and fine-tuning-based \ac{NIDS}. The scheme is for concept drift detection. $\mathrm{FAR}=0.19\%$.\\
\citep{li2019dart}&DART&UNSW-NBI5&93.9 &
 1.1&Uses accuracy metric and adaptation-based \ac{NIDS}. $\Delta P$ calculated with GAA-ADS ($K=10$) method. $\mathrm{FAR}=2.8\%$.\\

\citep{chadza2020learning}& \ac{HMM} & DARPA 2000&79.55&  59.95&Uses accuracy metric for current-state detection. Detects sequential network attacks using \ac{HMM}.\\

\citep{xiong2020anomaly}&Feature step 1&NSL-KDD&68.40&   01.20& Metric is accuracy. Uses \ac{TCA} mapping method and measured using \ac{MMD}. \ac{DTL}-based \ac{NIDS}.\\

\citep{fan2020iotdefender} & IoTDefender&CICIDS2017, and others&91.93&
  3.13& Uses accuracy metric. Federated \ac{DTL}-based \ac{IDS}. $\mathrm{FAR}=2\%$.\\

\citep{zhao2020network} &Net-S&UNSW-NB15&97.23&
 3.49& Uses accuracy metric. Federated \ac{DTL}-based \ac{IDS}.\\

\citep{sameera2020deep}& DNN &NSL-KDD&91.83&
 1.92& Domain adaptation with unsupervised \ac{DTL}. $\mathrm{FAR}=8.2\%$. \\

\citep{demertzis2019cyber} &OS-ELM, RBM &Gas\_Dataset&98.50& --&Multi-task learning is used with F1-score metric.\\
\citep{albelwi2022intrusion} & DNN & UNSW-NB15, CICIDS2017 &87.50&  1.09& Multi-task learning is used with combined datasets. Uses accuracy metric.\\
\citep{singh2021novel}&WideDeep&KDDCup99&99.916&  19.91&\ac{NIDS}-based \ac{IDS}, deep stacked GRU model multi-class classification, fine-tuning \ac{DTL}. Uses accuracy metric.\\
\citep{phan2020q} & \ac{MLP} &Many &99& 43.58&DDoS is the key study. Uses accuracy metric. Supervised \ac{DTL}-based \ac{NIDS}.\\
\citep{yang2021wpd}& WPD-ResNeSt & KDD99 &98.95&  3.25& Accuracy metric, $\mathrm{FAR}=3.1\%$, for CICDDoS2019 dataset, obtains $\mathrm{accuracy} =98.73\%$ and $\Delta P=2.72\%$. \ac{DTL}-based \ac{NIDS}. \\
\citep{taghiyarrenani2018transfer}&Transfer+SVM&N,Probe&98.43&   32.57& N,Dos is the target dataset. Uses accuracy metric. Domain adaptation in \ac{DTL}-based \ac{NIDS} is used.\\

\citep{singla2020preparing}&GAN &NSL-KDD&94.89&
  17.9&  Uses adversarial \ac{DA} in \ac{NIDS} to solve the scarcity of labeled data, UNSW-NB15 is used as the target dataset. Comparisons are made with a fine-tuning approach in terms of accuracy.\\
\hline
\end{tabular}
\end{table*}



\subsection{Partial \ac{DTL}-based \ac{IDS}}
The concept of \ac{DA} involves identifying the classes present in both the \ac{SD} and \ac{TD}, which are referred to as shared label spaces; the classes present only in the \ac{SD} are referred to as private label spaces. It is not advisable to directly transmit training data comprising both private label spaces (such as web attacks in the \ac{IoT}) and shared label spaces (such as \ac{DDoS} attacks in the \ac{IoT}) to the \ac{TD}, as this is likely to result in a negative transfer. The partial-\ac{DA} approach has been proposed as a solution to address this issue of inconsistent label spaces. Partial \ac{DA} is an approach that helps to locate the common sections (shared label spaces) and constrain the private label spaces of the \ac{SD} that have little association with the destination domain to enhance transfer performance \citep{himeur2023video,fan2021intrusion}.

By mapping two domains to a domain-invariant feature space, Fan et~al. in \citep{fan2021intrusion} suggested a framework employing a weighted adversarial nets-based partial-\ac{DA} approach to solve the issue of inconsistent label spaces. By transferring information from the large publicly labeled dataset of the conventional internet to the unlabeled dataset of the \ac{IoT}, this method can be used to build a highly accurate intrusion model. The suggested approach may also identify unknown \ac{IoT} threats with the use of information from the conventional internet. The experiments were conducted on the CIC-IDS2017 dataset (the old version of the CIC-IDS2018 dataset). The proposed approach was found to have average accuracy values of around 87.42\% and 98.19\% for known and unknown attacks, respectively. Similarly, by incorporating an auxiliary domain, the scheme described by Wu et~al. \citep{wu2022joint} can facilitate domain alignment by partly masking scenario heterogeneities across domains. The goal is to protect fundamental semantic features to aid \ac{TD} intrusion detection. As a result, it lessens the source--target disparity, resulting in a shared feature space that is discriminative and domain invariant. A 10.3\% accuracy boost was achieved in comparison with the \ac{SOTA} methods using the proposed method.

\subsection{Federated \ac{DTL}-based \ac{IDS}}
\begin{figure*}[!ht]
    \centering
    \includegraphics[scale=0.7]{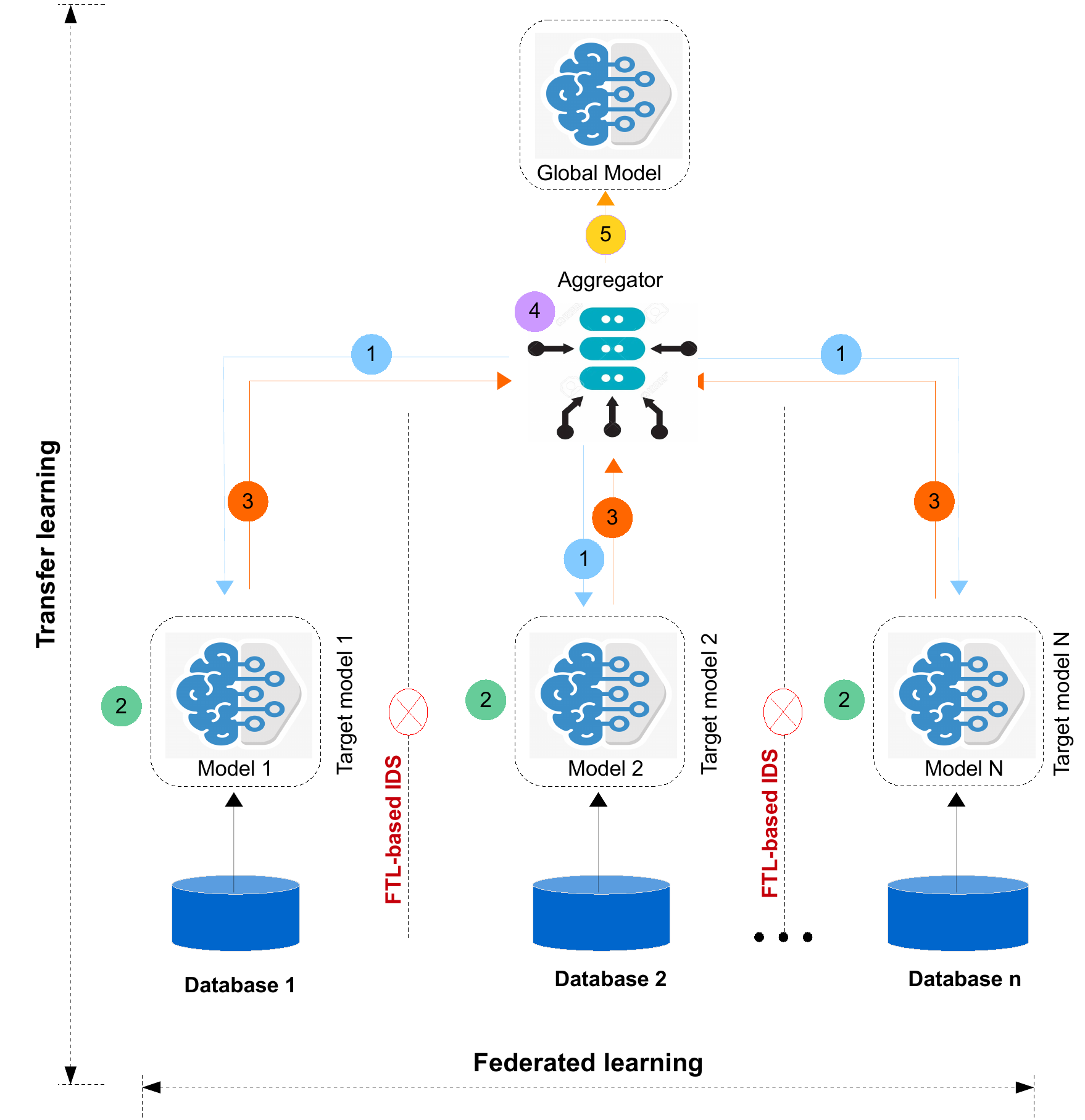}
    \caption{\ac{FTL}-based \ac{IDS} workflow.}
    \label{fig:ftl}
\end{figure*}

A global \ac{DL} model can be built and reused with local data on the edge side using the \ac{FL} technique without sharing the local data of any user. Once a newly local model has been aggregated with a global model, a combined and tested global model can be returned to edge-connected devices. \Ac{FTL} is used to increase training effectiveness via \ac{TL} while employing \ac{FL} to protect privacy by using a local training model. Unreliable end users should not be chosen for \ac{FL} to conduct the training. Fig.~\ref{fig:ftl} illustrates the principle of an \ac{FTL}-based \ac{IDS} workflow. The training process in \ac{FTL} is as follows. (1)~Parties train their local models independently on their datasets, minimizing task-specific loss functions to reduce prediction-label discrepancies. (2)~Models are communicated and aligned to create a joint architecture, sharing parameters or performing feature extraction. (3)~Knowledge transfer occurs through parameter sharing or feature extraction. (4)~Trusted aggregation methods, such as secure multiparty computation or homomorphic encryption, are used to aggregate encrypted gradients. (5)~The aggregated gradients are decrypted, and the global model parameters are updated, preserving data privacy while benefiting from collective knowledge. Because of this, researchers are looking toward using \ac{FTL}-based \ac{IDS} to protect edge devices.

The effectiveness of employing \ac{FTL} trending approaches with \ac{DL} algorithms to power \acp{IDS} to protect \ac{IoT} applications was shown by Otoum et~al. \citep{otoum2022federated}, along with a thorough analysis of their use. Their work considered the \ac{IoMT} as a use case. They found that the best performance was obtained when an \ac{IDS}-based FTL-\ac{CNN} model was used, and its accuracy reached up to 99.6\%. Other works have also used \ac{FTL} to solve many \ac{IDS} issues \citep{cheng2022federated,otoum2021federated,fan2020iotdefender,zhao2020network}. For example, Otoum et~al. \citep{otoum2021federated} suggested an \ac{FTL}-based \ac{IDS} for the purpose of protecting patients' connected healthcare devices. The \ac{DNN} technique is used by the model to train the network, transfer information from connected edge models, and create an aggregated global model that is tailored to each linked edge device without sacrificing data privacy. Testing was conducted using the CICIDS2017 dataset, and the accuracy obtained was found to be up to 95.14\%. Similarly, integration of \ac{FTL} and \ac{DRL}-based client selection was exploited by Cheng et~al. \citep{cheng2022federated}; the number of participating clients was restricted using a \ac{DRL}-based client-selection technique. Their results show that the accuracy may be greatly increased and converge to 73\% by excluding malicious clients from the model training. Fan et~al. \citep{fan2020iotdefender} suggested IoTDefender, an \ac{FTL}-based \ac{IDS} for \ac{5G} \ac{IoT}. The layered and distributed architecture of IoTDefender is perfectly supported by \ac{5G} edge computing. IoTDefender aggregates data using \ac{FL} and creates unique detection models using \ac{DTL}. It allows all \ac{IoT} networks to exchange data without compromising user privacy. The authors' test findings showed that IoTDefender is more successful than conventional methods, with a detection accuracy of 91.93\% and a reduced percentage of false positives when compared to a single unified model. Similarly, Zhao et~al. \citep{zhao2020network} used \ac{FTL}-based \ac{IDS}, to build an \ac{IDS} framework and exploit the aforementioned advantages of combining both \ac{FL} and \ac{TL}. Their experiments were conducted on the UNSW-NB15 dataset, and performance was found to reach an accuracy of 97.23\%. In contrast to the work of Zhao et~al., Otoum et~al. \citep{otoum2022feasibility} compared both \ac{FL} and \ac{DTL} for the same \ac{IDS} method and concluded that \ac{TL}-based \ac{IDS} achieves the highest detection of $\approx$94\%, followed by \ac{FL}-based \ac{IDS} with $\approx$92\%.

\subsection{DTL-based adversarial \ac{IDS}}
\Ac{DL} networks, such as \acp{CNN} and \acp{AE}, have been extensively investigated and used in the field of intrusion detection in recent years, and they have been found to obtain high detection performance. \Ac{DL} techniques, however, sometimes depend on a huge quantity of data, making it challenging for them to efficiently identify new or variant encrypted ransomware attacks. The primary goal of a \ac{GAN} is to maximize the distinction between real and adversarial samples by having the discriminator create adversarial samples (pseudo samples) as realistically as is feasible \citep{lee2017generative}. Under certain game mechanics, the discriminator may be able to differentiate actual samples with a high degree of accuracy using an unsupervised learning approach. As a result, if a normal sample is assumed to be the true sample, the \ac{GAN} discriminator can correctly identify it. However, whether knowingly or unknowingly, it may identify an aberrant sample.

\begin{figure}[b!]
    \centering
    \includegraphics[scale=0.8]{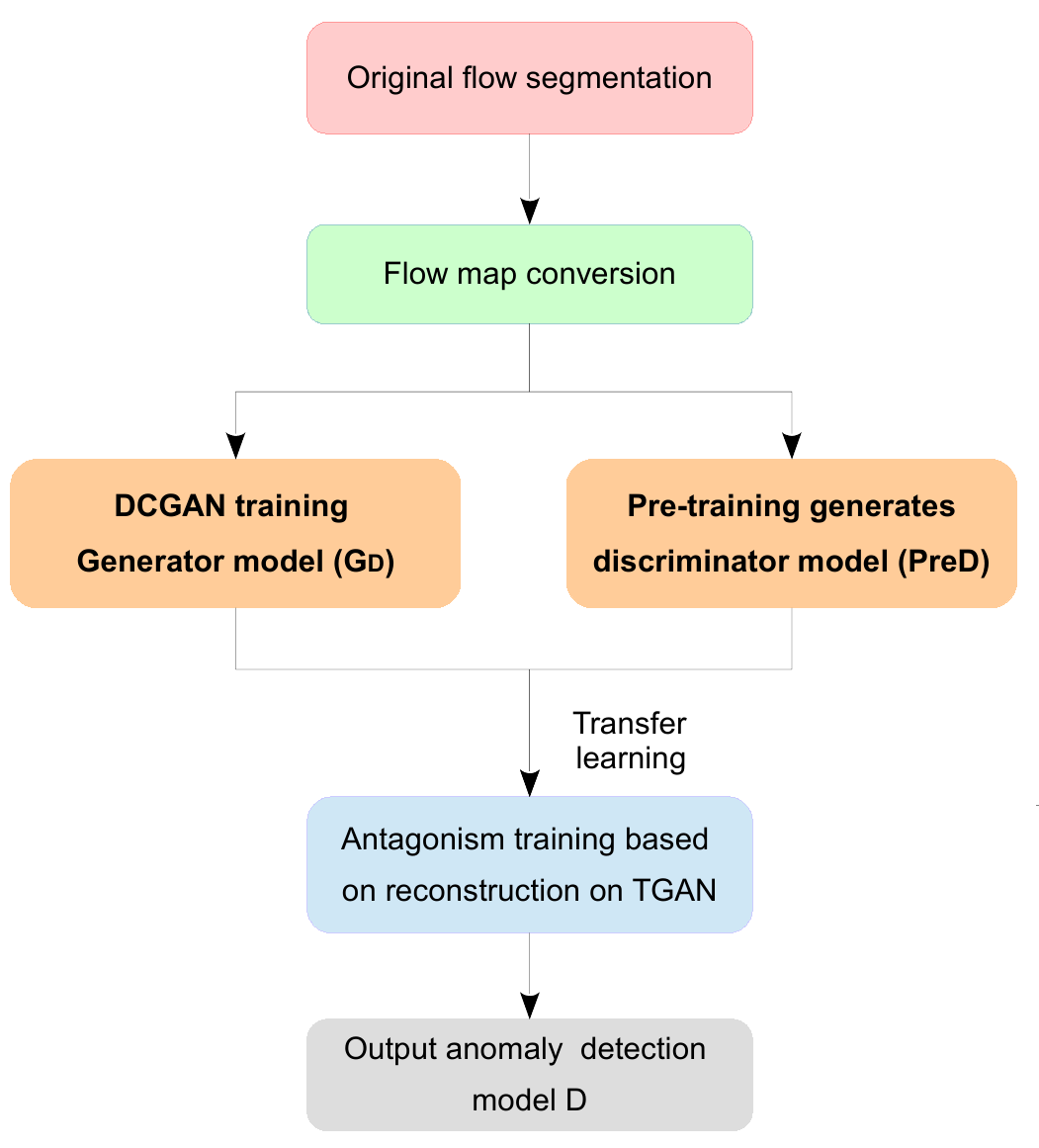}
    \caption{Framework for \ac{IDS} based on TGAN-\ac{IDS} \citep{zhang2020two}. }
    \label{fig:tganids}
\end{figure}

Based on this idea, Zhang et~al. \citep{zhang2021dual} proposed a transferred \ac{GAN}-\ac{IDS} (TGAN-\ac{IDS}) algorithm framework based on dual \acp{GAN} to identify unknown or variant encrypted ransomware attacks. The knowledge is transferred from a pre-trained \ac{CNN} model named PreD to the discriminator of TGAN. The structure and parameters of the generator in a deep convolutional \ac{GAN} are transferred to the generator of TGAN. Fig.~\ref{fig:tganids} presents a flowchart of the TGAN-\ac{IDS} framework. Experiments were conducted on a mixed dataset created using CICIDS2017 and other ransomware datasets. It was found that TGAN-\ac{IDS} works well in terms of the metrics of detection accuracy, recall, F1 score, etc. Studies with the KDD99, \ac{SWaT} and \ac{WADI} datasets also showed that TGAN-\ac{IDS} is acceptable for detection of additional unencrypted unknown network attacks.

\ac{ML}/\ac{DL} techniques assume that testing and training data follow similar or comparable data distributions, but this may not hold in dynamic time-varying systems such as a smart grid. As operational points may vary drastically over time, the ensuing data-distribution changes might lead to reduced detection performance and delayed incident responses. To address this difficulty, Zhang and Yan \citep{zhang2020semi} provided a semi-supervised framework based on domain-adversarial training to transfer the knowledge of past attack occurrences to identify returning attacks at various hours and load patterns. The performance of the suggested classification algorithm was assessed against well-studied fake-data-insertion assaults generated on the IEEE 30-bus system, and the findings demonstrated the superiority of the framework against persistent threats repeating in the highly dynamic environment of a smart grid. The best-case increase reached 36.0\% compared to the \ac{ANN} algorithm.

\section{\ac{DTL}-based \acp{IDS} for communication networks}
\label{sec5}
This comprehensive section covers a wide range of \ac{IDS} applications within the domain, including \ac{DTL}-based \acp{IDS} for device fingerprinting, wireless \acp{IDS} targeting \ac{IoT} and \ac{IoV} environments, host \acp{IDS}, network \acp{IDS} employing unsupervised, supervised, and semi-supervised techniques, as well as \acp{IDS} tailored for smart-grid systems. By examining these diverse areas, this section aims to highlight the effectiveness and adaptability of \ac{DTL}-based \acp{IDS} in fortifying the security posture of communication networks. The insights gained from this exploration can pave the way for innovative solutions to detect and mitigate potential threats in various communication network settings, ultimately contributing to enhanced network resilience and protection.

\subsection{DTL-based device-fingerprint \ac{IDS}}
Anomalies in traffic statistics are typically indications to intrusions such as \acp{DDoS} that generate unusually high volumes of traffic. Emulation attacks, on the other hand, are a different kind of threat that cannot be seen in traffic data. In a primary-user emulation attack, for instance, the attacker impersonates a primary user to keep other secondary users from competing for the same spectrum band, increasing the attacker's spectrum utilization. \ac{DL} or even \ac{ML} methods may be used for this type of attack to categorize the signals coming from the target device and the attacker's device. \ac{IDS}-based \ac{DTL} using device fingerprinting can be divided into two categories:

\vskip2mm
\noindent \textbf{(a)~Methods considering the impact of nearby devices.} For example, methods have been created to recognize emulation assaults based on device fingerprints (i.e., distinctive characteristics of each device) and environmental impacts (i.e., the influence of the physical environment on the fingerprint of the device) \citep{sharaf2016authentication,dabbagh2019authentication}. In particular, the suggested methodologies model and estimate the environmental impacts on the fingerprint of each device and look for anomalous discrepancies between the actual and predicted environmental effects. Since network devices that are physically adjacent to one another frequently exhibit comparable environmental effects, transductive \ac{TL} techniques have been suggested \citep{sharaf2016authentication,dabbagh2019authentication} as a way to share the knowledge of environmental impacts across neighboring devices. In particular, using a nearby device of a different kind, Sharaf-Dabbagh and Saad \citep{sharaf2016authentication} proposed a \ac{TL} technique that transfers the information acquired from a device such as a smartphone to another one such as a sensor. The idea is that similar environmental impacts may be shared by the two devices if they are physically adjacent. Higher transfer weights (i.e., the percentage of the \ac{SM}'s weight transmitted to the \ac{TM}) can produce better results, according to experimental data, which demonstrate that the used \ac{TL} technique can enhance detection performance by 70\%. As an alternative, the same authors \citep{dabbagh2019authentication} suggested identifying the alterations in the environs of close-by devices and using this knowledge to obviate their impacts on the fingerprints of the devices. The impacts of environmental changes are filtered out by the applied \ac{TL} system, according to simulation findings, which results in a lower \ac{FP} rate than the strategy without \ac{TL}.

\vskip2mm
\noindent \textbf{(b)~Methods not considering the impact of nearby devices.} Similar methods for identifying \ac{IoT} devices based on fingerprints have also been suggested \citep{zhao2018identification,sharaf2015transfer}. These methods, however, do not take environmental influences into account; instead, they make use of the information they have learned from prior encounters. Zhao et~al. \citep{zhao2018identification} presented an inductive \ac{TL} method for training an \ac{ML} model for the classification of \ac{IoT} devices by combining a large quantity of source data, gathered in the past, with a smaller target dataset, gathered from current data. Instead, they make use of what they have learned from prior knowledge. Sharaf-Dabbagh and Saad \citep{sharaf2015transfer} created a Bayesian model to categorize device fingerprints to identify intrusions. The authors initially created an abstract knowledge database made up of the device fingerprints from every time slot, and new fingerprints were obtained at a later period. In their method, to cluster and assign the new fingerprints to the devices, an inductive \ac{TL} method is used to merge the new fingerprints with the previous fingerprints contained in the abstract knowledge database. The primary distinction between the two methodologies is that the latter \citep{sharaf2015transfer} uses considerably shorter time frames, whereas the former \citep{zhao2018identification} considers changes in physical characteristics over a lengthy period of time (i.e., one year) and thus requires notably more time to complete. Table~\ref{tab:4} summarizes the aforementioned work relating to \ac{IDS}-based traditional \ac{DTL} and \ac{DTL}-based device fingerprinting.

\begin{table*}[H]
\caption{Summary of \ac{IDS}-based \ac{DTL}, auto-encoded \ac{DTL}-based \ac{IDS}, \ac{DTL}-based device fingerprinting, and \ac{DTL}-based wireless \ac{IDS}. Here, ``T'' indicates transductive \ac{DTL}, ``I'' indicates inductive \ac{DTL}, IGMM is an infinite Gaussian mixture model, DST-TL is deep neural network and adaptive self-taught-based \ac{TL}, and RFID is radio-frequency identification. When the dataset is not mentioned, this indicates that the authors used their own samples. If many metrics were used, only the ``improved'' one is mentioned.}
\label{tab:4}
\begin{tabular}{p{0.5cm}p{1.7cm}p{3cm}cp{1cm}p{8cm}}
\hline

Ref. & \ac{AI} used & How knowledge is transferred & \ac{DTL} & Improv. &  Comment\\
 \hline
 \citep{zhou2020indoor}&Multi-CNN &Adaptation&T&   23.54\%&Uses \ac{KNN} classifier, accuracy metric.\\

\citep{vu2020deep} &AE&Fine-tuning&T &13.54\%&Uses \ac{AUC} metric, nine recent \ac{IoT} datasets, \ac{MMD} for adaptation; IoT-4 dataset's result is the best.\\

\citep{mehedi2021deep}&LeNet&Adaptation&T &3\%& Uses accuracy metric, \ac{MMD} for adaptation. Effective for \ac{CAN} bus protocol.\\
\citep{wen2019time}&CNN&Fine-tuning&T& 21\% & Uses \ac{IoU} metric, image dataset in \ac{IDS} domain. \\
\citep{xu2020intrusion}&LeNet&Fine-tuning& T& 5\%& Uses F1-score metric; KDDCUP99 is in image format.\\
\citep{tariq2020cantransfer} &ConvLSTM&One-shot learning&I &26.69\%& Transcends the \ac{SOTA} schemes for detection of new attacks.\\

\citep{kim2020transfer}&\ac{RVAE}& Fine-tuning&T &3.1\%& Uses AUROC metric, CTU-13 dataset; unsupervised learning.\\

\citep{dos2021reminiscent}&AE&Update AE&T &23.9\%& Uses \ac{RR} metric, only 22\% of labeled training data; decreases computational costs by 28\%.\\
\citep{qureshi2020intrusion} &DST-TL&Self-taught&I&4.7\%& Uses accuracy metric, KDDTest+ dataset.\\

\citep{sharaf2016authentication}&IGMM& Environmental impact &T& 8\%& Uses accuracy metric, Bhattacharyya distance.\\

\citep{dabbagh2019authentication}&\ac{DTL}&Environmental impact of nearby devices&T& 40\% & Uses accuracy metric, Bhattacharyya distance, RFID dataset. Can provide up to 70\% improvement for device authentication.\\
\citep{zhao2018identification} &SVM&Previous device data &I&  10\%& More time-consuming than a previous work \citep{sharaf2015transfer}.\\
\citep{sharaf2015transfer}&Bayesian &Past fingerprint &I&  3.5\%& Uses recognition rate metric, \ac{DTL} AdaBoost algorithm. \\
\citep{yu2019individual} &\ac{KNN}& Signal's feature weights &I&  13.2\%&Uses recognition-rate metric, TrAdaBoost \ac{DTL} algorithm\\

\citep{wang2022intrusion}&KnTrELM &Adaptation&T&  18.78\%&Uses \ac{KNN} classifier, accuracy metric. R2L attack.\\

\citep{khoa2021deep}&Federated network A&Fine-tuning&T&  40\%&Uses N-BaIoT, KDD, NSL-KDD, and UNSW datasets, unsupervised learning, accuracy metric.\\
\citep{guan2021deep}&Many&Fine-tuning&T&  Non & Uses 10\% of USTC-TFC2016 dataset; EfcientNet-B0 model is the best compared to LeNet-5 and BiT models.\\
\citep{otoum2021signature}&DBN&Self-taught &I&  6.25\%&Uses accuracy metric, NSL-KDD dataset. Reduces training time by up to 14\%.\\
 \hline
\end{tabular}
\end{table*}


The compilation of methods applied in \ac{DTL} for \acp{IDS} presents a diverse landscape of \ac{AI} techniques and their outcomes. Zhou et~al. \citep{zhou2020indoor} used a multi-CNN model; this proved effective, resulting in a 23.54\% improvement, but it faced challenges due to the complex architecture and risk of overfitting, highlighting an avenue for future research in model simplification and regularization techniques. Meanwhile, the \ac{AE} method of Vu et~al. \citep{vu2020deep} resulted in a 13.54\% improvement, optimally on the IoT-4 dataset, suggesting the need for research into the method's applicability across different datasets and problem domains. Employing LeNet for fine-tuning, Mehedi et~al. \citep{mehedi2021deep} and Xu et~al. \citep{xu2020intrusion} achieved modest improvements, hinting at the potential benefits of exploring more complex and modern architectures. Tariq et~al. \citep{tariq2020cantransfer} used ConvLSTM to obtain a remarkable 26.69\% improvement, surpassing \ac{SOTA} schemes, albeit with limitations in handling complex, multidimensional data. The self-taught deep neural network and adaptive self-taught-based \ac{TL} method of Qureshi et~al. \citep{qureshi2020intrusion} showed a 4.7\% improvement, calling for wider application and testing on various \ac{IDS} problems and datasets. Sharaf et~al. \citep{sharaf2016authentication} harnessed Bayesian methods, yielding a 3.5\% improvement, but this was potentially limited by model assumptions, indicating a need for more flexible, nonparametric models. Khoa et~al. \citep{khoa2021deep} employed a federated network, achieving an impressive 40\% improvement, yet its complexity and context-specific application suggest future research should aim for improved efficiency and practicality. Lastly, Otoum et~al. \citep{otoum2021signature} used a self-taught \ac{DBN} for a 6.25\% improvement and training-time reduction, with future work potentially focused on enhancing \ac{DBN} training or investigating newer architectures.

\subsection{\ac{DTL}-based wireless \ac{IDS}}
Examining transient signals---which contain amplitude, phase, and frequency---can also identify wireless devices, and this has lower hardware requirements than examining steady-state signals. A transient is defined as a sudden change in a signal, such as may occur during an intrusion. However, because the wireless devices that need to be identified are frequently uncooperative, wireless-device recognition may suffer from a limited amount of transient-signal data. For this reason, Yu et~al. \citep{yu2019individual} discussed an inductive \ac{TL} technique for individual identification of wireless devices with little sample data. In particular, a wireless device acting as the reference \ac{AP} initially collects transient signals from other wireless devices before classifying the signals into \ac{SD} and \ac{TD}. Then, using an approach based on transient envelope feature extraction with entropy weighting, the feature weights of transient signals in the \ac{SD} are retrieved and transferred to the \ac{TD}. A new \ac{KNN}-based model can be trained using the transferred information by adjusting the feature weights of the \ac{SM} samples. The suggested \ac{TL} technique may increase the classification accuracy by up to 13.2\% compared to that of normal \ac{KNN} without the \ac{TL} method.

\begin{figure*}[!t]
    \centering
    \includegraphics[scale=0.9]{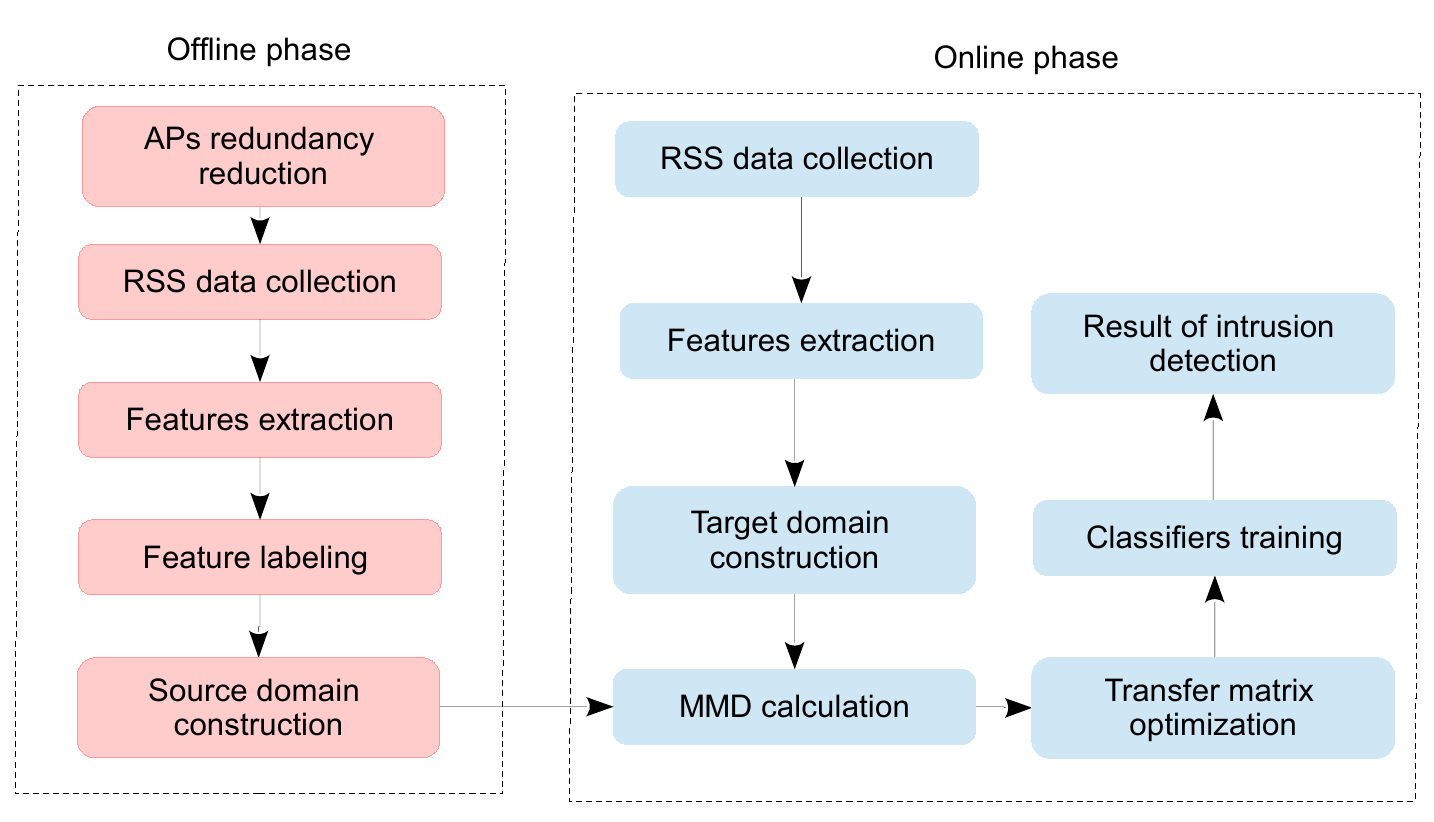}
    \caption{Example of \ac{MMD} calculation-based offline to online architecture.}
    \label{fig:RSS}
\end{figure*}

In the \ac{5G} wireless communication network environment, indoor \acp{IDS} based on \ac{WLAN} have been extensively employed for security monitoring, smart homes, entertainment gaming, and many other areas. Using deep signal feature fusion and minimized multiple kernel \ac{MMD}, Zhou et~al. \citep{zhou2020indoor} suggested a unique \ac{WLAN} indoor intrusion-detection technique. The dimensionality reduction and feature fusion of the \ac{RSS} is carried out using a multi-branch deep \ac{CNN}, and the tags are obtained in accordance with the offline and online \ac{RSS} fusion features that correspond to the silence and intrusion states; an example is shown in Fig.~\ref{fig:RSS}. Based on this, the \ac{SD} and \ac{TD} are then constructed. Other works \citep{zhou2019maximum,zhou2019indoor,zhou2021indoor} have employed the same \ac{DTL} technique for wireless \ac{IDS}; however, one took the time-variant nature of the \ac{RSS} into consideration \citep{zhou2019maximum}, another used intra-class \ac{DTL} to reduce the gap between the \ac{RSS} data in the \ac{SD} and \ac{TD} at the same locations \citep{zhou2019indoor}, and a third used \ac{GAN}-based supervised learning as a data-augmentation technique to create false \ac{RSS} data that were identical to the real data \citep{zhou2021indoor}. The latter can be used to identify unauthorized intrusions with little effort. Li et~al. \citep{li2019integrated} used fuzzy rough set-based joint decision criteria \citep{beaubouef2000fuzzy} to remove duplicate mobile \acp{AP} before proposing to build the \ac{SD} and \ac{TD} using, respectively, link-layer information labeled in the online phase and link-layer information unlabeled in the offline phase. The information in these two domains may then be transferred into the same subspace by computing the \ac{MMD} to generate the transfer matrix (Fig.~\ref{fig:RSS}).

A unique approach suggested by Xia et~al. \citep{xia2021wireless} consists of mixing \ac{DRL} with the concept of \ac{DTL} (fine-tuning) for the first time to address the issue of continually training original complicated models on both old and fresh anomalous traffic datasets. In this work, the \ac{SM} was trained on the NSL-KDD dataset, while the \ac{TM} was fine-tuned and trained on the Aegean Wi-Fi Intrusion Dataset. This involved using \ac{DRL} with a new dueling double deep Q-network model as its agent to improve its ability to detect abnormal traffic. The authors claimed that the proposed scheme is the best compared to \ac{SOTA} methods in terms of real-time detection.

\begin{itemize}
    \item \textbf{DTL-based \ac{IDS} for \ac{IoT}.} Some \ac{IoT} devices may have their functionality altered or hindered by malicious actors, and this will cause infected \ac{IoT} devices to behave differently when defending against attacks as well as participating in them. Many solutions have been proposed to these problems \citep{ge2021towards,yilmaz2021transfer,ullah2021design,mehedi2022dependable,khoa2021deep,guan2021deep}. To evade the attack-oriented feature-selection process in \ac{IoT} devices, generic features might be generated based on the header-field data in individual \ac{IP} packets. The source has a feed-forward neural network model with multi-class classification; \ac{DTL} is then used in the \ac{TD} to encode high-dimensional category information to build a binary classifier \citep{ge2021towards}. The application of \ac{DTL} to create an \ac{IDS} for such a continuously evolving \ac{IoT} environment was investigated by Yılmaz et~al. \citep{yilmaz2021transfer}. Here, \ac{DTL} was used in two contexts: knowledge transfer for creating appropriate intrusion algorithms for new devices, and knowledge transfer for identifying novel attack types. A routing protocol for low-power and lossy networks, which is designed for resource-constrained wireless networks, was used in this research as an example protocol, and specific attacks were made against it. Ullah and Mahmoud \citep{ullah2021design} described a way to design an \ac{IDS} scheme that employs \ac{CNN}, and this was tested on many intrusion datasets for \ac{IoT} environments; \ac{DTL} was used to build a target binary and multi-class classification using a multi-class pre-trained model as a \ac{SM}. Mehedi et~al. \citep{mehedi2022dependable} suggested a \ac{DTL}-based residual neural network (P-ResNet) \ac{IDS} that can be well trained with only a minimal amount of \ac{TD} data. The authors guaranteed its effectiveness by taking dependability performance-analysis factors into account, such as availability, efficacy, and scalability. The proposed P-ResNet-based \ac{IDS} was found to reach an average detection rate of 87\%, outperforming the schemes against which it was compared. The novelty of Guan et~al.'s scheme \citep{guan2021deep} lies in it employing \ac{DTL} on \ac{5G} \ac{IoT} scenarios to train the \ac{TD} without labels while retaining only 10\% of it. The goal of this approach is to reach an accuracy closer to the results of a fully trained \ac{5G} \ac{IoT} dataset.

\item \textbf{DTL-based \ac{IDS} for \ac{IoV}.} Many \ac{TL} strategies have been presented in recent years. Li et~al. \citep{li2020transfer} considered intrusion detection with various forms of assaults in an \ac{IoV} system. According to their experimental findings, this model greatly increased detection accuracy: by at least 23\% when compared to the traditional \ac{ML} and \ac{DL} techniques currently in use. The performance of \ac{TL} models has recently been improved using the deep computational intelligence system \citep{lu2015transfer}. The in-vehicle network new-generation labeled dataset presented by Kang et~al. \citep{kang2021car} is well suited to the application of \ac{DTL} models; this is because \ac{DTL} techniques have been found to perform better for time-series classification than other traditional \ac{ML} or \ac{DL} models \citep{li2020deep,kimura2020convolutional}. The distinctive contributions of Mehedi et~al. \citep{mehedi2021deep} include creating a \ac{DTL}-based LeNet model, evaluation taking into account real-world data, and the selection of effective attributes that are best suited to identifying harmful \ac{CAN} signals and effectively detecting normal and abnormal behaviors. The output \ac{DTL} models have demonstrated improved performance for real-time in-vehicle security. Another scheme presented by Otoum an Nayak \citep{otoum2021signature} was developed to secure external networks and an \ac{IoV} system. In this scheme, each network-connected vehicle acts as a packet inspector to help the \ac{DTL}-based \ac{IDS}; the \ac{SM} is \ac{DBN} and the \ac{TM} is a \ac{DNN} model, and the attacks are discovered, logged, and added to a cloud-based signature database. The packet inspector is supported by a blacklist of intrusion signatures that are installed in the connected vehicles.
\end{itemize}

\subsection{DTL-based host \ac{IDS}}
\Acp{HIDS} are a type of \ac{IDS} operating at the host level rather than at the network level. \Acp{HIDS} monitor and analyze the activity of servers, workstations, and laptops to detect signs of malicious activity. They typically use a combination of software and hardware components to collect and analyze data from the host, such as system logs, process and system activity, and file-system changes. \Acp{HIDS} are used to detect a variety of different types of attack, including malware infections, unauthorized access, and attempts to modify or delete critical data. They can also detect internal threats, such as malicious insiders or accidental data breaches. \Acp{HIDS} are typically deployed in addition to network-based \acp{IDS}, which operate at a higher level in the network and provide a more comprehensive view of network activity.

Instead of using a completely fine-tuning-based approach, Ünal and Dağ \citep{unal2022anomalyadapters} suggested an \ac{HIDS} method using \ac{AA} as a \ac{TL} technique for identifying visual representation. \ac{AA} is a flexible framework for multi-anomaly task identification; it employs adapters to learn a log structure and anomaly categories, and pre-trained transformers' versions are applied to encode log sequences. Contextual data is gathered using an adapter-based technique, which also prevents information loss during learning and learns anomaly-detection activities from various log sources without overusing parameters. To identify intrusions across several domains for cybersecurity, Ajayi and Gangopadhyay \citep{ajayi2021dahid} developed a \ac{DA} host-based \ac{IDS} framework. They specifically designed a \ac{DL} model that uses much less \ac{TD} data for host-based intrusion detection. To achieve this, they employed ADFA-WD as the \ac{SD} dataset and ADFA-WD:SAA as the \ac{TD} dataset. Their results demonstrate that \ac{DTL} may reduce the demand for large domain-specific datasets for creating host-based \ac{IDS} models. By using a transformer architecture with a multi-head attention mechanism, Logsy \citep{nedelkoski2020self} drifts away from the necessity for log-processing tools to minimize information loss in producing templates. This is accomplished by semantic-based efforts using pre-trained embedding to transfer information into the process of anomaly identification.

\subsection{DTL-based network \ac{IDS}}
By detecting illegal access to a computer network and looking into potential security breaches, an \ac{NIDS} plays a crucial role in its protection. Traditional \acp{NIDS} have trouble defending against modern, complex, and unexpected security assaults. Consequently, there is a growing demand for autonomous \acp{NIDS} that can identify harmful activity more precisely and reduce the \ac{FAR}.

\begin{figure*}[ht!]
    \centering
    \includegraphics[scale=0.8]{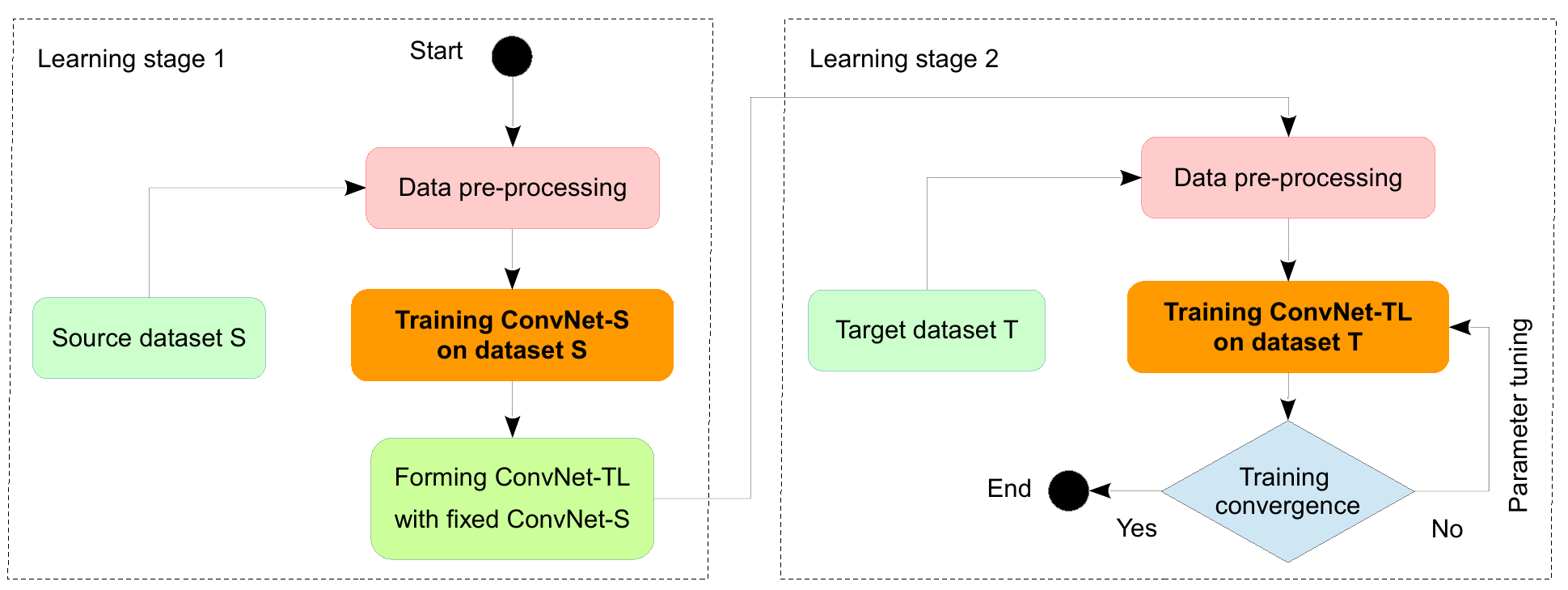}
    \caption{Example of \ac{TL}-based \ac{NIDS} approach using pre-trained \ac{CNN} model.}
    \label{fig:TL-ConvNet}
\end{figure*}

With the introduction of the \ac{TL}-based approach, the \ac{TL}-ConvNet scheme presented by Wu et~al. \citep{wu2019transfer} uses the knowledge gained from a base dataset and transfers it to the learning process of the target dataset. The UNSW-NB15 and NSL-KDD datasets were used by \ac{TL}-ConvNet as its source (base) and target datasets, respectively (Fig.~\ref{fig:TL-ConvNet}). The experimental findings in this study demonstrate that adopting \ac{DTL} in place of standard \ac{CNN} can significantly enhance performance. \ac{TL}-ConvNet's accuracy on the KDDTest+ and KDDTest-21 datasets was 87.30\% and 81.9\% (with improvements ($\Delta$ accuracy) of around to 2.68\% and 22.02\%), respectively. It should be noted that \ac{TL}-ConvNet has high \ac{FAR} values on the KDDTest+ (21.38\%) and KDDTest-21 (98.65\%) datasets, despite the fact that it can greatly increase overall accuracy.

In the same context, ResNet and GoogLeNet \citep{li2017intrusion} also employ data conversion; however, to use the existing ConvNet trained on an image dataset, the two designs convert the network-traffic data into pictures. The knowledge gained from the image dataset is then transferred to the final model for the \ac{IDS}, which is based on the trained ConvNet. A deep neural network built on the pre-trained VGG-16 architecture was suggested by Masum and Shahriar \citep{masum2020tl} as part of a novel network \ac{IDS} framework. The framework uses a two-stage method in which features are first extracted from VGG-16, which has been pre-trained on the ImageNet dataset, and a deep neural network is then applied to the extracted features for NSL-KDD dataset classification. Applying \ac{DTL} was found to lead to the highest accuracy rate and a competitive \ac{FAR} when compared to other \ac{SOTA} methods.

Conventional \ac{DL}-based anomalous network-traffic detection methods are greatly impacted by traffic-data-distribution issues and innovative network attacks. Because it is difficult to gather enough labeled data and generalize it to all traffic, a \ac{TL} method is used to remove the need for identical independent training and testing dataset distributions. By doing this, models trained on an original dataset with a different probability distribution can be more effectively used to improve detection performance in new scenarios. For example, Taghiyarrenani et~al. \citep{taghiyarrenani2018transfer} suggested a technique that uses the immediately available labeled samples from a separate network within a target network. These various samples are processed to serve as a training dataset for learning models that can then be used to predict anomalous samples in the designated network. The results of their experiments demonstrated that this technology successfully transmits knowledge between various networks.

Conventional \ac{DL}-based \ac{NIDS} methods are greatly impacted by traffic-data distribution issues and innovative network attacks, because it is difficult to gather enough labeled data and generalize it to all traffic. Three approaches have been used in the \ac{SOTA} methods to solve this problem:
\begin{itemize}
\item \textbf{Unsupervised techniques.} Numerous studies have considered using an unsupervised technique \citep{xiong2021anomaly,niu2019abnormal,zhao2017feature,zhao2019transfer}. The goal of these research studies is to enhance the performance of an anomalous network traffic detection model that is trained using open-source datasets and applied to a real-world target network. Researchers make the assumption that the target network does not include any labeled network traffic. Xiong et~al. \citep{xiong2021anomaly} used unsupervised \ac{DTL} to remove the restriction regarding the need for identical independent distributions for the training and testing datasets. By doing this, models trained on original datasets with different probability distributions are able to improve their detection performance in new scenarios. The scheme is realized in two steps: the first is feature mapping between the source and target unbalanced network traffic data, and the second is training a classifier based on the mapped features. The approaches of both Xiong et~al. \citep{xiong2021anomaly} and Nui et~al. \citep{niu2019abnormal} were based on using \ac{TCA} as a \ac{DTL} algorithm, as depicted in Fig.~\ref{fig:TCA}, to reduce the distance between the \ac{SD} and \ac{TD} distributions.

The HeTL approach was developed by Zhao et~al. \citep{zhao2017feature}; it employs a matrix to translate the \ac{SD} dataset and the \ac{TD} dataset to a high-dimensional space, and it then incorporates their distance into the optimization objectives. The mapped characteristics are trained using a fundamental \ac{ML} method such as \ac{SVM}, and this resulted in significant progress. The CeHTL algorithm is an improvement of the HeTL algorithm developed by Zhao et~al. \citep{zhao2019transfer}. In this latter work, they include a clustering stage before feature mapping and enhance the automated parameter selection. The matrix mapping approach, however, naturally lacks expressiveness when compared to a neural-network methods. PF-TL \citep{jung2021pf} is also an improvement over the optimization method of HeTL. It pulls features from an intrusion-detection event's payload and transfers them to the \ac{TM} to overcome the lack of unlabeled target data. The accuracy improvement was found to be 30\% when compared to the non-\ac{TL} method.

\begin{figure}[!t]
    \centering
    \includegraphics[scale=0.85]{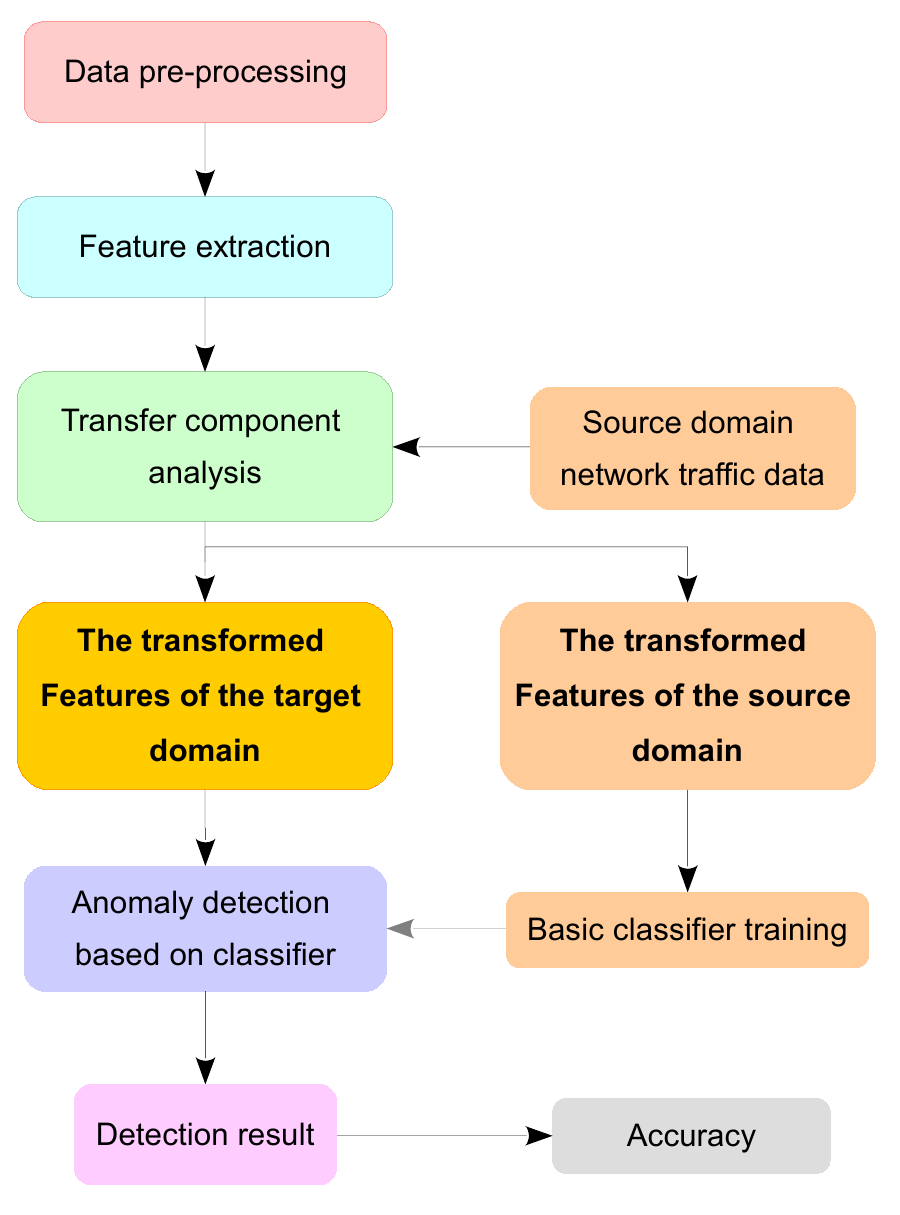}
    \caption{\ac{TCA}-based approach for detection of problematic network traffic.}
    \label{fig:TCA}
\end{figure}

In their case study, Catillo et~al. \citep{catillo2022transferability} explored the application of \ac{ML} to two open datasets with a particular emphasis on \ac{DoS} assaults. The CICIDS2017 and USB-IDS-1 datasets were used for training and testing, respectively. They realized that the detectors functioned very poorly in the case of slow attacks, but using fine-tuning as a \ac{DTL} strategy, they successfully transferred to just one attack, hulk (``no defense''). For attacks with mitigations, detection becomes even more confusing.

To extract the attack invariant from a current attack dataset and transmit the information to a target network system, Li et~al. \citep{li2021transfer} suggested an invariant extraction and \ac{seq2seq} generation-based network flow for \ac{DL}-based intrusion detection. Two components make up the \ac{DTL}-based attack traffic generation framework. A vector-embedding model IP2Vect, which converts the different fields in the protocol into vectors, makes up the first part; a \ac{seq2seq} model, the second component of their approach, tries to achieve end-to-end traffic creation. They created the \ac{seq2seq} model with several decoders for unsupervised training because the target and original networks lacked matched data. They also incorporated a component for adversarial training into the network to obtain the best representation of the attack invariant. Similarly, Wang et~al. \citep{wang2021network} developed a method called BYOL, providing a brand-new data augmentation technique, which consists of using use a multi-head attention technique that raises the characteristics that contribute more to the classification and decreases the features that contribute less to the classification in the intrusion detection. Excellent performance was achieved using linear evaluation when the learning was transferred to other datasets such as NSK-KDD and CIC IDS2017, making the proposed scheme comparable to the \ac{SOTA} supervised-learning methods. The TrAdaBoost algorithm---an extended version of AdaBoost for \ac{DTL}, as described in Algorithm~\ref{algTLBoost}---was used by Li et~al. \citep{li2020cross} to present a cross-domain knowledge-based anomalous traffic detection approach. Using different-distribution training data, a foundation model was built. As a result, samples whose predictions were incorrect had their weights decreased to have less of an influence. Comparing the suggested approach to \ac{LSTM} alone, the \ac{FAR} was reduced.

\begin{algorithm*}[ht]
 \textbf{Input:}
 - $a$ labeled samples set $T_a$ in \ac{SD},\\
 \hspace{1cm} - $b$ labeled sample set $T_b$ in \ac{TD}, such that $T = T_a \cup T_b, x_i \in T$ , \\
 \hspace{1cm} - $c(x_i)$ is the real class label of the sample, \Comment{\ac{LSTM} can be used to capture the time behavior of the network traffic} \\
 \hspace{1cm} - unlabeled testing dataset $S$, basic classifier $C$, number of iterations $N$.

\textbf{Steps:} \\

(1) Initial weight vector $w_i^1= (w_1^1,\dots, w_{a+b}^1 )$ , where:
\begin{equation*}
    w_i^{1}=\begin{cases} \frac{1}{a}, &  i=1\dots a
    \\ \frac{1}{b}, & \i=a+1\dots a+b \end{cases}
\end{equation*}

(2) Set $\beta=1/(1+\sqrt{2 \mathrm{ln}(n/N)})$  \\

(3) \For{$t\gets1$ \KwTo N}{

 (a)~Set $p^t=\frac{w^t}{\sum_{i=1}^{a+b} w_i^t}$. \\
 (b)~Call learning algorithm (i.e., \ac{LSTM} algorithm), providing it with the combined traffic set $D$ with the distribution $p^t$ over $S$ and the unlabeled traffic dataset $T$. Then, train an anomaly classifier with the hypothesis $ X \rightarrow Y$. \\
(c) Calculate the error on $T_b$:
\begin{equation*}
    \epsilon_t=\sum_{i=a+1}^{a+b}\frac{w_i^t |h_t(x_i)-c(x_i)|}{\sum_{i=a+1}^{a+b}w_i^t}
\end{equation*}

(d) Set $\beta_t=\frac{\epsilon_t}{(1-\epsilon_t)}$, weight coefficient of the weaker classifier $\alpha_t=\mathrm{ln} (\frac{1}{\beta_t})$ \\
(e) Update the weight vector:
\begin{equation*}
    w_i^{t+1}=\begin{cases} w_i^{t} \beta^{|h_t(x_i)-c(x_i)|}, & i=1,\dots, a \\ w_i^{t} \beta^{-|h_t(x_i)-c(x_i)|}, & i= a+1,\dots, a+b \end{cases}
\end{equation*}
}
\textbf{Output:}
\begin{equation*}
h_f(x) = \sum_{i=1}^N \alpha_t.h_t(x)
\end{equation*}

\caption{TrAdaBoost used for \ac{DTL}-based anomaly detection \citep{yu2019individual,li2020cross,li2020intrusion,sun2018network}.}
\label{algTLBoost}
\end{algorithm*}

\item \textbf{Supervised techniques.} The use of supervised procedures is the second approach. In these approaches, researchers investigate methods to minimize the dependency of the anomalous network traffic detection models on labeled data in the target network under the assumption that there is little labeled network traffic in the target network. To pre-train and fine-tune the model using the segmented UNSW-NB15 dataset, Singla et~al. \citep{singla2019overcoming} designed a \ac{DNN} and evaluated \ac{DTL}'s suitability for \ac{DL} model training with limited new labeled data for an \ac{NIDS}. They discovered that when there is very little training data available, using the capability of the fine-tuning allows detection models to perform much better in recognizing new intrusions. Sun et~al. \citep{sun2018network} investigated the issue of classification of anomalous network traffic in the case of small samples from the standpoint of \ac{SD} sample selection. They employed the maximum-entropy model Maxent as the fundamental classifier of the TrAdaBoost \ac{DTL} method (Algorithm~\ref{algTLBoost}). The research application possibilities are, however, constrained due to the high labor costs associated with labeling the traffic data in the target network and the high time and computation costs associated with the secondary training model.

The scheme of Dhillon and Haque \citep{dhillon2020towards} uses an \ac{SM} with a mixed architecture that consists of \ac{CNN} with \ac{LSTM}. The latter units can learn and detect patterns over time, in contrast to a format unit, which can only do so within the confines of an input space. Their studies revealed that adding learnt weights from the proposed \ac{SM} enhanced the performance and speed of models in the \ac{TD}, reaching an accuracy of up to 98.43\%. Andresini et~al. \citep{andresini2021network} introduced a \ac{DL}-based \ac{NIDS} approach that uses the Page--Hinkley test to identify concept-drift events in the monitored network-traffic data stream. They employed fine-tuning as a supervised \ac{DTL} strategy to the pre-trained MINDFUL model to adjust the \ac{DL} architecture to the drifted data. Their experimental study, conducted on the CICIDS2017 dataset, validated the efficacy of the suggested approach. Zhang et~al. \citep{zhang2019autonomous} suggested an automated model-update strategy that uses a statistical discriminator to filter the data packets of a new application from active network traffic to create a new training dataset; the \ac{DTL} technique was applied to update the current network traffic classifier. Their results show that using \ac{DTL} is twice as fast as not using \ac{DTL}.

Luo et~al. \citep{luo2018tinet} presented the analogy that the invariant network is a graph network between different computer-system entities such as processes, files, and internet sockets. They calculated the probability that each \ac{SD} object would be usable in the final invariant network of the \ac{TD}, a knowledge-transfer-based model for speeding up invariant-network building. For \ac{DDoS} attacks, He et~al. \citep{he2020small} determined that when there are only a few attack samples, this decreases the performance of network classification. They suggested using a fine-tuned eight-layer \ac{ANN} for \ac{DTL} from a \ac{SD} to a \ac{TD}; the proposed scheme was found to raise the performance by up to 20\%. Phan et~al. \citep{phan2020q} proposed a scheme employing both \ac{DTL} and reinforcement learning. In this scheme, \ac{DTL} is used for identification of the most useful knowledge gained from the \ac{SD} for handling the task in the \ac{TD}; reinforcement learning, specifically the Q-learning algorithm, was found to achieve the optimal results.

The concept of mimic learning involves employing models that were trained on original sensitive data to label previously unlabeled public data. This idea is the fundamental contribution of the research of Shafee et~al. \citep{shafee2020mimic}. They introduce a practical evaluation of a mimic-learning strategy for intrusion-detection datasets that efficiently converts knowledge from a private, non-shareable dataset to a public, shareable predictive model.

\item \textbf{Semi-supervised techniques.} Semi-supervised techniques combine the strengths of both supervised and unsupervised learning (cf. Section~\ref{TDTL}) by using a small amount of labeled data with a larger amount of unlabeled data to train the model. This approach can be particularly useful when labeled data is scarce or expensive to obtain. This approach has been applied in many suggested frameworks related to \ac{DTL}-based \ac{IDS} \citep{zhang2020semi,li2019dart,ning2021novel}. The research of Zhang and Yan \citep{zhang2020semi} offers a semi-supervised framework based on domain-adversarial training to address the same label-scarcity challenge by transferring the knowledge of existing assault incidences to detect returning threats at various load patterns and hourly intervals in smart-grid networks. Other schemes have been suggested to address the issue of unknown malware-variant classification \citep{li2019dart,ning2021novel}. Li et~al. \citep{li2019dart} suggest DART, a semi-supervised system for identifying harmful network traffic based on the ``adaptation regularization transfer learning'' module, which transfers knowledge for regularization. In particular, the DART method trains an adaptive classifier-based \ac{DTL} by simultaneously optimizing three variables: (i)~the structural risk functions; (ii)~the joint distribution between the known malware and unseen malware variants domains; and (iii)~the manifold consistency underlying marginal distribution. The approach was tested on the UNSW-NB15 dataset and evaluated against other cutting-edge \acp{IDS}. It was found to outperform traditional traffic-categorization techniques; for detecting unseen malware, it achieved an F-measure of over 90\% and a recall of 93\%.
\end{itemize}

\begin{figure*}[!t]
    \centering
    \includegraphics[width=1.0\textwidth]{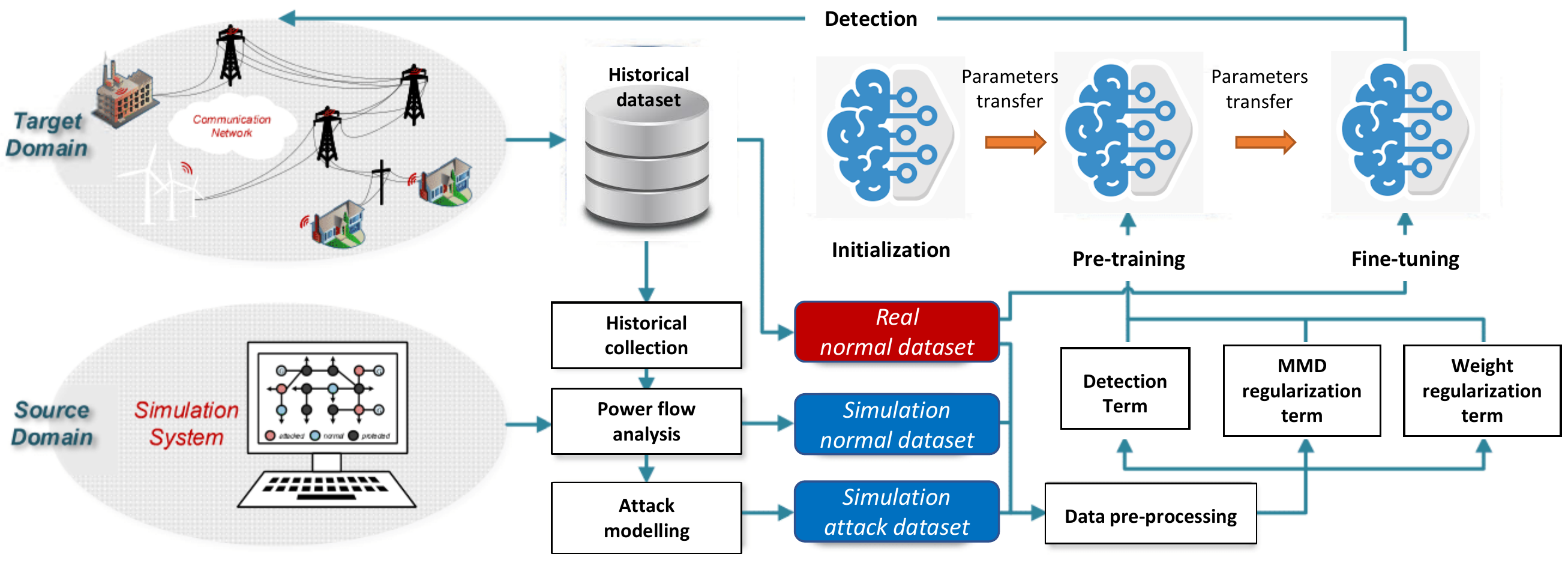}
    \caption{Example of \ac{DTL}-based smart-grid \ac{IDS} \citep{xu2021detecting}.}
    \label{fig:sgrid}
\end{figure*}
\subsection{\ac{DTL}-based smart-grid \acp{IDS}}
Due to their susceptibility to cyber-physical threats, a number of useful applications in smart grids, such as state estimation, have received significant attention recently. The high-profile \ac{FDIA}, one of the most dangerous assaults, tries to trick operators into performing incorrect actions by inserting false measurements, leading to financial losses and security problems. Xu et~al. \citep{xu2021detecting} provided a \ac{FDIA} detection method from the standpoint of \ac{DTL}. In their work, the simulated power system was specifically addressed as a \ac{SD}, as this offers a variety of simulated normal and attack data. The operating system in the actual world, whose transmission-line characteristics are unknown, was chosen as the \ac{TD}, and enough real normal data were gathered to follow the most recent system states online. The intended transfer approach comprises two steps of optimization, and it tries to fully use all the available data (Fig.~\ref{fig:sgrid}). A \ac{DNN} is initially constructed by concurrently maximizing the number of properly selected objective terms using both simulated and actual data. In the second stage, it is fine-tuned using real data along with the \ac{MMD} algorithm to measure the distribution discrepancy between the sample means of the \ac{SD} and the \ac{TD}.

In the same domain, two works of Zhang and Yan \citep{zhang2019domain,zhang2020semi} presented frameworks to solve the problems of \ac{FDIA} attacks. Both proposed methods use domain-specific adversarial training to map \ac{SD} and \ac{TD} labels. Threats are detected by transferring knowledge of prior attack instances. However, one of these works \citep{zhang2020semi} employs semi-supervised training, while the other \citep{zhang2019domain} is a fully unsupervised method.

\subsection{DTL-based \ac{IDS} for cloud security}
\begin{figure*}[!t]
    \centering
    \includegraphics[width=0.8\textwidth]{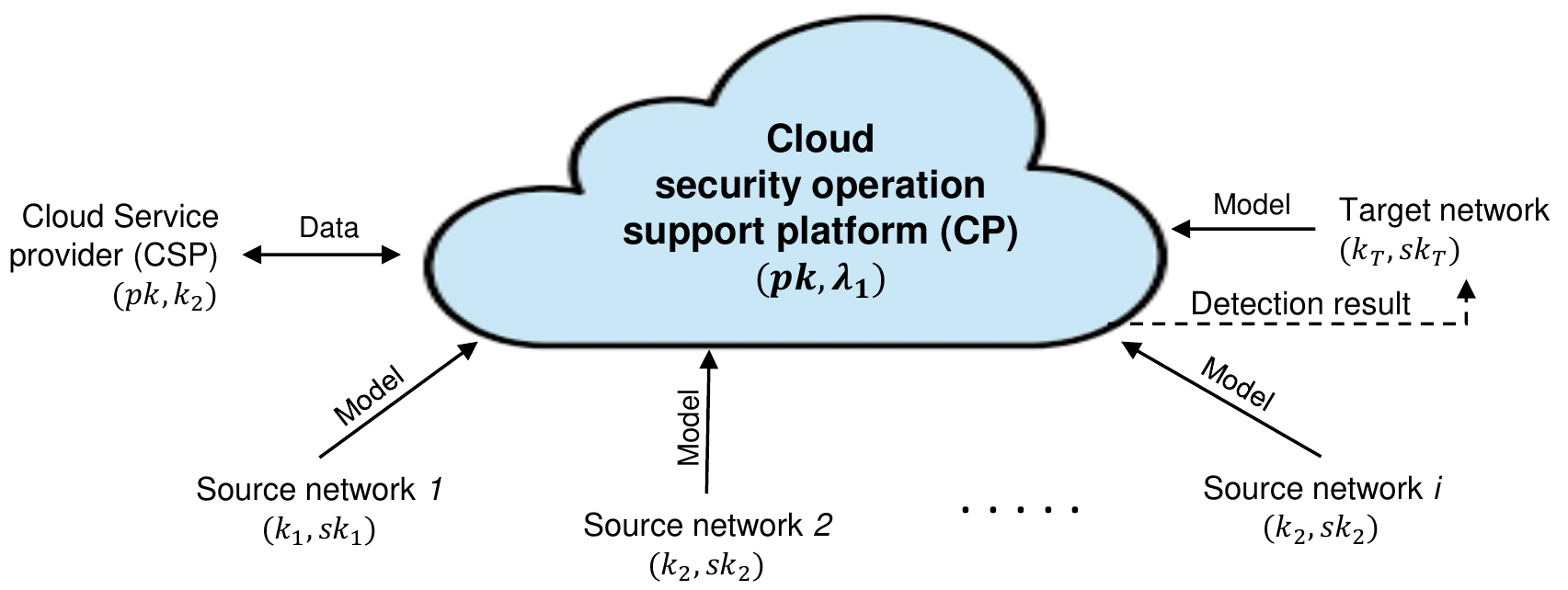}
    \caption{Cloud-system model. Here, ``CP'' is a partially honest cloud server with enough storage to enable safe model transfer from the source network to the target one and conduct security detection on recently created data from the target network.}
    \label{fig:cloudmodel}
\end{figure*}

Cloud computing, also known as on-demand computing, provides customers with a variety of services. Due to its rising popularity, it is susceptible to many intruders who can threaten the confidentiality and security of data stored in the cloud (Fig.~\ref{fig:cloudmodel}). Because of their dispersed nature, the most difficult aspect of cloud-based solutions is security. The biggest issues of on-demand services are privacy and security, yet they are open to intrusion from any kind of assault. Existing \acp{IDS} face a number of difficulties as a result of the expanding size of cloud networks and the need to protect data privacy. These problems include a high computational cost, protracted training periods, and feature distributions that are not uniform, which results in poor model performance. These issues have been resolved via \ac{DTL}.

However, privacy cannot be protected during data processing because current \ac{DTL}-based techniques can only function in plaintext when separate domains and clouds are untrusted entities. Consequently, Xu et~al. \citep{xu2021privacy} developed a multi-source \ac{IDS}-based \ac{DTL} method that protects privacy. In their approach, first, the models from several \acp{SD} are encrypted and uploaded to the cloud using Paillier homomorphic encryption. Next, a \ac{DTL}-based multi-source \ac{IDS} based on encrypted XGBoost (E-XGBoost) for privacy-preserving technique is applied. The experimental findings demonstrate that the suggested method can effectively move the encryption models from a number of \acp{SD} to the \ac{TD}, and the accuracy rate can reach 93.01\% in ciphertext with no appreciable loss in detection performance compared to works in plaintext. The model's training period is substantially shorter, at the minute level, than it is at the more customary hour level. Similarly, the purpose of the study of Ahmadi et~al. \citep{ahmadi2018intrusion} was to investigate whether a simplistic \ac{ML} classifier with a modest common set of features, trained on a non-cloud dataset with a packet-based structure, can be applied to a cloud dataset incorporating time-based traffic using \ac{DTL} to detect particular intrusions. This allows for analysis of the differences and similarities between assaults on cloud-based and non-cloud datasets, as well as recommendations for future research.

To increase the overall security of a cloud-based computing environment, the work of Sreelatha et~al. \citep{sreelatha2022improved} builds an effective cloud \ac{IDS} employing \ac{SOA}-based feature selection and \ac{EEDTL} classification. Based on the \ac{SOA}, the number of features from the provided incursion dataset is decreased with the least amount of information lost. The \ac{EEDTL} model is then used to categorize various threats according to the best qualities that were chosen for them. The \ac{TL} employs a pre-trained network called AlexNet for fine-tuning the characteristics in the convolution layers. To update the network weights, the \ac{EEO} is also used. The suggested cloud \ac{IDS} was found to successfully distinguish between normal network traffic behavior and an attack.

\begin{figure}[!t]
    \centering
    \includegraphics[width=0.5\textwidth]{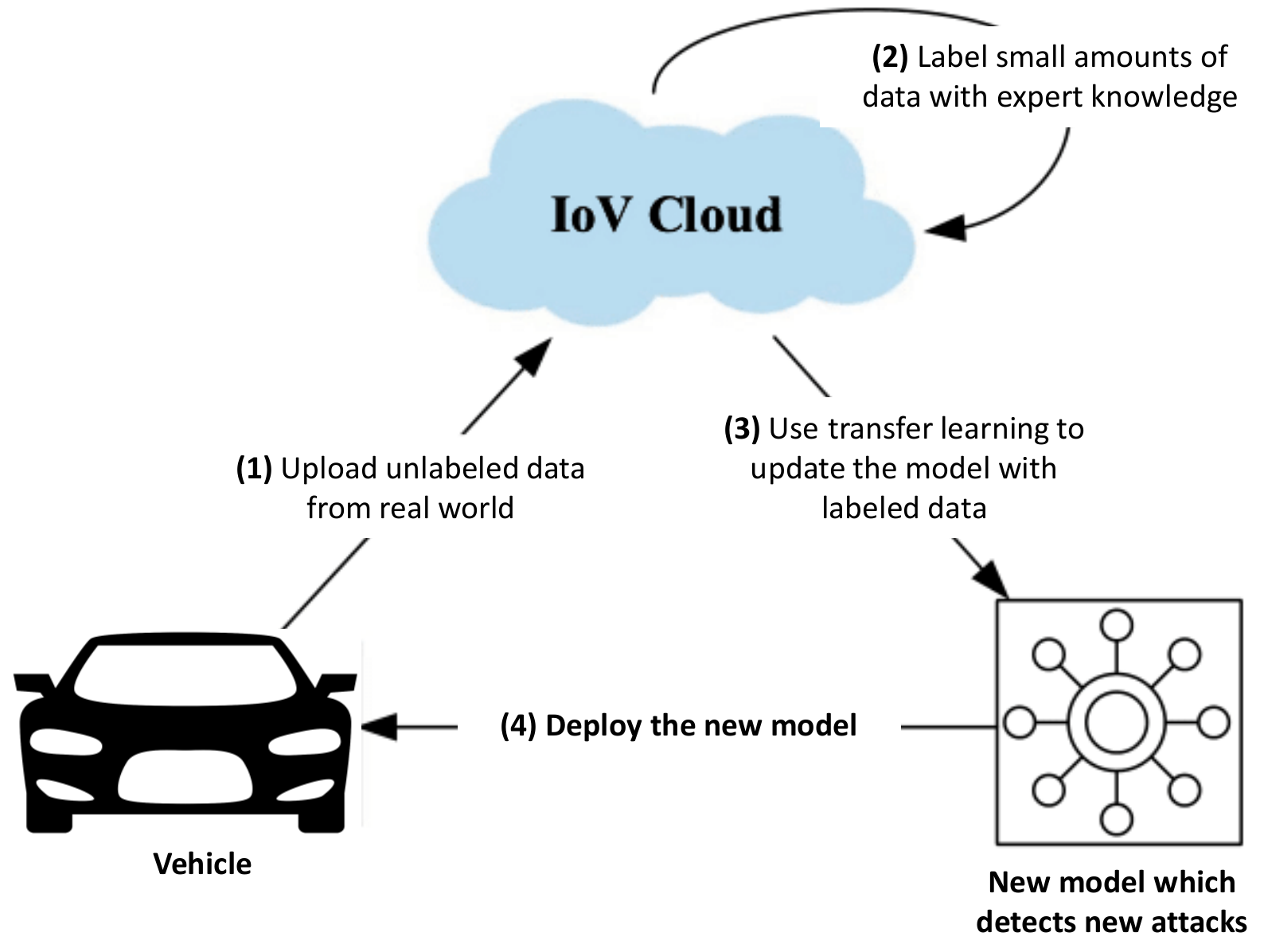}
    \caption{Example of cloud-assisted update scheme using \ac{DTL} \citep{li2021transfer}.}
    \label{fig:clouddtl2}
\end{figure}

Li et~al. \citep{li2021transfer} introduced a learning method and constructed two model-update techniques based on whether the \ac{IoV} cloud can rapidly provide a small quantity of labeled data for a fresh attack. The first was a cloud-assisted upgrading technique that allowed the \ac{IoV} cloud to contribute a small quantity of data. The \ac{IoV} cloud was unable to provide any tagged data in a timely way using the second method, known as the local-update technique. However, the local-update technique can be used in new ways to acquire fictional labels for unlabeled data using pre-classification, and the pseudo-labeled data is then used for several iterations of learning algorithms. The car can finish the upgrade without having to obtain any tagged data from the \ac{IoV} cloud (Fig.~\ref{fig:clouddtl2}).

This study of Ranjithkumar and Pandian \citep{ranjithkumar2022fuzzy} focused on a semi-supervised learning model for cloud-intrusion detection called the fuzzy-based semi-supervised approach through latent Dirichlet allocation (F-LDA). This aids with resolving the failure of supervised and unsupervised \acp{IDS} to detect attack patterns. The latent Dirichlet allocation initially offers superior generalization potential for labeled data training, and the fuzzy model is used to analyze the dataset and manage the unlabeled data.

\section{Open challenges and future directions} \label{sec6}

The research examined in this review highlights the efficacy of DTL-based \ac{IDS} approaches in diverse applications. This is because they are computationally efficient and have the potential to surpass traditional \ac{ML} and \ac{DL} algorithms. Nevertheless, there are other critical challenges that require consideration to improve the overall performance and adaptability of DTL-based IDS systems. This section intends to shed light on the most daunting unresolved problems and the pressing issues currently receiving significant attention in this field.

\subsection{Open challenges}
\ac{DTL}-based \acp{IDS} face several challenges. \ac{DL} models require a large amount of data for training, but in \acp{IDS}, the number of normal samples is usually much larger than the number of intrusion samples, leading to a data-imbalance problem. This can cause the model to become biased toward normal samples and can negatively affect intrusion-detection performance. Moreover, \acp{IDS} deal with a large amount of diverse data, such as network traffic, system logs, and system calls, which can be difficult to represent as feature vectors suitable for \ac{DL} models. Effective feature extraction is crucial to the performance of \ac{DTL} in \acp{IDS}. Additionally, overfitting is a common problem in \ac{DL}, especially when dealing with a limited amount of data. In \acp{IDS}, overfitting can lead to poor generalization performance, whereby the model becomes too specialized to the training data and fails to detect new intrusions.

Moving forward, it is important to note that \ac{DL} models are vulnerable to adversarial attacks, in which malicious actors manipulate the input data to mislead the model and evade detection. This presents a major challenge for \acp{IDS} based on \ac{DTL}, as attackers can deliberately target the \ac{DL} models to evade detection \citep{li4254447eifdaa}. In addition, \ac{IoT} models must take into account the connectivity of many smart devices. As a result, sophisticated and difficult intrusions may be carried out by cyberattackers. Finding a balance between detection rate and resource consumption is thus required, and more realistic assault models need to be developed \citep{liu2019adversarial,papadopoulos2021launching}. An ongoing worry for \ac{IoT} systems is finding ways to safeguard \ac{IDS} communication routes. Several networks assume control to protect network-wide communication between \ac{IDS} nodes and component parts \citep{alazab2023routing}. As a result, in \ac{IoT} scenarios, inadequate protection techniques are often used to secure communication between nodes and sensors, making it simple for an attacker to see and decode network data. There is thus a need for \ac{IoT} protection with a powerful \ac{IDS} communication system \citep{aldaej2019enhancing}. Malicious code poses a further threat since it often targets applications with \ac{DoS} and worm assaults, as well as targeting routers and security cameras. These kinds of assault are capable of taking advantage of software flaws. A developing computer system like \ac{IoT} security, which incorporates a detection mechanism for individual \acp{IoT} risks, represents a typical attack method. \citep{sadhu2022internet}.

It is also worth noting that \ac{DL} models are often difficult to interpret, making it challenging to understand why a model is making certain decisions. This can be a problem in \acp{IDS}, as it can be difficult to determine the causes of false alarms or missed detections. Furthermore, with the growth in the numbers of \ac{IoT} devices, it should be borne in mind that connecting multiple heterogeneous networks to the internet might put them at risk. This interaction with other systems carries with it weaker security requirements, which may make building confidence in their security and privacy difficult. With the expansion of \ac{IoT} devices, the trust system must adapt to and fulfill new requirements \citep{singh2020deep}. Even though a number of scholars have suggested evaluating good reputation and engagement, further study is necessary.

Although \ac{FTL} techniques have many advantages, including better model performance, shorter training times, and data privacy, they can also present some security challenges. For example, server trust issues can arise when aggregating models, and if the server is attacked, the entire system could be put at risk \citep{otoum2022federated}.

\subsection{Future directions}
Overall, the challenges outlined above highlight the need for continued research and development in \acp{IDS} based on \ac{DTL} to improve their performance, robustness, and interpretability. The future research directions of \ac{DTL}-based \acp{IDS} are likely to be influenced by several trends. In this regard, as adversarial attacks become more sophisticated, the development of methods to make \ac{DL} models more robust to such attacks will be a key focus. This may include the development of new \ac{DL} models that are specifically designed to be more resistant to adversarial attacks, as well as new methods for detecting and mitigating such attacks \citep{apruzzese2022modeling}. There is growing interest in developing \ac{DL} models that are more transparent and interpretable, especially in the context of security-sensitive applications such as intrusion detection. Future work in this area may focus on developing methods to better understand how \ac{DL} models make decisions and on making these decisions more interpretable to humans \citep{nwakanma2023explainable,liu2021faixid}.

Additionally, \acp{IDS} often deal with multiple types of data, such as network traffic, system logs, and system calls. Future work may thus also focus on developing \ac{DTL} models that can effectively process and analyze multiple types of data, improving the overall performance and accuracy of \acp{IDS}. Anomaly detection is also a key component of \acp{IDS}, and future work may focus on developing new \ac{DTL} models that can better detect and identify anomalous behavior in large and complex datasets. This may involve incorporating unsupervised learning methods, such as clustering and \acp{AE}, into \acp{IDS} \citep{alhakami2019network}.

\acp{IDS} are just one component of a larger security infrastructure, and future work may focus on integrating \ac{DTL}-based \acp{IDS} with other security tools, such as firewalls and intrusion-prevention systems, to provide a more comprehensive security solution. Moreover, applying blockchain technology to the \ac{FTL} technique will ensure a high capacity for data analytics to be used in collaborative learning with minimum response times and costs, and this may help to solve the issues mentioned above. \Acp{FTL} models can be extended using blockchain technology to provide a high level of security and privacy for training on sensitive and critical information, as well as to analyze sensitive data. Lastly, to show how \ac{FTL} can function with wireless-connection technology provided by \ac{5G} or \ac{6G} networks, the performance of other \ac{FTL} approaches can also be assessed using fresh datasets for various domains in \ac{IoT} applications.

Using generative chatbots and models such as ChatGPT for \ac{DTL}-based \ac{IDS} might be unconventional, as these \ac{AI} models are typically used for natural language processing tasks \citep{sohail2023future}. However, they could potentially be adapted for \ac{IDS} applications. Typically, ChatGPT can recognize patterns in textual data, and it could potentially be used to analyze logs for pattern recognition. For instance, if network logs were converted into a textual format that encapsulated the necessary information for intrusion detection, ChatGPT could potentially analyze these logs, learn from them, and then predict whether or not future logs indicate an intrusion. It could also be used to identify anomalous patterns in a sequence of logs, functioning as an anomaly-based \ac{IDS} \citep{tian2023cldtlog}. As noted above, network security involves dealing with a vast amount of unstructured data. Logs, user and system behaviors, and other network activities could be interpreted as textual information. Natural language processing models such as ChatGPT could be applied to process these data, extracting useful features for \ac{IDS} tasks. Furthermore, such pre-trained models, which are trained on a vast amount of internet text, are able to learn broad representations of data. These representations could potentially be adapted for \ac{IDS} tasks via \ac{DTL}, where the pre-trained model is fine-tuned on specific \ac{IDS} tasks \citep{sohail2023using}. Additionally, chatbots can interact with system users in real time. They can thus be used to alert system administrators to potential threats or anomalies and receive feedback that can be used to adjust their detection algorithms. This real-time feedback loop could enhance the effectiveness and precision of an \ac{IDS}. Lastly, generative chatbots could also be used to generate simulated network traffic or even cyberattack patterns for training \ac{IDS} models. This could be especially useful for creating diverse and robust datasets that represent a wide range of possible intrusion scenarios.

\section{Conclusion}\label{sec7}
The use of \acp{IDS} based on \ac{DTL} is a growing area of research and development, and it has significant potential to improve the ability to detect and prevent security threats. By leveraging the power of \ac{DL} models, these systems can analyze large, complex datasets and identify anomalous behavior that may indicate an intrusion. This review has examined the most recent \ac{AI} techniques used in many different types of \ac{ICN}, specifically \ac{IDS}-based \ac{DTL} approaches. By dividing selected publications issued after 2015 into three categories, we have provided a comprehensive review of the current state of \ac{DTL} approaches used in \acp{IDS}. This review paper has also covered other useful information such as the datasets used, types of \ac{DTL} employed, pre-trained networks, \ac{IDS} techniques, evaluation metrics, and the improvements gained. The algorithms and methods used in several studies or illustrated in any \ac{DTL}-based \ac{IDS} subcategory have been presented clearly and in depth to enable researchers to better understand the current state of the use of \ac{DTL} approaches in \acp{IDS} in different types of network. Overall, this work provides valuable insights that could help to enhance the security of \acp{ICS} and safeguard critical infrastructure from malicious attacks.

\ac{DTL}-based \acp{IDS} face several challenges, including data imbalance, difficulties in effective feature extraction, overfitting, vulnerability to adversarial attacks, and difficulty in interpretation. In addition, the connectivity of many smart devices in the \ac{IoT} model presents a risk of sophisticated and difficult intrusions by cyberattackers. Protecting the \ac{IDS} communication routes and developing a detection mechanism specifically for \ac{IoT} that focuses on individual detection risks are necessary steps for \ac{IoT} security. Malicious code poses a further threat, and there is a need for a mechanism to detect \ac{IoT} attacks.

The challenges faced by \acp{IDS} based on \ac{DTL} suggest a need for continued research and development in this field to improve performance, robustness, and interpretability. The future direction of this technology is likely to be influenced by several trends, including developing methods to make models more resistant to adversarial attacks, increasing transparency and interpretability, effectively processing and analyzing multiple types of data, incorporating unsupervised learning methods, integrating \acp{IDS} with other security tools, applying blockchain technology to improve data analytics, and evaluating the performance of \ac{FTL} approaches using fresh datasets in various domains of \ac{IoT} applications.

\printcredits

\section*{Declaration of competing interest}
The authors declare that they have no known competing financial interests or personal relationships that could have appeared to influence the work reported in this paper.

\section*{Data availability}
No data was used for the research described in the article.

\section*{Acknowledgments}
We would like to express our sincere gratitude to the anonymous reviewers for their valuable feedback and suggestions, which have improved the quality of this work. The first author acknowledges that the study was partially funded by the Algerian Ministry of Higher Education and Scientific Research (Grant No. PRFU--A25N01UN260120230001).

\bibliographystyle{elsarticle-num}

\bibliography{references}

\begin{thebibliography}{100}
\expandafter\ifx\csname url\endcsname\relax
  \def\url#1{\texttt{#1}}\fi
\expandafter\ifx\csname urlprefix\endcsname\relax\def\urlprefix{URL }\fi
\expandafter\ifx\csname href\endcsname\relax
  \def\href#1#2{#2} \def\path#1{#1}\fi

\bibitem{himeur2022ai}
Y.~Himeur, M.~Elnour, F.~Fadli, N.~Meskin, I.~Petri, Y.~Rezgui, F.~Bensaali, A.~Amira, Ai-big data analytics for building automation and management systems: a survey, actual challenges and future perspectives, Artificial Intelligence Review (2022) 1--93.

\bibitem{sayed2022artificial}
A.~Sayed, Y.~Himeur, F.~Bensaali, A.~Amira, Artificial intelligence with iot for energy efficiency in buildings, in: Emerging Real-World Applications of Internet of Things, CRC Press, 2022, pp. 233--252.

\bibitem{himeur2022two}
Y.~Himeur, F.~Fadli, A.~Amira, A two-stage energy anomaly detection for edge-based building internet of things (biot) applications, in: 2022 5th International Conference on Signal Processing and Information Security (ICSPIS), IEEE, 2022, pp. 180--185.

\bibitem{atalla2023iot}
S.~Atalla, S.~Tarapiah, A.~Gawanmeh, M.~Daradkeh, H.~Mukhtar, Y.~Himeur, W.~Mansoor, K.~F.~B. Hashim, M.~Daadoo, Iot-enabled precision agriculture: Developing an ecosystem for optimized crop management, Information 14~(4) (2023) 205.

\bibitem{adou2022modeling}
Y.~Adou, E.~Markova, Y.~Gaidamaka, Modeling and analyzing preemption-based service prioritization in {5G} networks slicing framework, Future Internet 14~(10) (2022) 299.

\bibitem{kheddar2022all}
H.~Kheddar, Y.~Himeur, S.~Atalla, W.~Mansoor, An efficient model for horizontal slicing in 5g network using practical simulations, in: 2022 5th International Conference on Signal Processing and Information Security (ICSPIS), IEEE, 2022, pp. 158--163.

\bibitem{zhang2022artificial}
Z.~Zhang, F.~Wen, Z.~Sun, X.~Guo, T.~He, C.~Lee, Artificial intelligence-enabled sensing technologies in the {5G}/internet of things era: From virtual reality/augmented reality to the digital twin, Advanced Intelligent Systems (2022) 2100228.

\bibitem{zhao2022secure}
Z.~Zhao, F.~Guo, G.~Wu, W.~Susilo, B.~Wang, Secure infectious diseases detection system with {IoT}-based {e-Health} platforms, IEEE Internet of Things Journal (2022).

\bibitem{yu2022construction}
M.~Yu, Construction of regional intelligent transportation system in smart city road network via {5G} network, IEEE Transactions on Intelligent Transportation Systems (2022).

\bibitem{ullah2022optimization}
I.~Ullah, M.~Fayaz, M.~Aman, D.~Kim, An optimization scheme for {IoT} based smart greenhouse climate control with efficient energy consumption, Computing 104~(2) (2022) 433--457.

\bibitem{himeur2022latest}
Y.~Himeur, S.~S. Sohail, F.~Bensaali, A.~Amira, M.~Alazab, Latest trends of security and privacy in recommender systems: a comprehensive review and future perspectives, Computers \& Security (2022) 102746.

\bibitem{gallenmuller20205g}
S.~Gallenm{\"u}ller, J.~Naab, I.~Adam, G.~Carle, 5g urllc: A case study on low-latency intrusion prevention, IEEE Communications Magazine 58~(10) (2020) 35--41.

\bibitem{onate2023analysis}
W.~O{\~n}ate, R.~Sanz, Analysis of architectures implemented for {IIoT}, Heliyon (2023) e12868.

\bibitem{himeur2022recent}
Y.~Himeur, A.~Alsalemi, F.~Bensaali, A.~Amira, A.~Al-Kababji, Recent trends of smart nonintrusive load monitoring in buildings: A review, open challenges, and future directions, International Journal of Intelligent Systems 37~(10) (2022) 7124--7179.

\bibitem{alshamrani2019survey}
A.~Alshamrani, S.~Myneni, A.~Chowdhary, D.~Huang, A survey on advanced persistent threats: Techniques, solutions, challenges, and research opportunities, IEEE Communications Surveys \& Tutorials 21~(2) (2019) 1851--1877.

\bibitem{himeur2022blockchain}
Y.~Himeur, A.~Sayed, A.~Alsalemi, F.~Bensaali, A.~Amira, I.~Varlamis, M.~Eirinaki, C.~Sardianos, G.~Dimitrakopoulos, Blockchain-based recommender systems: Applications, challenges and future opportunities, Computer Science Review 43 (2022) 100439.

\bibitem{kaiser2023review}
J.~Kaiser, D.~Mcfarlane, G.~Hawkridge, P.~Andr{\'e}, P.~Leit{\~a}o, A review of reference architectures for digital manufacturing: Classification, applicability and open issues, Computers in Industry 149 (2023) 103923.

\bibitem{umer2022machine}
M.~A. Umer, K.~N. Junejo, M.~T. Jilani, A.~P. Mathur, Machine learning for intrusion detection in industrial control systems: Applications, challenges, and recommendations, International Journal of Critical Infrastructure Protection (2022) 100516.

\bibitem{jiang2019machine}
D.~Jiang, J.~Zhao, Machine learning in industrial control system security: A survey, in: Chinese Intelligent Systems Conference, Springer, 2019, pp. 310--317.

\bibitem{liu2020web}
C.~Liu, J.~Yang, J.~Wu, Web intrusion detection system combined with feature analysis and svm optimization, EURASIP Journal on Wireless Communications and Networking 2020~(1) (2020) 1--9.

\bibitem{park2018anomaly}
S.~Park, M.~Kim, S.~Lee, Anomaly detection for {HTTP} using convolutional autoencoders, IEEE Access 6 (2018) 70884--70901.

\bibitem{guizani2020network}
N.~Guizani, A.~Ghafoor, A network function virtualization system for detecting malware in large {IoT} based networks, IEEE Journal on Selected Areas in Communications 38~(6) (2020) 1218--1228.

\bibitem{elsayed2021novel}
M.~S. ElSayed, N.-A. Le-Khac, M.~A. Albahar, A.~Jurcut, A novel hybrid model for intrusion detection systems in {SDNs} based on cnn and a new regularization technique, Journal of Network and Computer Applications 191 (2021) 103160.

\bibitem{kao2022novel}
M.-T. Kao, D.-Y. Sung, S.-J. Kao, F.-M. Chang, A novel two-stage deep learning structure for network flow anomaly detection, Electronics 11~(10) (2022) 1531.

\bibitem{zhou2020indoor}
M.~Zhou, X.~Li, Y.~Wang, Y.~Li, A.~Ren, Indoor intrusion detection based on deep signal feature fusion and minimized-mkmmd transfer learning, Physical Communication 42 (2020) 101164.

\bibitem{xu2021detecting}
B.~Xu, F.~Guo, C.~Wen, R.~Deng, W.-A. Zhang, Detecting false data injection attacks in smart grids with modeling errors: A deep transfer learning based approach, arXiv preprint arXiv:2104.06307 (2021).

\bibitem{tama2022systematic}
B.~A. Tama, S.~Y. Lee, S.~Lee, A systematic mapping study and empirical comparison of data-driven intrusion detection techniques in industrial control networks, Archives of Computational Methods in Engineering (2022) 1--28.

\bibitem{asharf2020review}
J.~Asharf, N.~Moustafa, H.~Khurshid, E.~Debie, W.~Haider, A.~Wahab, A review of intrusion detection systems using machine and deep learning in internet of things: Challenges, solutions and future directions, Electronics 9~(7) (2020) 1177.

\bibitem{hu2018survey}
Y.~Hu, A.~Yang, H.~Li, Y.~Sun, L.~Sun, A survey of intrusion detection on industrial control systems, International Journal of Distributed Sensor Networks 14~(8) (2018) 1550147718794615.

\bibitem{al2020federated}
N.~A. A.-A. Al-Marri, B.~S. Ciftler, M.~M. Abdallah, Federated mimic learning for privacy preserving intrusion detection, in: 2020 IEEE International Black Sea Conference on Communications and Networking (BlackSeaCom), IEEE, 2020, pp. 1--6.

\bibitem{wu2022rtids}
Z.~Wu, H.~Zhang, P.~Wang, Z.~Sun, Rtids: a robust transformer-based approach for intrusion detection system, IEEE Access 10 (2022) 64375--64387.

\bibitem{caville2022anomal}
E.~Caville, W.~W. Lo, S.~Layeghy, M.~Portmann, Anomal-e: A self-supervised network intrusion detection system based on graph neural networks, Knowledge-Based Systems 258 (2022) 110030.

\bibitem{nguyen2022transfer}
C.~T. Nguyen, N.~Van~Huynh, N.~H. Chu, Y.~M. Saputra, D.~T. Hoang, D.~N. Nguyen, Q.-V. Pham, D.~Niyato, E.~Dutkiewicz, W.-J. Hwang, Transfer learning for wireless networks: A comprehensive survey, Proceedings of the IEEE (2022).

\bibitem{vu2020deep}
L.~Vu, Q.~U. Nguyen, D.~N. Nguyen, D.~T. Hoang, E.~Dutkiewicz, Deep transfer learning for {IoT} attack detection, IEEE Access 8 (2020) 107335--107344.

\bibitem{lu2015transfer}
J.~Lu, V.~Behbood, P.~Hao, H.~Zuo, S.~Xue, G.~Zhang, Transfer learning using computational intelligence: A survey, Knowledge-Based Systems 80 (2015) 14--23.

\bibitem{agarwal2021transfer}
N.~Agarwal, A.~Sondhi, K.~Chopra, G.~Singh, Transfer learning: Survey and classification, in: Smart Innovations in Communication and Computational Sciences, Springer, 2021, pp. 145--155.

\bibitem{himeur2023video}
Y.~Himeur, S.~Al-Maadeed, H.~Kheddar, N.~Al-Maadeed, K.~Abualsaud, A.~Mohamed, T.~Khattab, Video surveillance using deep transfer learning and deep domain adaptation: Towards better generalization, Engineering Applications of Artificial Intelligence 119 (2023) 105698.

\bibitem{alanazi2022scada}
M.~Alanazi, A.~Mahmood, M.~J.~M. Chowdhury, {SCADA} vulnerabilities and attacks: A review of the state-of-the-art and open issues, Computers \& Security (2022) 103028.

\bibitem{williams1994purdue}
T.~J. Williams, The purdue enterprise reference architecture, Computers in industry 24~(2-3) (1994) 141--158.

\bibitem{xu2022improved}
W.~Xu, Y.~Gao, C.~Yang, H.~Chen, An improved purdue enterprise reference architecture to enhance cybersecurity, in: Proceedings of the 2022 5th International Conference on Blockchain Technology and Applications, 2022, pp. 104--109.

\bibitem{sicato2020comprehensive}
J.~C.~S. Sicato, S.~K. Singh, S.~Rathore, J.~H. Park, A comprehensive analyses of intrusion detection system for {IoT} environment, Journal of Information Processing Systems 16~(4) (2020) 975--990.

\bibitem{keshk2019integrated}
M.~Keshk, E.~Sitnikova, N.~Moustafa, J.~Hu, I.~Khalil, An integrated framework for privacy-preserving based anomaly detection for cyber-physical systems, IEEE Transactions on Sustainable Computing 6~(1) (2019) 66--79.

\bibitem{keshk2019privacy}
M.~Keshk, B.~Turnbull, N.~Moustafa, D.~Vatsalan, K.-K.~R. Choo, A privacy-preserving-framework-based blockchain and deep learning for protecting smart power networks, IEEE Transactions on Industrial Informatics 16~(8) (2019) 5110--5118.

\bibitem{masdari2020survey}
M.~Masdari, H.~Khezri, A survey and taxonomy of the fuzzy signature-based intrusion detection systems, Applied Soft Computing 92 (2020) 106301.

\bibitem{erlacher2018fixids}
F.~Erlacher, F.~Dressler, Fixids: A high-speed signature-based flow intrusion detection system, in: NOMS 2018-2018 IEEE/IFIP Network Operations and Management Symposium, IEEE, 2018, pp. 1--8.

\bibitem{li2019designing}
W.~Li, S.~Tug, W.~Meng, Y.~Wang, Designing collaborative blockchained signature-based intrusion detection in {IoT} environments, Future Generation Computer Systems 96 (2019) 481--489.

\bibitem{aldweesh2020deep}
A.~Aldweesh, A.~Derhab, A.~Z. Emam, Deep learning approaches for anomaly-based intrusion detection systems: A survey, taxonomy, and open issues, Knowledge-Based Systems 189 (2020) 105124.

\bibitem{kaouk2019review}
M.~Kaouk, J.-M. Flaus, M.-L. Potet, R.~Groz, A review of intrusion detection systems for industrial control systems, in: 2019 6th International Conference on Control, Decision and Information Technologies (CoDIT), IEEE, 2019, pp. 1699--1704.

\bibitem{kumar2021research}
S.~Kumar, S.~Gupta, S.~Arora, Research trends in network-based intrusion detection systems: A review, IEEE Access 9 (2021) 157761--157779.

\bibitem{malek2019user}
Z.~S. Malek, B.~Trivedi, A.~Shah, User behavior-based intrusion detection using statistical techniques, in: Advanced Informatics for Computing Research: Second International Conference, ICAICR 2018, Shimla, India, July 14--15, 2018, Revised Selected Papers, Part II 2, Springer, 2019, pp. 480--489.

\bibitem{mohammadi2021comprehensive}
M.~Mohammadi, T.~A. Rashid, S.~H.~T. Karim, A.~H.~M. Aldalwie, Q.~T. Tho, M.~Bidaki, A.~M. Rahmani, M.~Hosseinzadeh, A comprehensive survey and taxonomy of the svm-based intrusion detection systems, Journal of Network and Computer Applications 178 (2021) 102983.

\bibitem{vaddi2020dynamic}
P.~K. Vaddi, M.~C. Pietrykowski, D.~Kar, X.~Diao, Y.~Zhao, T.~Mabry, I.~Ray, C.~Smidts, Dynamic bayesian networks based abnormal event classifier for nuclear power plants in case of cyber security threats, Progress in Nuclear Energy 128 (2020) 103479.

\bibitem{ap2019secure}
H.~AP, et~al., {Secure-MQTT}: an efficient fuzzy logic-based approach to detect {DoS} attack in mqtt protocol for internet of things, EURASIP Journal on Wireless Communications and Networking 2019~(1) (2019) 1--15.

\bibitem{lansky2021deep}
J.~Lansky, S.~Ali, M.~Mohammadi, M.~K. Majeed, S.~H.~T. Karim, S.~Rashidi, M.~Hosseinzadeh, A.~M. Rahmani, Deep learning-based intrusion detection systems: a systematic review, IEEE Access 9 (2021) 101574--101599.

\bibitem{olufowobi2019saiducant}
H.~Olufowobi, C.~Young, J.~Zambreno, G.~Bloom, Saiducant: Specification-based automotive intrusion detection using controller area network (can) timing, IEEE Transactions on Vehicular Technology 69~(2) (2019) 1484--1494.

\bibitem{le2021machine}
L.~Le~Jeune, T.~Goedeme, N.~Mentens, Machine learning for misuse-based network intrusion detection: overview, unified evaluation and feature choice comparison framework, Ieee Access 9 (2021) 63995--64015.

\bibitem{yang2021mth}
L.~Yang, A.~Moubayed, A.~Shami, {MTH-IDS}: a multitiered hybrid intrusion detection system for internet of vehicles, IEEE Internet of Things Journal 9~(1) (2021) 616--632.

\bibitem{yao2019hybrid}
H.~Yao, P.~Gao, P.~Zhang, J.~Wang, C.~Jiang, L.~Lu, Hybrid intrusion detection system for edge-based {IIoT} relying on machine-learning-aided detection, IEEE Network 33~(5) (2019) 75--81.

\bibitem{koroniotis2017towards}
N.~Koroniotis, N.~Moustafa, E.~Sitnikova, J.~Slay, Towards developing network forensic mechanism for botnet activities in the {IoT} based on machine learning techniques, in: International Conference on Mobile Networks and Management, Springer, 2017, pp. 30--44.

\bibitem{soni2021systematic}
M.~Soni, M.~Singhal, R.~Katarya, et~al., A systematic survey on recent deep learning trends in intrusion detection system, in: 2021 3rd International Conference on Advances in Computing, Communication Control and Networking (ICAC3N), IEEE, 2021, pp. 1927--1932.

\bibitem{sharafaldin2018towards}
I.~Sharafaldin, A.~Gharib, A.~H. Lashkari, A.~A. Ghorbani, Towards a reliable intrusion detection benchmark dataset, Software Networking 2018~(1) (2018) 177--200.

\bibitem{CICIDS2018}
A realistic cyber defense dataset (cse-cic-ids2018), \url{https://registry.opendata.aws/cse-cic-ids2018}, accessed: 2023-01-02 (2018).

\bibitem{sharafaldin2019developing}
I.~Sharafaldin, A.~H. Lashkari, S.~Hakak, A.~A. Ghorbani, Developing realistic distributed denial of service ({DDoS}) attack dataset and taxonomy, in: 2019 International Carnahan Conference on Security Technology ({ICCST}), IEEE, 2019, pp. 1--8.

\bibitem{ferrag2022edge}
M.~A. Ferrag, O.~Friha, D.~Hamouda, L.~Maglaras, H.~Janicke, {Edge-IIoTset}: A new comprehensive realistic cyber security dataset of {IoT} and {IIoT} applications for centralized and federated learning, IEEE Access 10 (2022) 40281--40306.

\bibitem{goh2017dataset}
J.~Goh, S.~Adepu, K.~N. Junejo, A.~Mathur, A dataset to support research in the design of secure water treatment systems, in: Critical Information Infrastructures Security: 11th International Conference, CRITIS 2016, Paris, France, October 10--12, 2016, Revised Selected Papers 11, Springer, 2017, pp. 88--99.

\bibitem{ahmed2017wadi}
C.~M. Ahmed, V.~R. Palleti, A.~P. Mathur, Wadi: a water distribution testbed for research in the design of secure cyber physical systems, in: Proceedings of the 3rd international workshop on cyber-physical systems for smart water networks, 2017, pp. 25--28.

\bibitem{creech2013semantic}
G.~Creech, J.~Hu, A semantic approach to host-based intrusion detection systems using contiguousand discontiguous system call patterns, IEEE Transactions on Computers 63~(4) (2013) 807--819.

\bibitem{ramirez2019learning}
P.~Z. Ramirez, A.~Tonioni, S.~Salti, L.~D. Stefano, Learning across tasks and domains, in: Proceedings of the IEEE/CVF International Conference on Computer Vision, 2019, pp. 8110--8119.

\bibitem{lu2021general}
Y.~Lu, Z.~Tian, R.~Zhou, W.~Liu, A general transfer learning-based framework for thermal load prediction in regional energy system, Energy 217 (2021) 119322.

\bibitem{Kheddar023ASR}
H.~Kheddar, Y.~Himeur, S.~Al-Maadeed, A.~Amira, F.~Bensaali, Deep transfer learning for automatic speech recognition: Towards better generalization, Knowledge-Based Systems 277 (2023) 110851.

\bibitem{niu2020decade}
S.~Niu, Y.~Liu, J.~Wang, H.~Song, A decade survey of transfer learning (2010--2020), IEEE Transactions on Artificial Intelligence 1~(2) (2020) 151--166.

\bibitem{alyafeai2020survey}
Z.~Alyafeai, M.~S. AlShaibani, I.~Ahmad, A survey on transfer learning in natural language processing, arXiv preprint arXiv:2007.04239 (2020).

\bibitem{wang2020transfer}
J.~Wang, Y.~Chen, W.~Feng, H.~Yu, M.~Huang, Q.~Yang, Transfer learning with dynamic distribution adaptation, ACM Transactions on Intelligent Systems and Technology (TIST) 11~(1) (2020) 1--25.

\bibitem{ganin2016domain}
Y.~Ganin, E.~Ustinova, H.~Ajakan, P.~Germain, H.~Larochelle, F.~Laviolette, M.~Marchand, V.~Lempitsky, Domain-adversarial training of neural networks, The journal of machine learning research 17~(1) (2016) 2096--2030.

\bibitem{bousmalis2016domain}
K.~Bousmalis, G.~Trigeorgis, N.~Silberman, D.~Krishnan, D.~Erhan, Domain separation networks, Advances in neural information processing systems 29 (2016).

\bibitem{chen2019joint}
C.~Chen, Z.~Chen, B.~Jiang, X.~Jin, Joint domain alignment and discriminative feature learning for unsupervised deep domain adaptation, in: Proceedings of the AAAI conference on artificial intelligence, Vol.~33, 2019, pp. 3296--3303.

\bibitem{long2017deep}
M.~Long, H.~Zhu, J.~Wang, M.~I. Jordan, Deep transfer learning with joint adaptation networks, in: International conference on machine learning, PMLR, 2017, pp. 2208--2217.

\bibitem{zhang2018collaborative}
W.~Zhang, W.~Ouyang, W.~Li, D.~Xu, Collaborative and adversarial network for unsupervised domain adaptation, in: Proceedings of the IEEE conference on computer vision and pattern recognition, 2018, pp. 3801--3809.

\bibitem{he2016deep}
K.~He, X.~Zhang, S.~Ren, J.~Sun, Deep residual learning for image recognition, in: Proceedings of the IEEE conference on computer vision and pattern recognition, 2016, pp. 770--778.

\bibitem{li2017intrusion}
Z.~Li, Z.~Qin, K.~Huang, X.~Yang, S.~Ye, Intrusion detection using convolutional neural networks for representation learning, in: International conference on neural information processing, Springer, 2017, pp. 858--866.

\bibitem{zhang2021dual}
X.~Zhang, J.~Wang, S.~Zhu, Dual generative adversarial networks based unknown encryption ransomware attack detection, IEEE Access 10 (2021) 900--913.

\bibitem{szegedy2015going}
C.~Szegedy, W.~Liu, Y.~Jia, P.~Sermanet, S.~Reed, D.~Anguelov, D.~Erhan, V.~Vanhoucke, A.~Rabinovich, Going deeper with convolutions, in: Proceedings of the IEEE conference on computer vision and pattern recognition, 2015, pp. 1--9.

\bibitem{simonyan2014very}
K.~Simonyan, A.~Zisserman, Very deep convolutional networks for large-scale image recognition, arXiv preprint arXiv:1409.1556 (2014).

\bibitem{masum2020tl}
M.~Masum, H.~Shahriar, Tl-nid: Deep neural network with transfer learning for network intrusion detection, in: 2020 15th International Conference for Internet Technology and Secured Transactions (ICITST), IEEE, 2020, pp. 1--7.

\bibitem{lecun1998gradient}
Y.~LeCun, L.~Bottou, Y.~Bengio, P.~Haffner, Gradient-based learning applied to document recognition, Proceedings of the IEEE 86~(11) (1998) 2278--2324.

\bibitem{lin2018using}
W.-H. Lin, H.-C. Lin, P.~Wang, B.-H. Wu, J.-Y. Tsai, Using convolutional neural networks to network intrusion detection for cyber threats, in: 2018 IEEE International Conference on Applied System Invention ({ICASI}), IEEE, 2018, pp. 1107--1110.

\bibitem{mehedi2021deep}
S.~T. Mehedi, A.~Anwar, Z.~Rahman, K.~Ahmed, Deep transfer learning based intrusion detection system for electric vehicular networks, Sensors 21~(14) (2021) 4736.

\bibitem{krizhevsky2017imagenet}
A.~Krizhevsky, I.~Sutskever, G.~E. Hinton, Imagenet classification with deep convolutional neural networks, Communications of the ACM 60~(6) (2017) 84--90.

\bibitem{sreelatha2022improved}
G.~Sreelatha, A.~V. Babu, D.~Midhunchakkaravarthy, Improved security in cloud using sandpiper and extended equilibrium deep transfer learning based intrusion detection, Cluster Computing (2022) 1--16.

\bibitem{huang2017densely}
G.~Huang, Z.~Liu, L.~Van Der~Maaten, K.~Q. Weinberger, Densely connected convolutional networks, in: Proceedings of the IEEE conference on computer vision and pattern recognition, 2017, pp. 4700--4708.

\bibitem{wen2019time}
T.~Wen, R.~Keyes, Time series anomaly detection using convolutional neural networks and transfer learning, arXiv preprint arXiv:1905.13628 (2019).

\bibitem{xu2020intrusion}
Y.~Xu, Z.~Liu, Y.~Li, Y.~Zheng, H.~Hou, M.~Gao, Y.~Song, Y.~Xin, Intrusion detection based on fusing deep neural networks and transfer learning, in: International Forum on Digital TV and Wireless Multimedia Communications, Springer, 2020, pp. 212--223.

\bibitem{raghavendra2017transferable}
R.~Raghavendra, S.~Venkatesh, K.~B. Raja, C.~Busch, Transferable deep convolutional neural network features for fingervein presentation attack detection, in: 2017 5th International Workshop on Biometrics and Forensics ({IWBF}), IEEE, 2017, pp. 1--5.

\bibitem{xia2022pedestrian}
L.~Xia, A pedestrian intrusion detection method based on improved mask r-cnn model, in: Genetic and Evolutionary Computing: Proceedings of the Fourteenth International Conference on Genetic and Evolutionary Computing, October 21-23, 2021, Jilin, China, Vol. 833, Springer Nature, 2022, p. 370.

\bibitem{li2020efficient}
Y.~Li, Y.~Qin, Z.~Xie, Z.~Cao, L.~Jia, Z.~Yu, J.~Zheng, E.~Zhang, Efficient {SSD}: a real-time intrusion object detection algorithm for railway surveillance, in: 2020 International Conference on Sensing, Diagnostics, Prognostics, and Control ({SDPC}), IEEE, 2020, pp. 391--395.

\bibitem{nayak2019video}
R.~Nayak, M.~M. Behera, U.~C. Pati, S.~K. Das, Video-based real-time intrusion detection system using deep-learning for smart city applications, in: 2019 IEEE International Conference on Advanced Networks and Telecommunications Systems (ANTS), IEEE, 2019, pp. 1--6.

\bibitem{tariq2020cantransfer}
S.~Tariq, S.~Lee, S.~S. Woo, Cantransfer: Transfer learning based intrusion detection on a controller area network using convolutional {LSTM} network, in: Proceedings of the 35th annual ACM symposium on applied computing, 2020, pp. 1048--1055.

\bibitem{kim2020transfer}
J.~Kim, A.~Sim, J.~Kim, K.~Wu, J.~Hahm, Transfer learning approach for botnet detection based on recurrent variational autoencoder, in: Proceedings of the 3rd International Workshop on Systems and Network Telemetry and Analytics, 2020, pp. 41--47.

\bibitem{dos2021reminiscent}
R.~R. DoS~Santos, E.~K. Viegas, A.~O. Santin, A reminiscent intrusion detection model based on deep autoencoders and transfer learning, in: 2021 IEEE Global Communications Conference ({GLOBECOM}), IEEE, 2021, pp. 1--6.

\bibitem{qureshi2020intrusion}
A.~S. Qureshi, A.~Khan, N.~Shamim, M.~H. Durad, Intrusion detection using deep sparse auto-encoder and self-taught learning, Neural Computing and Applications 32~(8) (2020) 3135--3147.

\bibitem{wang2019cooperative}
W.~Wang, Q.~Chen, X.~He, L.~Tang, Cooperative anomaly detection with transfer learning-based hidden markov model in virtualized network slicing, IEEE Communications Letters 23~(9) (2019) 1534--1537.

\bibitem{chadza2020learning}
T.~Chadza, K.~G. Kyriakopoulos, S.~Lambotharan, Learning to learn sequential network attacks using hidden markov models, IEEE Access 8 (2020) 134480--134497.

\bibitem{huang2018automatic}
H.~Huang, H.~Deng, J.~Chen, L.~Han, W.~Wang, Automatic multi-task learning system for abnormal network traffic detection., International Journal of Emerging Technologies in Learning 13~(4) (2018).

\bibitem{demertzis2019cyber}
K.~Demertzis, L.~Iliadis, P.~Kikiras, N.~Tziritas, Cyber-typhon: an online multi-task anomaly detection framework, in: Artificial Intelligence Applications and Innovations: 15th IFIP WG 12.5 International Conference, AIAI 2019, Hersonissos, Crete, Greece, May 24--26, 2019, Proceedings 15, Springer, 2019, pp. 19--36.

\bibitem{albelwi2022intrusion}
S.~A. Albelwi, An intrusion detection system for identifying simultaneous attacks using multi-task learning and deep learning, in: 2022 2nd International Conference on Computing and Information Technology (ICCIT), IEEE, 2022, pp. 349--353.

\bibitem{wang2018deep}
M.~Wang, W.~Deng, Deep visual domain adaptation: A survey, Neurocomputing 312 (2018) 135--153.

\bibitem{zhao2017feature}
J.~Zhao, S.~Shetty, J.~W. Pan, Feature-based transfer learning for network security, in: MILCOM 2017-2017 IEEE Military Communications Conference (MILCOM), IEEE, 2017, pp. 17--22.

\bibitem{shi2010transfer}
X.~Shi, Q.~Liu, W.~Fan, S.~Y. Philip, R.~Zhu, Transfer learning on heterogenous feature spaces via spectral transformation, in: 2010 IEEE international conference on data mining, IEEE, 2010, pp. 1049--1054.

\bibitem{sun2016return}
B.~Sun, J.~Feng, K.~Saenko, Return of frustratingly easy domain adaptation, in: Proceedings of the AAAI conference on artificial intelligence, Vol.~30, 2016.

\bibitem{taghiyarrenani2018transfer}
Z.~Taghiyarrenani, A.~Fanian, E.~Mahdavi, A.~Mirzaei, H.~Farsi, Transfer learning based intrusion detection, in: 2018 8th International Conference on Computer and Knowledge Engineering ({ICCKE}), IEEE, 2018, pp. 92--97.

\bibitem{sameera2020deep}
N.~Sameera, M.~Shashi, Deep transductive transfer learning framework for zero-day attack detection, ICT Express 6~(4) (2020) 361--367.

\bibitem{hu2022deep}
X.~Hu, C.~Zhu, G.~Cheng, R.~Li, H.~Wu, J.~Gong, A deep subdomain adaptation network with attention mechanism for malware variant traffic identification at an {IoT} edge gateway, IEEE Internet of Things Journal (2022).

\bibitem{zhu2020deep}
Y.~Zhu, F.~Zhuang, J.~Wang, G.~Ke, J.~Chen, J.~Bian, H.~Xiong, Q.~He, Deep subdomain adaptation network for image classification, IEEE transactions on neural networks and learning systems 32~(4) (2020) 1713--1722.

\bibitem{ning2021malware}
J.~Ning, G.~Gui, Y.~Wang, J.~Yang, B.~Adebisi, S.~Ci, H.~Gacanin, F.~Adachi, Malware traffic classification using domain adaptation and ladder network for secure industrial internet of things, IEEE Internet of Things Journal (2021).

\bibitem{wu2022joint}
J.~Wu, Y.~Wang, B.~Xie, S.~Li, H.~Dai, K.~Ye, C.~Xu, Joint semantic transfer network for {IoT} intrusion detection, IEEE Internet of Things Journal (2022).

\bibitem{yao2021multisource}
Y.~Yao, X.~Li, Y.~Zhang, Y.~Ye, Multisource heterogeneous domain adaptation with conditional weighting adversarial network, IEEE Transactions on Neural Networks and Learning Systems (2021).

\bibitem{yao2019heterogeneous}
Y.~Yao, Y.~Zhang, X.~Li, Y.~Ye, Heterogeneous domain adaptation via soft transfer network, in: Proceedings of the 27th ACM international conference on multimedia, 2019, pp. 1578--1586.

\bibitem{yao2020discriminative}
Y.~Yao, Y.~Zhang, X.~Li, Y.~Ye, Discriminative distribution alignment: A unified framework for heterogeneous domain adaptation, Pattern Recognition 101 (2020) 107165.

\bibitem{singla2020preparing}
A.~Singla, E.~Bertino, D.~Verma, Preparing network intrusion detection deep learning models with minimal data using adversarial domain adaptation, in: Proceedings of the 15th ACM Asia Conference on Computer and Communications Security, 2020, pp. 127--140.

\bibitem{zhang2019domain}
Y.~Zhang, J.~Yan, Domain-adversarial transfer learning for robust intrusion detection in the smart grid, in: 2019 IEEE International Conference on Communications, Control, and Computing Technologies for Smart Grids ({SmartGridComm}), IEEE, 2019, pp. 1--6.

\bibitem{unal2022anomalyadapters}
U.~{\"U}nal, H.~Da{\u{g}}, Anomalyadapters: Parameter-efficient multi-anomaly task detection, IEEE Access 10 (2022) 5635--5646.

\bibitem{ajayi2021dahid}
O.~Ajayi, A.~Gangopadhyay, Dahid: Domain adaptive host-based intrusion detection, in: 2021 IEEE International Conference on Cyber Security and Resilience (CSR), IEEE, 2021, pp. 467--472.

\bibitem{nedelkoski2020self}
S.~Nedelkoski, J.~Bogatinovski, A.~Acker, J.~Car{DoS}o, O.~Kao, Self-attentive classification-based anomaly detection in unstructured logs, in: 2020 IEEE International Conference on Data Mining (ICDM), IEEE, 2020, pp. 1196--1201.

\bibitem{niu2019abnormal}
J.~Niu, Y.~Zhang, D.~Liu, D.~Guo, Y.~Teng, Abnormal network traffic detection based on transfer component analysis, in: 2019 IEEE International Conference on Communications Workshops (ICC Workshops), IEEE, 2019, pp. 1--6.

\bibitem{zhao2019transfer}
J.~Zhao, S.~Shetty, J.~W. Pan, C.~Kamhoua, K.~Kwiat, Transfer learning for detecting unknown network attacks, EURASIP Journal on Information Security 2019~(1) (2019) 1--13.

\bibitem{li2021transfer}
Y.~Li, T.~Liu, D.~Jiang, T.~Meng, Transfer-learning-based network traffic automatic generation framework, in: 2021 6th International Conference on Intelligent Computing and Signal Processing (ICSP), IEEE, 2021, pp. 851--854.

\bibitem{fan2020iotdefender}
Y.~Fan, Y.~Li, M.~Zhan, H.~Cui, Y.~Zhang, Iotdefender: A federated transfer learning intrusion detection framework for {5G IoT}, in: 2020 IEEE 14th International Conference on Big Data Science and Engineering ({BigDataSE}), IEEE, 2020, pp. 88--95.

\bibitem{singh2021novel}
N.~B. Singh, M.~M. Singh, A.~Sarkar, J.~K. Mandal, A novel wide \& deep transfer learning stacked gru framework for network intrusion detection, Journal of Information Security and Applications 61 (2021) 102899.

\bibitem{phan2020q}
T.~V. Phan, S.~Sultana, T.~G. Nguyen, T.~Bauschert, $ q $-transfer: A novel framework for efficient deep transfer learning in networking, in: 2020 International Conference on Artificial Intelligence in Information and Communication (ICAIIC), IEEE, 2020, pp. 146--151.

\bibitem{wu2019transfer}
P.~Wu, H.~Guo, R.~Buckland, A transfer learning approach for network intrusion detection, in: 2019 IEEE 4th international conference on big data analytics (ICBDA), IEEE, 2019, pp. 281--285.

\bibitem{wang2021network}
Z.~Wang, Z.~Li, J.~Wang, D.~Li, Network intrusion detection model based on improved byol self-supervised learning, Security and Communication Networks 2021 (2021).

\bibitem{li2020cross}
Y.~Li, X.~Ji, C.~Li, X.~Xu, W.~Yan, X.~Yan, Y.~Chen, W.~Xu, Cross-domain anomaly detection for power industrial control system, in: 2020 IEEE 10th International Conference on Electronics Information and Emergency Communication (ICEIEC), IEEE, 2020, pp. 383--386.

\bibitem{singla2019overcoming}
A.~Singla, E.~Bertino, D.~Verma, Overcoming the lack of labeled data: Training intrusion detection models using transfer learning, in: 2019 IEEE International Conference on Smart Computing (SMARTCOMP), IEEE, 2019, pp. 69--74.

\bibitem{sun2018network}
G.~Sun, L.~Liang, T.~Chen, F.~Xiao, F.~Lang, Network traffic classification based on transfer learning, Computers \& electrical engineering 69 (2018) 920--927.

\bibitem{dhillon2020towards}
H.~Dhillon, A.~Haque, Towards network traffic monitoring using deep transfer learning, in: 2020 IEEE 19th International Conference on Trust, Security and Privacy in Computing and Communications (TrustCom), IEEE, 2020, pp. 1089--1096.

\bibitem{andresini2021network}
G.~Andresini, A.~Appice, C.~Loglisci, V.~Belvedere, D.~Redavid, D.~Malerba, A network intrusion detection system for concept drifting network traffic data, in: International Conference on Discovery Science, Springer, 2021, pp. 111--121.

\bibitem{li2019dart}
H.~Li, Z.~Chen, R.~Spolaor, Q.~Yan, C.~Zhao, B.~Yang, Dart: Detecting unseen malware variants using adaptation regularization transfer learning, in: ICC 2019-2019 IEEE International Conference on Communications (ICC), IEEE, 2019, pp. 1--6.

\bibitem{xiong2020anomaly}
P.~Xiong, B.~Cui, Z.~Cheng, Anomaly network traffic detection based on deep transfer learning, in: International Conference on Innovative Mobile and Internet Services in Ubiquitous Computing, Springer, 2020, pp. 384--393.

\bibitem{zhao2020network}
Y.~Zhao, J.~Chen, Q.~Guo, J.~Teng, D.~Wu, Network anomaly detection using federated learning and transfer learning, in: International Conference on Security and Privacy in Digital Economy, Springer, 2020, pp. 219--231.

\bibitem{yang2021wpd}
T.~Yang, Y.~Hou, Y.~Liu, F.~Zhai, R.~Niu, Wpd-resnest: Substation station level network anomaly traffic detection based on deep transfer learning, CSEE Journal of Power and Energy Systems (2021).

\bibitem{fan2021intrusion}
Y.~Fan, Y.~Li, H.~Cui, H.~Yang, Y.~Zhang, W.~Wang, An intrusion detection framework for {IoT} using partial domain adaptation, in: International Conference on Science of Cyber Security, Springer, 2021, pp. 36--50.

\bibitem{otoum2022federated}
Y.~Otoum, V.~Chamola, A.~Nayak, Federated and transfer learning-empowered intrusion detection for {IoT} applications, IEEE Internet of Things Magazine 5~(3) (2022) 50--54.

\bibitem{cheng2022federated}
Y.~Cheng, J.~Lu, D.~Niyato, B.~Lyu, J.~Kang, S.~Zhu, Federated transfer learning with client selection for intrusion detection in mobile edge computing, IEEE Communications Letters 26~(3) (2022) 552--556.

\bibitem{otoum2021federated}
Y.~Otoum, Y.~Wan, A.~Nayak, Federated transfer learning-based ids for the internet of medical things ({IoMT}), in: 2021 IEEE Globecom Workshops (GC Wkshps), IEEE, 2021, pp. 1--6.

\bibitem{otoum2022feasibility}
S.~Otoum, N.~Guizani, H.~Mouftah, On the feasibility of split learning, transfer learning and federated learning for preserving security in its systems, IEEE Transactions on Intelligent Transportation Systems (2022).

\bibitem{lee2017generative}
H.~Lee, S.~Han, J.~Lee, Generative adversarial trainer: Defense to adversarial perturbations with gan, arXiv preprint arXiv:1705.03387 (2017).

\bibitem{zhang2020two}
Y.~Zhang, Y.~Song, J.~Liang, K.~Bai, Q.~Yang, Two sides of the same coin: White-box and black-box attacks for transfer learning, in: Proceedings of the 26th ACM SIGKDD International Conference on Knowledge Discovery \& Data Mining, 2020, pp. 2989--2997.

\bibitem{zhang2020semi}
Y.~Zhang, J.~Yan, Semi-supervised domain-adversarial training for intrusion detection against false data injection in the smart grid, in: 2020 International Joint Conference on Neural Networks (IJCNN), IEEE, 2020, pp. 1--7.

\bibitem{sharaf2016authentication}
Y.~Sharaf-Dabbagh, W.~Saad, On the authentication of devices in the internet of things, in: 2016 IEEE 17th International Symposium on A World of Wireless, Mobile and Multimedia Networks (WoWMoM), IEEE, 2016, pp. 1--3.

\bibitem{dabbagh2019authentication}
Y.~S. Dabbagh, W.~Saad, Authentication of wireless devices in the internet of things: Learning and environmental effects, IEEE Internet of Things Journal 6~(4) (2019) 6692--6705.

\bibitem{zhao2018identification}
C.~Zhao, Z.~Cai, M.~Huang, M.~Shi, X.~Du, M.~Guizani, The identification of secular variation in {IoT} based on transfer learning, in: 2018 International Conference on Computing, Networking and Communications (ICNC), IEEE, 2018, pp. 878--882.

\bibitem{sharaf2015transfer}
Y.~Sharaf-Dabbagh, W.~Saad, Transfer learning for device fingerprinting with application to cognitive radio networks, in: 2015 IEEE 26th Annual International Symposium on Personal, Indoor, and Mobile Radio Communications (PIMRC), IEEE, 2015, pp. 2138--2142.

\bibitem{yu2019individual}
Z.~Yu, R.~Zong, S.~Li, Individual recognition based on transfer learning for wireless network devices, in: Proceedings of the 2019 3rd International Conference on Advances in Image Processing, 2019, pp. 45--49.

\bibitem{wang2022intrusion}
K.~Wang, J.~Li, An intrusion detection method integrating {KNN} and transfer extreme learning machine, in: 2022 2nd Asia-Pacific Conference on Communications Technology and Computer Science (ACCTCS), IEEE, 2022, pp. 221--226.

\bibitem{khoa2021deep}
T.~V. Khoa, D.~T. Hoang, N.~L. Trung, C.~T. Nguyen, T.~T.~T. Quynh, D.~N. Nguyen, N.~V. Ha, E.~Dutkiewicz, Deep transfer learning: A novel collaborative learning model for cyberattack detection systems in {IoT} networks, arXiv preprint arXiv:2112.00988 (2021).

\bibitem{guan2021deep}
J.~Guan, J.~Cai, H.~Bai, I.~You, Deep transfer learning-based network traffic classification for scarce dataset in {5G IoT} systems, International Journal of Machine Learning and Cybernetics 12~(11) (2021) 3351--3365.

\bibitem{otoum2021signature}
Y.~Otoum, A.~Nayak, Signature-over-the-air with transfer learning ids for intelligent connected vehicles (icv), in: 2021 IEEE Globecom Workshops (GC Wkshps), IEEE, 2021, pp. 1--6.

\bibitem{zhou2019maximum}
M.~Zhou, Y.~Li, L.~Xie, W.~Nie, Maximum mean discrepancy minimization based transfer learning for indoor wlan personnel intrusion detection, IEEE Sensors Letters 3~(8) (2019) 1--4.

\bibitem{zhou2019indoor}
M.~Zhou, Y.~Li, X.~Huang, Q.~Pu, H.~Yuan, Indoor wlan intrusion detection using intra-class transfer learning with low effort, in: 2019 IEEE 30th Annual International Symposium on Personal, Indoor and Mobile Radio Communications (PIMRC), IEEE, 2019, pp. 1--6.

\bibitem{zhou2021indoor}
M.~Zhou, Y.~Li, H.~Yuan, J.~Wang, Q.~Pu, Indoor wlan personnel intrusion detection using transfer learning-aided generative adversarial network with light-loaded database, Mobile Networks and Applications 26~(3) (2021) 1024--1042.

\bibitem{li2019integrated}
X.~Li, M.~Zhou, Y.~Li, H.~Yuan, Z.~Tian, Integrated redundant aps reduction and transfer learning for indoor wlan intrusion detection via link-layer data transformation, in: China Conference on Wireless Sensor Networks, Springer, 2019, pp. 251--262.

\bibitem{beaubouef2000fuzzy}
T.~Beaubouef, F.~E. Petry, Fuzzy rough set techniques for uncertainty processing in a relational database, International Journal of Intelligent Systems 15~(5) (2000) 389--424.

\bibitem{xia2021wireless}
Y.~Xia, S.~Dong, T.~Peng, T.~Wang, Wireless network abnormal traffic detection method based on deep transfer reinforcement learning, in: 2021 17th International Conference on Mobility, Sensing and Networking (MSN), IEEE, 2021, pp. 528--535.

\bibitem{ge2021towards}
M.~Ge, N.~F. Syed, X.~Fu, Z.~Baig, A.~Robles-Kelly, Towards a deep learning-driven intrusion detection approach for internet of things, Computer Networks 186 (2021) 107784.

\bibitem{yilmaz2021transfer}
S.~Y{\i}lmaz, E.~Aydogan, S.~Sen, A transfer learning approach for securing resource-constrained {IoT} devices, IEEE Transactions on Information Forensics and Security 16 (2021) 4405--4418.

\bibitem{ullah2021design}
I.~Ullah, Q.~H. Mahmoud, Design and development of a deep learning-based model for anomaly detection in {IoT} networks, IEEE Access 9 (2021) 103906--103926.

\bibitem{mehedi2022dependable}
S.~T. Mehedi, A.~Anwar, Z.~Rahman, K.~Ahmed, R.~Islam, Dependable intrusion detection system for {IoT}: A deep transfer learning based approach, IEEE Transactions on Industrial Informatics 19~(1) (2022) 1006--1017.

\bibitem{li2020transfer}
S.~Li, T.~T. Cai, H.~Li, Transfer learning in large-scale gaussian graphical models with false discovery rate control, arXiv preprint arXiv:2010.11037 (2020).

\bibitem{kang2021car}
H.~Kang, B.~Kwak, Y.~Lee, H.~Lee, H.~Lee, H.~Kim, Car hacking: Attack and defense challenge 2020 dataset, IEEE Dataport (2021).

\bibitem{li2020deep}
F.~Li, K.~Shirahama, M.~A. Nisar, X.~Huang, M.~Grzegorzek, Deep transfer learning for time series data based on sensor modality classification, Sensors 20~(15) (2020) 4271.

\bibitem{kimura2020convolutional}
N.~KIMURA, D.~BABA, Convolutional neural network (cnn)-based transfer learning implemented to time-series flood forecast, in: Proceedings of the 22nd IAHR APD Congress (Sapporo, 2020), 2020.

\bibitem{xiong2021anomaly}
P.~Xiong, B.~Cui, Z.~Cheng, Anomaly network traffic detection based on deep transfer learning, in: International Conference on Innovative Mobile and Internet Services in Ubiquitous Computing, Springer, 2021, pp. 384--393.

\bibitem{jung2021pf}
I.~Jung, J.~Lim, H.~K. Kim, Pf-tl: Payload feature-based transfer learning for dealing with the lack of training data, Electronics 10~(10) (2021) 1148.

\bibitem{catillo2022transferability}
M.~Catillo, A.~Del~Vecchio, A.~Pecchia, U.~Villano, Transferability of machine learning models learned from public intrusion detection datasets: the cicids2017 case study, Software Quality Journal (2022) 1--27.

\bibitem{li2020intrusion}
J.~Li, W.~Wu, D.~Xue, An intrusion detection method based on active transfer learning, Intelligent Data Analysis 24~(2) (2020) 363--383.

\bibitem{zhang2019autonomous}
J.~Zhang, F.~Li, H.~Wu, F.~Ye, Autonomous model update scheme for deep learning based network traffic classifiers, in: 2019 IEEE Global Communications Conference (GLOBECOM), IEEE, 2019, pp. 1--6.

\bibitem{luo2018tinet}
C.~Luo, Z.~Chen, L.-A. Tang, A.~Shrivastava, Z.~Li, H.~Chen, J.~Ye, Tinet: Learning invariant networks via knowledge transfer, in: Proceedings of the 24th ACM SIGKDD International Conference on Knowledge Discovery \& Data Mining, 2018, pp. 1890--1899.

\bibitem{he2020small}
J.~He, Y.~Tan, W.~Guo, M.~Xian, A small sample {DDoS} attack detection method based on deep transfer learning, in: 2020 International Conference on Computer Communication and Network Security ({CCNS}), IEEE, 2020, pp. 47--50.

\bibitem{shafee2020mimic}
A.~Shafee, M.~Baza, D.~A. Talbert, M.~M. Fouda, M.~Nabil, M.~Mahmoud, Mimic learning to generate a shareable network intrusion detection model, in: 2020 IEEE 17th Annual Consumer Communications \& Networking Conference (CCNC), IEEE, 2020, pp. 1--6.

\bibitem{ning2021novel}
J.~Ning, Y.~Wang, J.~Yang, H.~Gacanin, S.~Ci, A novel malware traffic classification method using semi-supervised learning, in: 2021 IEEE 94th Vehicular Technology Conference (VTC2021-Fall), IEEE, 2021, pp. 1--5.

\bibitem{xu2021privacy}
M.~Xu, X.~Li, Y.~Wang, B.~Luo, J.~Guo, Privacy-preserving multisource transfer learning in intrusion detection system, Transactions on Emerging Telecommunications Technologies 32~(5) (2021) e3957.

\bibitem{ahmadi2018intrusion}
R.~Ahmadi, R.~D. Macredie, A.~Tucker, Intrusion detection using transfer learning in machine learning classifiers between non-cloud and cloud datasets, in: International Conference on Intelligent Data Engineering and Automated Learning, Springer, 2018, pp. 556--566.

\bibitem{ranjithkumar2022fuzzy}
S.~Ranjithkumar, S.~C. Pandian, Fuzzy based latent dirichlet allocation for intrusion detection in cloud using ml, CMC-COMPUTERS MATERIALS \& CONTINUA 70~(3) (2022) 4261--4277.

\bibitem{li4254447eifdaa}
S.~Li, J.~Wang, Y.~Wang, G.~Zhou, Y.~Zhao, Eifdaa: Evaluation of an ids with function-discarding adversarial attacks in the iiot, Heliyon 9~(2) (2023) e13520.

\bibitem{liu2019adversarial}
X.~Liu, X.~Du, X.~Zhang, Q.~Zhu, H.~Wang, M.~Guizani, Adversarial samples on android malware detection systems for iot systems, Sensors 19~(4) (2019) 974.

\bibitem{papadopoulos2021launching}
P.~Papadopoulos, O.~Thornewill~von Essen, N.~Pitropakis, C.~Chrysoulas, A.~Mylonas, W.~J. Buchanan, Launching adversarial attacks against network intrusion detection systems for iot, Journal of Cybersecurity and Privacy 1~(2) (2021) 252--273.

\bibitem{alazab2023routing}
A.~Alazab, A.~Khraisat, S.~Singh, S.~Bevinakoppa, O.~A. Mahdi, Routing attacks detection in 6lowpan-based internet of things, Electronics 12~(6) (2023) 1320.

\bibitem{aldaej2019enhancing}
A.~Aldaej, Enhancing cyber security in modern internet of things (iot) using intrusion prevention algorithm for iot (ipai), IEEE Access (2019).

\bibitem{sadhu2022internet}
P.~K. Sadhu, V.~P. Yanambaka, A.~Abdelgawad, Internet of things: Security and solutions survey, Sensors 22~(19) (2022) 7433.

\bibitem{singh2020deep}
S.~K. Singh, Y.-S. Jeong, J.~H. Park, A deep learning-based {IoT}-oriented infrastructure for secure smart city, Sustainable Cities and Society 60 (2020) 102252.

\bibitem{apruzzese2022modeling}
G.~Apruzzese, M.~Andreolini, L.~Ferretti, M.~Marchetti, M.~Colajanni, Modeling realistic adversarial attacks against network intrusion detection systems, Digital Threats: Research and Practice (DTRAP) 3~(3) (2022) 1--19.

\bibitem{nwakanma2023explainable}
C.~I. Nwakanma, L.~A.~C. Ahakonye, J.~N. Njoku, J.~C. Odirichukwu, S.~A. Okolie, C.~Uzondu, C.~C. Ndubuisi~Nweke, D.-S. Kim, Explainable artificial intelligence (xai) for intrusion detection and mitigation in intelligent connected vehicles: A review, Applied Sciences 13~(3) (2023) 1252.

\bibitem{liu2021faixid}
H.~Liu, C.~Zhong, A.~Alnusair, S.~R. Islam, Faixid: A framework for enhancing ai explainability of intrusion detection results using data cleaning techniques, Journal of network and systems management 29~(4) (2021) 40.

\bibitem{alhakami2019network}
W.~Alhakami, A.~ALharbi, S.~Bourouis, R.~Alroobaea, N.~Bouguila, Network anomaly intrusion detection using a nonparametric bayesian approach and feature selection, IEEE Access 7 (2019) 52181--52190.

\bibitem{sohail2023future}
S.~S. Sohail, F.~Farhat, Y.~Himeur, M.~Nadeem, D.~{\O}. Madsen, Y.~Singh, S.~Atalla, W.~Mansoor, The future of gpt: A taxonomy of existing chatgpt research, current challenges, and possible future directions, Current Challenges, and Possible Future Directions (April 8, 2023) (2023).

\bibitem{tian2023cldtlog}
G.~Tian, N.~Luktarhan, H.~Wu, Z.~Shi, Cldtlog: System log anomaly detection method based on contrastive learning and dual objective tasks, Sensors 23~(11) (2023) 5042.

\bibitem{sohail2023using}
S.~S. Sohail, D.~{\O}. Madsen, Y.~Himeur, M.~Ashraf, Using chatgpt to navigate ambivalent and contradictory research findings on artificial intelligence, Available at SSRN 4413913 (2023).

\end{thebibliography}

\end{document}